\documentclass[sigconf,10pt]{acmart}

\usepackage[T1]{fontenc}
\usepackage[dvipsnames]{xcolor}
\usepackage[utf8]{inputenc}
\usepackage{amsmath,bbold}
\usepackage{amsthm}
\usepackage{listings}
\usepackage{array}
\usepackage{graphicx}
\graphicspath{ {./figures/} }
\usepackage{tikz}
\usepackage{varwidth}
\usepackage{graphicx}
\usepackage[ruled, linesnumbered]{algorithm2e}
\usepackage{enumitem}
\usepackage[skins, fitting]{tcolorbox}
\usepackage{subcaption}
\tcbuselibrary{breakable}
\usepackage{makecell}
\usepackage{chngcntr}
\usepackage[normalem]{ulem}
\theoremstyle{acmdefinition}

\newif\iflong 
 \longtrue

\iflong
    \renewcommand\footnotetextcopyrightpermission[1]{}
\fi

\newtcbox{\emphbox}{breakable,
arc=0pt,outer arc=0pt,colback=red!10!white, boxsep=2pt, boxrule=1pt,left=1pt,right=1pt,top=1pt,bottom=1pt,bottomrule=0pt,toprule=0pt,leftrule=0pt, rightrule=0pt,}
\definecolor{dkgreen}{rgb}{0,0.6,0}
\definecolor{dkgray}{rgb}{0.75,0.75,0.75}

\newcommand{\edit}[1]{#1}
\newcommand{\todo}[1]{}
\newcommand{\note}[1]{}

\newcommand{\reviewone}[1]{#1}
\newcommand{\reviewtwo}[1]{#1}
\newcommand{\reviewthree}[1]{#1}
\newcommand{\onethree}[1]{#1}

\setlist{  
  listparindent=\parindent,
  parsep=0pt,
}

\title{Version Reconciliation for Collaborative Databases}

\lstdefinestyle{inenum}{belowskip=-7pt}

\newcommand*\cir[1]{\raisebox{2pt}{\tikz[baseline=(char.base)]{
            \node[shape=circle,draw,inner sep=.1pt] (char) {#1};}}}
\newcommand{\cirdown}[1]{{\textcircled{{#1}}}}
\newtcbox{\logicbox}[2]{on line,detach title,before upper={\textcolor{black}{\tcbtitletext}},frame hidden,
colback=#1!20!white,
boxsep=1pt,left=0pt,right=0pt,top=0pt,bottom=0pt,boxrule=0pt,
title={\scriptsize{}#2}}

\newcommand{\dvds}{DVDS\xspace}
\newcommand{\sys}{MindPalace\xspace}
\newcommand{\update}{modification\xspace}
\newcommand{\updates}{modifications\xspace}
\newcommand{\amle}{auto-mergeable\xspace}

\newcommand{\amty}{auto-mergeability\xspace}
\newcommand{\Amty}{Auto-mergeability\xspace}

\newcommand{\namle}{non-auto-mergeable\xspace}

\newcommand{\inter}{interleaving\xspace}
\newcommand{\inters}{interleavings\xspace}
\newcommand{\pref}[2]{\text{\texttt{pref}}_#1(#2)}
\newcommand{\ser}{H_1H_2}
\newcommand{\inlinesql}[1]{\texttt{\small{#1}}}

\newcommand{\colA}{\texttt{A}}
\newcommand{\colB}{\texttt{B}}
\newcommand{\valA}{\texttt{a}}
\newcommand{\valB}{\texttt{b}}

\newcommand{\colK}{\texttt{K}}
\newcommand{\colL}{\texttt{L}}
\newcommand{\valK}{\texttt{k}}
\newcommand{\valL}{\texttt{l}}
\newcommand{\colS}{\texttt{S}}
\newcommand{\valS}{\texttt{s}}
\newcommand{\colC}{\texttt{C}}

\copyrightyear{2021}
\acmYear{2021}
\setcopyright{acmlicensed}\acmConference[SoCC '21]{ACM Symposium on Cloud Computing}{November 1--4, 2021}{Seattle, WA, USA}
\acmBooktitle{ACM Symposium on Cloud Computing (SoCC '21), November 1--4, 2021, Seattle, WA, USA}
\acmPrice{15.00}
\acmDOI{10.1145/3472883.3486980}
\acmISBN{978-1-4503-8638-8/21/11}

\setlist[itemize]{align=parleft,left=0pt..1.2em}
\setlist[enumerate]{align=parleft,left=0pt..1.2em}

\newcommand{\stitle}[1]{\vspace{.5ex} \noindent{\bf #1}}

\begin{document}
\author{Nalin Ranjan}
\affiliation{%
  \institution{Princeton University}
}
\email{nranjan@princeton.edu}

\author{Zechao Shang}
\authornote{Work was done mainly when the author was at the University of
Chicago.}
\email{zechao.shang@snowflake.com}
\affiliation{%
  \institution{Snowflake Inc.}
}

\author{Aaron J. Elmore}
\email{aelmore@cs.uchicago.edu}
\author{Sanjay Krishnan}
\email{skr@uchicago.edu}
\affiliation{%
  \institution{University of Chicago}
}

\begin{abstract}
We propose \sys, a \reviewtwo{prototype of a versioned database} for efficient collaborative data management.
\sys supports \textit{offline} collaboration, where users work independently without real-time correspondence. The core of \sys is a critical step of offline collaboration: reconciling divergent branches made by simultaneous data manipulation. We formalize the concept of \amty, a condition under which branches may be reconciled without human intervention, and propose an efficient framework for determining whether two branches are \amle and identifying particular records for manual reconciliation.
\end{abstract}

\maketitle


\setlength{\abovedisplayskip}{1pt}
\setlength{\belowdisplayskip}{1pt}

\section{Introduction}
\label{intro}
\noindent Data analytics is never a one-person job. Analysts working on the same dataset collaborate~\cite{pachyderm,dolt,daff,noms,dvc} in cleaning, transforming, and manipulating data. \
However, conventional wisdom from collaborative software development (e.g., Git) suggests that collaboration is better in an \textit{offline} way, i.e., mainly on disconnected copies of data without any real-time coordination with a central server. In the same spirit, we argue collaborative data science should also be offline. Beyond the obvious scalability benefits from the elimination of centralized coordination~\cite{terry1995managing}, a user can work at her own pace and use any data wrangling/analytical software that she prefers, without interruption from other concurrent users, high latency from excessive real-time data transmission, or potential availability issues (e.g. those associated with keeping a live connection).
Like with Git, after a user is done, it is her responsibility to push changes to the primary repository and merge these changes with any  changes to the dataset that happened in the interim.
Unfortunately, we find that existing dataset versioning systems~\cite{bhardwaj2014datahub,maddox2016decibel,huang2017orpheus,wang2018forkbase} crucially lack support for \emph{offline versioning}.

  

We present a system, \sys, to facilitate such a workflow.
\sys is a collaborative  distributed versioned data system (DVDS) that implements such an offline versioning system. 
\sys supports a \textit{clone} operation that makes a physical copy of a desired dataset. Users can modify {(i.e., issue SQL \lstinline{UPDATE}s on individual or batches of tuples)} independent and disconnected copies \textit{offline}, and then later \textit{merge} branches (analogous to a branch merging in Git). 
A merge involves two major steps: automated \textit{detection} of conflicts (i.e., those data that cannot be merged automatically by the system) amongst different branches and \textit{reconciliation} of identified conflicts (which will require user interaction).
While our target application is collaborative analytics, the ideas are applicable to a variety of systems or applications that support partial asynchronous replication.

The key contribution of \sys is an algorithmic framework to implement the most critical step of offline \dvds, the \textit{merge} feature.
We identify problems (``conflicts'') a merge may encounter, and demonstrate a framework for identifying and resolving them based on a new formal property that we call \textit{\amty}.
Informally, two branches are \amle if the system can merge them without a user's intervention. Specifically, \amty requires that a final database state does not depend on the choice of \textit{\inter} of operations, or a total ordering of two branches' \updates that respects local orderings.
While users' or applications' demands may differ, we find that our formulation of identifying and dealing with conflicts is more than sufficient for a variety of use cases. We discuss the benefits of such a scheme in Section \ref{principles}.

We contribute an efficient algorithm to detect conflicts of \amty in two branches.
\reviewtwo{It 
considers a branch as a sequence of logical \updates (i.e., \lstinline{UPDATE} statements) instead of a set of physical record operations (i.e., `change line 357 from \textit{Revision} to \textit{Accept}'), which prior data versioning systems~\cite{maddox2016decibel,huang2017orpheus,wang2018forkbase} \reviewtwo{and \texttt{diff}-based method} rely on. Logical update tracking is highly efficient, as it avoids transferring large volumes of data across machines to perform a merge. Moreover, in analyzing logical semantics of \updates to identify precise conditions on conflicting data, our algorithm minimizes the number of time-consuming physical database operations.
Besides the efficiency gain, logical tracking allows for reasoning about scenarios where analysts wish to manipulate partial copies of a dataset (e.g., a random sample) and capture intentions via predicates (e.g., change all `US' to `USA'). 
While partial copies may not physically overlap, users' operations may still conflict on records in the primary dataset that were not in the partial replicas.  }

When two branches are not \amle, they need additional user input to resolve identified conflicts. However, it is almost infeasible to do so manually, as the number of valid \inters grows exponentially in the number of \updates. We propose an interaction framework that gradually reconciles versions, terminating once the remaining unreconciled portions of the histories are \amle. In every iteration, we only ask users to choose a preferred order between two \updates. Ultimately, the version reconciliation procedure proposed is designed to optimize for 1) clarity, by identifying precisely which records are in conflict and reproducing them for the user's reference; 2) conciseness, by aiming to minimize the amount of human input demanded; and 3) simplicity, by only presenting manageable decisions to users when input is required.

In summary, \sys provides users with a principled and scalable way to merge  data updates.
We present four major technical contributions.
    First, we analyze and formalize the problem of automatic merging. We define \amty as a necessary condition of merging without intervention.
    Second, we propose a scalable and efficient system, \sys, to detect \amty and identify \namle tuples if any. 
    Third, we propose a resolution framework that presents pairwise {prompts} to a user to reconcile \namle versions. It prioritizes resolution of ``early'' conflicts, thus avoiding potential redundancies.
    Fourth, we implement \sys as \reviewtwo{an auxiliary client} to PostgreSQL and compare it to alternative approaches. \reviewone{\sys supports a wide range of \lstinline{UPDATE}s, \lstinline{INSERT}s, and \lstinline{DELETE}s (see Section~\ref{more_general} for details). } In many diverse scenarios, \sys performs better, both in identifying a tight set of conflicting records and in total runtime.


\section{Motivation and Overview}

\noindent In this section, we start with a motivating example, discuss the desired properties of a version merging system, and introduce a proposed model of user interaction with a \dvds.

\vspace{-0.3cm}
\subsection{Motivating Example} \label{motivating_example}
We discuss a setting where two users collaborate in cleaning \reviewone{and preparing data}. Two data analysts, Alvarez and Bano, work on an energy dataset (Table~\ref{tab:example-1}) to build their ML model. The dataset records annual electricity consumption, population, city name, state, and other relevant information for every city. Upon downloading the data, Alvarez immediately notices that in all California cities there is a critical mistake: electricity consumption was reported in TWh instead of GWh. To fix this, he multiplies all data in this column by 1000 ($A_1$: \lstinline{UPDATE SET Electricity = Electricity * 1000 WHERE State = 'CA'}). He then decides that small suburbs are not the focus of their analysis and issues ($A_2$: \lstinline{DELETE FROM Table WHERE Population <= 0.2}). Finally, he uploads all his updates from this session (Table~\ref{tab:example-a}). Meanwhile, after downloading a local copy of the data Bano notices that several cities' electricity data are invalid as 0, and fills them with data from other sources ($B_1$: \lstinline{UPDATE SET Electricity = 9 WHERE City = 'San Jose'}; $B_2$: \lstinline{UPDATE SET Electricity = 0.4 WHERE City = 'Burbank'}). After that, she realizes that the cities with low energy consumption per capita would be outliers for their analytics and issues the command ($B_3$: \lstinline{DELETE FROM Table WHERE Electricity / Population < 10}). Upon uploading her \updates (Table~\ref{tab:example-b}), she notices that a merge was needed with Alvarez's committed \updates.

In this example, both users had two ``phases'' of work: each \reviewone{fixed} the data in the first few \updates ($A_1$, $B_1$, and $B_2$), and \reviewone{prepared the data} in the subsequent \updates ($A_2$ and $B_3$). Ideally, the data fixing operations should happen before the \reviewone{preparation} ones, so one of the desired ways to reconcile them is to apply operations on data in the order of $\{B_1, B_2, A_1, A_2, B_3\}$ (Table~\ref{tab:example-right}). We emphasize that if the operations are not performed in an ideal order, undesirable final states of data occur: for example, simply concatenating both users' \updates in a serial order, i.e. $\{A_1, A_2, B_1, B_2, B_3\}$ (or even $\{B_1, B_2, B_3, A_1, A_2\}$) incorrectly removes San Jose from the dataset (Table~\ref{tab:example-wrong}).

\begin{table}[t]
\caption{Visualization of Example \ref{motivating_example}. Merging two branches (\ref{tab:example-a} and \ref{tab:example-b}) requires careful inspection (\ref{tab:example-right}); arbitrary or default reconciliation may cause data corruption (\ref{tab:example-wrong}). Names of cities are abbreviated in subsequent tables. } 
\vspace*{-.3cm}
\small
\subcaptionbox{The Initial Dataset \label{tab:example-1}}{
\vspace*{-.35cm}
\begin{tcolorbox}[breakable,
arc=0pt,outer arc=0pt,colback=black!5!white, colframe=white!5!white, boxsep=1pt, boxrule=0pt,left=0pt,right=0pt,top=1pt,bottom=1pt,bottomrule=12pt,toprule=-4pt,leftrule=-3pt, width=8.3cm]
\centering
\begin{tabular}{  c   c   c   c } 
\hline
\textbf{Name} & \textbf{State} & \textbf{Population (M)} & \textbf{Electricity (GWh)}  \\
 \hline
 Los Angles & CA & 3.2 & 43  \\
 \hline
 Seattle & D.C. & 0.6 & 8,709 \\
 \hline
 Burbank & CA & 0.1 & 0  \\
 \hline
 San Jose & CA & 1.0 & 0  \\
 \hline
\end{tabular}
\end{tcolorbox}}
\setlength\tabcolsep{3pt}
\subcaptionbox{After Alvarez's Updates \label{tab:example-a}}{
\vspace*{.25cm}
\begin{tcolorbox}[breakable,
arc=0pt,outer arc=0pt,colback=black!5!white, colframe=white!5!white, boxsep=1pt, boxrule=0pt,left=0pt,right=0pt,top=1pt,bottom=1pt,toprule=-4pt,leftrule=0pt, bottomrule=-4pt, rightrule=0pt,width=4cm]
\centering
\begin{tabular}{  c   c   c   c } 
\hline
\textbf{Name} & \textbf{State} & \textbf{Pop} & \textbf{Elec}  \\
 \hline
 LA & CA & 3.2 & 43,000  \\
 \hline
 S & D.C. & 0.6 & 8,709 \\
 \hline
\end{tabular}
\end{tcolorbox}}
\subcaptionbox{After Bano's Updates \label{tab:example-b}}{
\begin{tcolorbox}[breakable,
arc=0pt,outer arc=0pt,colback=black!5!white, colframe=white!5!white, boxsep=1pt, boxrule=0pt,left=0pt,right=0pt,top=1pt,bottom=1pt,toprule=-4pt,bottomrule=-4pt,leftrule=0pt, rightrule=0pt,width=4cm]
\centering
\begin{tabular}{  c   c   c   c } 
\hline
\textbf{Name} & \textbf{State} & \textbf{Pop} & \textbf{Elec}   \\
 \hline
 LA & CA & 3.2 & 43  \\
 \hline
 S & D.C. & 0.6 & 8,709 \\
 \hline
\end{tabular}
\end{tcolorbox}}
\subcaptionbox{After $\{B_1, B_2, A_1, A_2, B_3\}$ \label{tab:example-right}}{
\begin{tcolorbox}[breakable,
arc=0pt,outer arc=0pt,colback=black!5!white, colframe=white!5!white, boxsep=1pt, boxrule=0pt,left=0pt,right=0pt,top=1pt,bottom=1pt,bottomrule=4pt,toprule=-4pt,leftrule=0pt, rightrule=0pt, width=4cm]
\centering
\begin{tabular}{  c   c   c   c } 
\hline
\textbf{Name} & \textbf{State} & \textbf{Pop} & \textbf{Elec}   \\
 \hline
 LA & CA & 3.2 & 43,000  \\
 \hline
 S & D.C. & 0.6 & 8,709 \\
 \hline
 SJ & CA & 1.0 & 9,000  \\
 \hline
 \end{tabular}
 \end{tcolorbox}}
\subcaptionbox{After $\{A_1, A_2, B_1, B_2, B_3\}$ \label{tab:example-wrong}}{
\vspace*{-.25cm}
\begin{tcolorbox}[breakable,
arc=0pt,outer arc=0pt,colback=black!5!white, colframe=white!5!white, boxsep=1pt, boxrule=0pt,left=0pt,right=0pt,top=1pt,bottom=1pt,bottomrule=4pt,toprule=-4pt,leftrule=0pt, rightrule=0pt, width=4cm]
\centering
\begin{tabular}{  c   c   c   c } 
\hline
\textbf{Name} & \textbf{State} & \textbf{Pop} & \textbf{Elec}   \\
 \hline
 LA & CA & 3.2 & 43,000  \\
 \hline
 S & D.C. & 0.6 & 8,709 \\
 \hline
\end{tabular}
\end{tcolorbox}}


\vspace*{-0.65cm}
\end{table}


\subsection{Merging and Conflicts: Principles}
\label{principles}

Like with Git, an offline versioning system for data crucially requires a methodology for merging the modifications of different users. Different version control systems may have different definitions of what constitutes a valid merge; indeed, this is a design choice that will depend on the use case. For example, a valid merging strategy in Git involves taking two sets of committed physical modifications and determining a (single) final physical state of each modified line of code, ignoring any lines that were not modified. 

Any version control system must be equipped to deal with conflicts should they arise. A conflict (relative to a set of valid merging strategies) occurs if not every valid merging strategy results in the same final state of the data. Keeping with the Git example, a conflict occurs when two users Alvarez and Bano modify the same line of code differently, because multiple valid merging strategies yielding different outcomes exist: merges could preserve Alvarez's updates over Bano's or Bano's over Alvarez's. In other words, the state of the data post-merge \textit{depends on which merging strategies were allowed~\cite{natacha2017} and which one was chosen}. Git automatically merges if there are no conflicting lines and demands manual line-by-line merging otherwise.

We argue that for datasets, defining a valid merge as a total ordering of all \updates \reviewone{while preserving the local orders among each user's \updates} is suitable for many settings (in data science and beyond). For instance, 
valid merges of the \updates in Example \ref{motivating_example} (total orderings) might include
$\{A_1, A_2, B_1, B_2, B_3\}$ or $\{B_1, A_1, A_2, B_2, B_3\}$,
but not 
$\{A_2, A_1, B_2, B_1, B_3\}$. 
This definition has multiple advantages.
\reviewtwo{First, modeling the \updates as logical operations (e.g., SQL \lstinline{UPDATE} statements) instead of physical records (i.e., reads and writes)  captures dependencies between different users' updates that may not be reflected just by examining the data itself. Unlike Git or Google Sheets, which just examines the physical writes, logical operations captures precise read-write dependencies. }
Second, by finding a total order for \textit{all} the updates (e.g., as opposed to a merging strategy that just deletes updates until no conflicts exist anymore), it represents the \updates of users as much as possible.
Third, by mandating that total orders respect local orders, it guarantees a user that her updates appear in the order she issued them. In doing so, the offline versioning system retroactively emulates an online centralized system with concurrent updates from multiple users.
Finally, this formulation allows for a lightweight, fine-grained resolution of conflicts. In the above example, if it turns out that Alvarez's and Bano's \updates conflict, often just specifying a few simple constraints such as ``$A_1$ must precede $B_1$'' suffices in resolving all conflicts (to be discussed in greater detail in future Section~\ref{resolution}).

We again emphasize that the merging framework outlined in this section need not be the only valid framework for merging updates. It is one such framework that provides guarantees commonly needed in a variety of settings. The rest of this paper will offer solutions to the challenges (namely detecting and resolving conflicts) under this framework.

\subsection{\dvds Overview}

\noindent A \dvds must support certain fundamental Git-like operations.
A \textbf{clone} creates a remote physical copy of a dataset repository locally. Since the dataset can be large, a user may specify which portions of the dataset of which he wishes to clone, such as a random sample. 
A user can perform a sequence of atomic changes, or \textbf{commits}, to a local dataset. For simplicity, in this paper we assume each commit contains one SQL \lstinline{UPDATE} statement. A commit could change individual tuple, or perform batch modifications, e.g., ``change all \textit{US} to \textit{USA}''. We emphasize that like Git, the local commits are visible to nobody except the local user until they are pushed. 
A \textbf{push} uploads one or more \textit{commits} to the primary repository. 
When pushing commits to the primary repository, a \textbf{merge} must occur if there are concurrent commits. \edit{If there are multiple concurrent branches, a merge must occur every time the user merges one branch to the master, or merges two branches.}
Depending on the context, this operation may be restricted to users with administrative privileges.

\section{Auto-Mergeability}
\label{sec-amty-def}

\noindent First, \edit{we formally define \amty, and explain how this definition is crucial for reliable reconciliation in a \dvds}. 

\begin{table}[!tbp]\small
\caption{Notations and corresponding descriptions.}
\vspace*{-.3cm}
\label{notationtable}
\begin{center}
\begin{tabular}{  r l  } 
 \hline
 \textbf{Symbol} & \textbf{Description and Use} \\
 \hline
 $H$ & A history as a sequence of \updates \\
 \hline
 $m, n$ & Lengths of histories \\
 \hline
 $\phi_i$, $\psi_j$ / $\phi$, $\psi$ & Modifications in a given history.  \\
 \hline
 $D$ (or $D_0$) & A (initial) version of a database \\
 \hline
 $D_H$ (or $D_{\phi}$) & State of DB after \updates $H$ (or $\phi$) \\
 \hline
 $U[Q]$ (or $U_[C]$) & $U$'s tuples that satisfy $Q$'s predicate (or $C$) \\
 \hline
 $t$ & A single tuple \\
 \hline
 \reviewthree{\colA, \colB, $\cdots$} & \reviewthree{Attributes of tuples} \\
 \hline
 \reviewthree{\valA, \valB, $\cdots$} & \reviewthree{Values of attributes} \\
 \hline
 pref$_k(H)$ & The first $k$ \updates from history $H$ \\
 \hline
 $H_1H_2$ & Concatenation of histories to form another \\
 \hline
 $\phi_i < \phi_j$ & $\phi_i$ precedes $\phi_j$ in some history $H$\\
 \hline
 $\Sigma^c$ & Complement of the set $\Sigma$ \\
 \hline
 $R \bigtriangleup S$ & Symmetric difference (=$R\backslash S\cup S\backslash R$) \\\hline
\end{tabular}
\end{center}\vspace*{-.5cm}
\end{table}

\subsection{Property Definition}
\edit{\Amty is a property of two users' update histories.
It stipulates that all possible interleavings of the operations in those histories result in the same final database state. }
Let $\phi(\cdot)$ denote a single \update operation to a dataset $D$.
\edit{We call an ordered sequence of \updates a \textbf{history} $H = \phi_1\phi_2\cdots\phi_m$}. 
\edit{The application of this sequence of updates to a dataset results in a final \emph{state} of a history $H$: 
\begin{equation}
    D_H = \phi_m(\dots\phi_2(\phi_1(D)))
\end{equation}   
Table \ref{notationtable} contains all of notation and terminology that we use throughout the paper.}

If we have two different users who are modifying the data repository at the same time without synchronization, they will naturally have two different histories $H_1$ and $H_2$ that form two branches. 
\edit{An \textbf{\inter} is a combined history that contains \emph{all} of the operations in $H_1$ and $H_2$ and orders them in a way that respects the local ordering.}
\edit{Namely, for two histories $H_1 = \phi_1\phi_2\cdots\phi_m$ and $H_2 = \psi_1\psi_2\cdots\psi_n $, an \textbf{\inter} is a combined history $H$ such that: for $i < j$, $\phi_i < \phi_j$ in $H$, and for $i < j$, $\psi_i < \psi_j$ in $H$~\footnote{Throughout this paper, $\phi$ will always refer to a \update in the first history, while $\psi$ will be for a \update in the second.}}


\edit{It is worth making a few remarks on this definition. First, \inter is a class of version reconciliation operations that combines all of the two users \updates into a single sequence of operations (another history), with the constraint that the sequence must respect each history's order of operations. This definition excludes any semantic forms of reconciliation that might combine operations from both histories or ignore duplicated ones. Under this model of reconciliation, \amty is a powerful concept. If it holds, manual reconciliation of two histories is not needed because \emph{any} \inter results in the same final state. Likewise, if it doesn't hold, some manual intervention is required because there is at least some ambiguity about the final state.}

\vspace{-0.05cm}
\begin{tcolorbox}[breakable,
arc=0pt,outer arc=0pt,colback=red!5!white,colframe=red!75!black, boxrule=0pt,left=0pt,right=0pt,top=0pt,bottom=0pt,bottomrule=0pt,toprule=0pt,leftrule=0pt, rightrule=0pt,title={\textbf{\Amty Definition}}]
  \edit{A pair of modification histories $H_1$ and $H_2$ are \emph{\amle} if and only if all \inters produce result in same final database state.}
\end{tcolorbox}
\vspace{0.5em}

\edit{Furthermore, while intuitive, it is actually unclear how to test for \amty. } The exhaustive solution to the \amty problem---simply permuting all valid \inters and verifying they produce the same result---has an impractical time complexity, with the number of \inters growing with $C(m+n, n)$, where $m$ and $n$ are the sizes of two \update histories. 
This number is $1.2\times10^{17}$ when $n$ and $m$ are just 30. 
\edit{Accordingly, the core contribution of this paper is an efficient algorithm that detects conflicts using pairwise comparisons between \updates.}

\subsection{Conflict Identification and Reconciliation}
\label{sec:conflict_id_and_rec}
\noindent \edit{The \amty definition in the previous section can be leveraged to design the reconciliation workflow of a \dvds. \Amty can alternatively be defined in terms of individual tuples in the original dataset.} We define a set of tuples $C_t \subset D_0$ as \emph{\amty conflicts} if $H_1$ and $H_2$ are \amle over $D_0 \setminus C_t$. Naturally, $H_1$ and $H_2$ are \amle if and only if no tuples are conflicting tuples. \edit{The definition of tuple-level \amty conflicts gives us a finer-grained objective than simply determining if $H_1$ and $H_2$ are \amle. It is these conflicts that will allow a user to determine how to reconcile two \namle histories because she can pinpoint the particular tuples whose final state is ambiguous. }
Therefore, an ideal \dvds should identify the conflicting tuples as precisely as possible. 

\edit{There are interesting precision and recall considerations in detecting conflicting tuples}. Two types of errors may manifest: false positives (identifying \amle tuples as \namle) and false negatives (identifying \namle tuples as \amle). False negatives must be avoided, as trusting a record by mistake could cause severe issues to following data analytics: \reviewone{for instance, in Example \ref{motivating_example}, a user will not even be aware of the loss of the San Jose record if the \dvds incorrectly believes the \updates are \amle and arbitrarily chooses to apply the serial order in Table \ref{tab:example-wrong}}. False positives do not compromise the integrity of a database or cause cascading errors, but they are not harmless: the system has to demand manual user reconciliation when it is in fact unnecessary.
\sys tries to identify $C_t$ efficiently without losing accuracy: it returns a $C_t'$ with a guarantee of no false negatives (i.e., $C_t \subseteq C_t'$) and significantly fewer false positives than baselines (i.e., the quantity $|C_t' \setminus C_t|$ is small). 

Any reconciliation process involving a human must be relatively lightweight. For this reason,
we believe that asking a user to determine a total order of all possible \updates is infeasible (the number of valid interleavings is exponential in the number of \updates), as is having her choose among a desired version for every conflict tuple. Instead, we ask a series of questions that only requires her to specify the relative order of two \updates from two histories. We discuss our reconciliation workflow in Section~\ref{resolution}.



%




\stitle{Integrity Constraints:}
\Amty is naturally compatible with database integrity constraints. 
If a set of histories is \amle, then a user may use integrity-checking mechanisms to verify that \textit{one} interleaving, e.g., a serial ordering, does preserve integrity constraints, and conclude that \textit{every} \inter preserves constraints. 
Similarly, if the arbitrarily chosen interleaving is not constraint-preserving, one can assert that \textit{none} of the possible \inter \edit{(of an-\amle history)} is constraint-preserving.

\stitle{Conflicts and Incompatible Reads/Writes:}
Locking schemes are a hallmark of transaction processing that detect ``conflicts'' in a concurrent environment and can be utilized to detect \namle tuples as those accessed by incompatible lock requests.
However, even with the finest granularity (i.e., cell-level locks), they still incur high false positives. 
\reviewthree{Consider Example~\ref{motivating_example}: while all California cities have incompatible locks (written by $A_1$ and read by $B_3$), only San Jose is in conflict. Los Angeles is \amle because $B_3$ will not delete it if its electricity is no less than 32 (population $\times$ 10), which is true before or after $A_1$. As with this example, a leading reason for why locking schemes are ineffective in the \amty problem is because they are only concerned with whether read/write happens and not the values read/written.
Advanced lock schemes that consider values, including semantic locks~\cite{semanticlock}, cannot completely mitigate the high false positive rate as they lack mechanisms to efficiently enumerate which \updates from the other history happen prior to the lock, and whether the final result would be the same. }

More importantly, locking schemes are highly inefficient. A necessary prerequisite of any physical locking schemes is materialization of \updates. Specifically, for a history $\phi_1\phi_2\dots\phi_m$, we need to execute $\phi_1\dots\phi_{i-1}$ on data $D$ before we know exactly which records are being locked by $\phi_i$. In order to compare lock compatibility from different users, we need to either ship locking records from users (which could be as huge as the data itself), or ship  modification logs and re-execute them (on a centralized server or one user's local environment). Both ways incur high overhead.

\section{Conflict Detection}
\label{detection}
\noindent \edit{In this Section, we propose an algorithm that determines whether two histories are \amle, and identifies conflicting tuples if not}.  

\subsection{Pairwise Conflict Detection}
\label{pairwise}
\noindent \edit{Our key insight is an important important sufficient condition of \amty. } If all pairs of \updates from two histories are commutative (\edit{we will be more precise in the next paragraph}), then two histories are \amle. This theorem opens up the opportunity for efficient conflict detection algorithms: instead of enumerating all possible \inters of two histories, we only need to examine all pairs of \updates from two histories, which has a polynomial complexity.

\edit{Commutativity means that the order of modifications does not matter (at least over the data present in the database). Let $\phi$ and $\psi$ be two \update operations. For any tuple $t$, we can define $\phi(\psi(t))$ (applying $\phi$ and $\psi$) and $\psi(\phi(t))$ (vice versa). We say that $\phi$ and $\psi$ are \textbf{commutative} if and only if $\phi(\psi(t)) \equiv \psi(\phi(t))$ over all tuples in a database $D$. Conversely, they are \textbf{commutative} (or ``conflict'') on a database version $D$ if $\phi(\psi(t)) \ne \psi(\phi(t))$ on at least a tuple $t\in{}D$. }

\edit{To leverage commutativity, we need to define some manipulations over histories. } We denote $\pref{k}{H}$ as the first $k$ \updates from $H$ (note that $\pref{0}{H}=\emptyset$), and denote concatenation of two histories by putting one after another, i.e., $H_1 H_2$ represents the \updates of $H_1$ followed by the \updates of $H_2$. \edit{Armed with these definitions, we arrive at the following theorem.}\\

\noindent \textbf{Theorem 1:} Two histories $H_1$ and $H_2$ are \amle if there does not exist $i, j$ such that $\phi_i \in H_1, \psi_j \in H_2$ conflict (are non-commutative) on the database version $D_{\pref{{i-1}}{H_1}\pref{{j - 1}}{H_2}}$. 

\iflong 
\noindent \textbf{Proof: } 
For $H_1$ and $H_2$, \amty means that for every pair of valid \inters $\Pi$ and $\Pi'$, we have $D_{\Pi} = D_{\Pi'}$.  We will prove this by showing that for any valid \inter $\Pi$, $D_{\Pi}=D_{\ser}$. 

We will prove that these exist a sequence of \inters $\Pi_1, \Pi_2, \cdots, \Pi_p$ such that $\Pi_1 = \Pi$, $\Pi_p=H_1H_2$, and $D_{\Pi_i}=D_{\Pi_{i+1}}$ for all $1\leq i\leq p-1$. This sequence is a result of ``swapping'' \updates. Intuitively speaking, we first find $\phi_1$ from $\Pi$, move it to left as much as possible: each step we swap it with its left neighbor from $H_2$. Then we find $\phi_2$ and repeat the swapping procedure. After all moves finish, the interleaving becomes $H_1 H_2$ (the serial ordering).

A critical step of this proof is to demonstrate that the two \inters before and after a swap produce the same result on $D$. Assume  $\phi_i\in{}H_1$ is the leftmost \update that is not on the left of all \updates from $H_2$. Formally speaking, suppose $\Pi \neq H_1H_2$ and  $$\Pi = \pref{{i-1}}{H_1}~\pref{{j - 1}}{H_2}~\psi_j\phi_i\Pi_S $$ where $\Pi_S$ is any arbitrary valid \inters of the suffix (i.e., remaining \updates in $H_1$ and $H_2$). By swapping $\phi_i$ and $\psi_j$, we have obtained a new \inter $$\Pi' = \pref{{i-1}}{H_1}~\pref{{j - 1}}{H_2}~\phi_i\psi_j\Pi_S $$ We know that $D_{\Pi}=D_{\Pi'}$, since from the assumption, we have $$D_{\pref{{i-1}}{H_1}~\pref{{j - 1}}{H_2}~\psi_j\phi_i}= D_{\pref{{i-1}}{H_1}~\pref{{j - 1}}{H_2}~\phi_i \psi_j}$$ and executing $\Pi_S$ after at the same database gives same result.~$\square$ 
\fi

\stitle{Remarks:} It is important to note that Theorem 1 is sufficient but not necessary. \reviewone{One extreme example where this is the case is when the last \updates of $H_1$ and $H_2$ delete the database. Here, $H_1$ and $H_2$ are \amle (the database is empty regardless of the specific interleaving) regardless of any prior pair of conflicting \updates.
In general, when two \updates conflict and Theorem 1 reports the histories as \namle, a late \update from one user may nullify the inconsistencies caused by conflict \updates and makes the database \amle.} However, we believe that false positives are uncommon, for a few reasons.
First, it is possible that there exist multiple pairs of \updates that conflict on $t$, so even if the effect of one is moot, the tuple still has attributes that are \namle.
Second, only the effect of a read-write conflict (see Section \ref{pairwise_2}) can be overwritten without triggering another conflict. 
Third, inconsistencies in one attribute may breed further inconsistencies in other attributes. Thus, multiple \updates may be needed to overwrite these inconsistencies into a consistent state, which is unlikely.
We empirically evaluate the amount of falsely reported conflicts under varying \edit{data and query} parameters in Section \ref{expr}.

\newcommand{\blueone}{\logicbox{blue}{\cir{1}}{$\colB = \colL$}}
\newcommand{\bluetwo}{\logicbox{blue}{\cir{2}}{$\valB \neq \valL$}}
\newcommand{\bluethreea}{\logicbox{blue}{\cir{3a}}{$t \in {D_\psi}[\phi] \cap {D_\phi}[\psi]$}}
\newcommand{\bluethreeb}{\logicbox{blue}{\cir{3b}}{$t \in ({D_\psi}[\phi]^c \cap {D_\phi}[\psi]^c) \cap (D[\phi] \cap D[\psi])$}}

\newcommand{\redone}{\logicbox{red}{\cir{1}}{$\colB \neq \colL$}}
\newcommand{\redtwo}{\logicbox{red}{\cir{2}}{$t.\colB \neq \valB$}}
\newcommand{\redthree}{\logicbox{red}{\cir{3}}{$t \in D[\phi] \bigtriangleup {D_\psi}[\phi]$}}

\subsection{Efficiently Detecting Pairwise Conflicts}
\label{sec_algorithm}


\noindent A straightforward implementation of Theorem 1 is still not efficient enough, due to the huge space/time cost of building temporary data views (all $D_{\pref{i}{H_1}\pref{j}{H_2}}$) and executing \updates. \edit{Instead of doing this check physically for each version, we reason about the conditions that conflicting tuples must satisfy. To do so, we first identify the first point in both histories at which conflicts arise} $D_{\pref{i}{H_1}\pref{j}{H_2}}$, then ``backtrack'' the \updates, maintaining an equivalent condition for every version (in Section \ref{virtualizing}) until one that can be queried on the latest common ancestor $D_0$ is obtained. Finally, we execute the condition on $D_0$ to find all conflicting tuples.

In this section, we assume for simplicity of explanation that \updates read and write a single table and that \updates are of the form \lstinline[mathescape]{UPDATE SET $\colB = \valB$ WHERE $\colA = \valA$}, where $\colA$ and $\colB$ refer to columns (attributes) and $\valA$ and $\valB$ are literal values. \reviewthree{Throughout the paper we use upper case letters for column names, and lower case letters for constant values.} We use the equivalent notation $Q: (\colA = \valA) \to (\colB = \valB)$ to represent the \update. When $C$ is a condition, $V[C]$ represents the tuples in $V$ that satisfy $C$.
We denote by $V[Q]$ the set of tuples in a database version $V$ that are impacted by $Q: (\colA = \valA) \to (\colB := \valB)$, i.e., $V[\colA=\valA]$. Much of the theory presented in this paper applies to general deterministic \update functions, and we will discuss how to extend it to more complex \updates in Section~\ref{more_general}.

\subsubsection{Detecting Pairwise Conflicts}
\label{pairwise_2}

\noindent We emphasize that the following discussion assumes that we have known the actual version of data we work on (i.e., $D_{\pref{i}{H_1}\pref{j}{H_2}}$). We relax this assumption in Section~\ref{virtualizing}.

We determine whether two \updates, $\phi$ and $\psi$, are commutative on a version of a database $D$. A na\"ive solution simply entails materializing the two possible \inters $\phi(\psi(D))$ and $\psi(\phi(D))$ and checking if there are differences. To avoid overheads with actually executing \updates and materialization, we instead deduce logical conditions that non-commutative tuples (i.e., tuples $t \in D$ such that $\phi(\psi(t)) \neq \psi(\phi(t))$) must satisfy. This provides two major advantages. First, it provides a succinct summary of what parts of the data are not commutative. Second, it allows for optimizations outlined in Section \ref{virtualizing}.

\edit{The names given to types of conflicts (\textit{e.g.} `write-write') that we look for are analogous to similar concepts defined in traditional transaction processing/locking theory.} However, we emphasize that the concepts we describe are more precise and identify a tuple as conflicting if and only if the tuple is non-commutative, while traditional definitions are \textit{sufficient but not necessary} in detecting non-commutativity. To revisit a previous example, if two \updates both conditionally write to the same tuple, traditional locking theory identifies the tuple as conflicting, whereas the methods we propose inspects whether the conditions are satisfied.

\begin{tcolorbox}[breakable,
arc=0pt,outer arc=0pt,colback=red!5!white,colframe=red!75!black, boxrule=0pt,left=0pt,right=0pt,top=0pt,bottom=0pt,bottomrule=0pt,toprule=0pt,leftrule=0pt, rightrule=0pt,title={\textbf{Commutative Condition Problem}}]
Find conditions on a tuple $t$ under which $\phi(\psi(t)) = \psi(\phi(t))$, given $\phi: \colA = \valA \to \colB := \valB$ and $\psi: \colK = \valK \to \colL := \valL$.
\end{tcolorbox}

In following discussions, we highlight the conditions of two types conflicts as \logicbox{blue}{}{blue} or \logicbox{red}{}{red}. These must be true regardless whether the \updates have simple predicates (e.g., $\colA = \valA$) or complex conditions (on $\colA$ or more attributes). We show how to directly check them when the predicates are in simple form, in \textbf{bold}.

\stitle{\logicbox{blue}{}{Write-write conflicts:}} We know that for the pair of \updates $\phi$ and $\psi$, a write-write conflict occurs on tuple $t$  if and only if we have \blueone~(otherwise the \updates cannot possibly write the same cell of data), \bluetwo~ (otherwise the \updates applied by both are equivalent, and order of application does not matter), and one of the following conditions: 
\begin{itemize}
    \item \bluethreea: If a tuple $t$ is affected by both \updates after the other has been applied, then attribute $\colB = \colL$ of $t$ will see different values depending on which order of \updates is executed.
    \item \bluethreeb: If both \updates are never executed after the other is applied (e.g. $\phi: (\colS = \valS_1) \to (\colS := \valS_2)$ and $\phi: (\colS = \valS_1) \to (\colS := \valS_3)$ with $\valS_1 \neq \valS_2 \neq \valS_3$), then again attribute $\colB = \colL$ will see different values depending on which order is executed. Note that this condition can often be neglected, since a tuple can only satisfy this condition if both \updates read and write the exact same single attribute (in this example, only if $\colA = \colB = \colK = \colL$).
\end{itemize}
The first two are straightforward to verify by checking query semantics, so we discuss the third. Checking if $t \in D[\phi]$ (resp. $t \in D[\psi]$) corresponds directly to finding tuples satisfying the predicate of $\phi$ (resp. $\psi$). For the sake of exposition, we will show how to check $t \in {D_\psi}[\phi]$. The symmetrical case is similar. 
\begin{enumerate}
    \item If $\boldsymbol{\colA \neq \colL}$, then $\psi$ does not alter the tuples touched by $\phi$. Namely, this means ${D_\psi}[\phi] = D[\phi]$. Then  $t \in {D_\psi}[\phi]$ if and only if $t \in D[\phi]$, which in turn means $\boldsymbol{t.\colA = \valA}$. 
    \item If $\boldsymbol{\colA = \colL}$, then $\psi$ does change tuples touched by $\phi$, and we know:
    \begin{enumerate}
        \item If $\boldsymbol{\valA = \valL}$, all $t$ satisfying $(t.\colK = \valK)$ on the original version $D$---regardless of whether they satisfied $(t.\colA = \valA)$ on $D$---will satisfy $(t.\colA = \valA)$ on version $D_\psi$. Specifically, we can conclude ${D_\psi}[\phi] = D[\phi] \cup D[\psi]$, so $t \in {D_\psi}[\phi]$ means $\boldsymbol{(t.\colA = \valA)} \boldsymbol{ \vee (t.\colK = \valK)}$.
        \item  If $\boldsymbol{\valA \neq \valL}$, all $t$ satisfying $(t.\colK = \valK)$ on the original version $D$---regardless of whether they satisfied \\ $(t.\colA = \valA)$ on $D$---\textit{must not} satisfy $(t.\colA = \valA)$ on version $D_\psi$. More specifically, we can conclude ${D_\psi}[\phi] = D[\phi] \setminus D[\psi]$, and therefore $t \in {D_\psi}[\phi]$ is equivalent to having $\boldsymbol{(t.\colA = \valA) \wedge}$ $\boldsymbol{\lnot(t.\colK = \valK)}$.
    \end{enumerate}
\end{enumerate}

\begin{example}
Consider the following two queries issued on the same database $D$ from Example \ref{motivating_example}:
\begin{lstlisting}[mathescape=true]
$\phi$: UPDATE db SET State = 'WA' WHERE City = 'Seattle' -- not in DC
$\psi$: UPDATE db SET State = 'DC' WHERE State = 'D.C.'
\end{lstlisting}
To find tuples where there is a write-write conflict, we have \blueone $ \text{ } = $ \inlinesql{`State'}, and \logicbox{blue}{\cir{2}}{$\valB = $ \inlinesql{`WA'} $\neq$ \inlinesql{`DC'} $= \valL$}, so we must check that \bluethreea $\text{ }$ holds. (Note that we do not have to check  \logicbox{blue}{\cir{3b}}{} since $\phi$ does not read and write the same single attribute.) For $t \in {D_\psi}[\phi]$, by rule (1) described previously, this is equivalent to checking the condition (\inlinesql{t.City = `Seattle'}). For $t \in {D_\phi}[\psi]$, by rule (2)(b), we check the condition ({\inlinesql{t.State = `D.C.'}}) $\wedge \lnot$  (\inlinesql{t.City = `Seattle'}). We conjoin the two conditions to obtain $(\inlinesql{t.City = `Seattle'}) \text{ } \wedge (\inlinesql{t.State = 'D.C.'} \wedge \lnot \inlinesql{t.City = `Seattle'})$,
which is not satisfiable; hence, no data is in conflict. $\square$
\end{example}

\stitle{\logicbox{red}{}{Read-write conflicts:}} We define a read-write conflict between $\phi: \colA = \valA \to \colB := \valB$ and $\psi: \colK = \valK \to \colL := \valL$ as a conflict that occurs  because $\phi$ reads an attribute that $\psi$ writes. We will refer to the reverse case hereinafter as a write-read conflict, which can be handled similarly. For a read-write conflict between $\phi$ and $\psi$ to occur on some tuple $t$, we must have that \redone~and \redtwo~(otherwise $\phi$ does not actually change the value of $t.B$), and \redthree (i.e., $\psi$ changes whether $t$ will be updated by $\phi$; $\bigtriangleup$ denotes the set symmetric difference. Condition \logicbox{red}{}{\cir{1}} can be checked against query semantics, and \logicbox{red}{}{\cir{2}} is already in the form apt for database querying. We demonstrate how to evaluate \logicbox{red}{}{\cir{3}}:
\begin{enumerate}
    \item We must have $\boldsymbol{t.\colK = \valK}$, otherwise $t \notin D[\psi]$, so the \update applied by $\psi$ does not affect $t$: namely, $\psi(t) = t$. In other words, whether the \update $\phi$ applies to $t$ doesn't depend on when $\psi$ is executed. (i.e., $t \in D[\phi] \cap D_\psi[\phi]$).
    \item Logic dictates that \redthree $\text{ }$ implies either: 
    \begin{enumerate}
        \item $\boldsymbol{t \in D[\phi] \setminus {D_\psi}[\phi]}$, or in other words, the application of $\psi$ removed $t$ from the set of tuples that $\phi$ affects. This is only possible if $\boldsymbol{t.\colA = \valA$ and $\valA \neq \valL}$.
        \item Or $\boldsymbol{t \in {D_\psi}[\phi] \setminus D[\phi]}$, or in other words, the application of $\psi$ included $t$ in the set of tuples that $\phi$ affects. This is only possible if $\boldsymbol{t.\colA \neq \valA$ and $\valA = \valL}$.
    \end{enumerate}
\end{enumerate}


\begin{example}
Consider the following two queries issued on the same database $D$ from Example \ref{motivating_example}:
\begin{lstlisting}[mathescape=true]
$\phi$: UPDATE db SET Population = 5 WHERE City = 'Los Angeles' -- new data
$\psi$: UPDATE db SET City = 'Los Angeles' WHERE City = 'Los Angles' -- fix a typo
\end{lstlisting}
Upon inspection of semantics, we note \redone~is satisfied. Any tuple $t$ in conflict must have \redtwo, i.e., \inlinesql{Population $\neq$ 5}. To check if there is any \redthree, according to rule (1) we must have \inlinesql{t.City = 'Los Angles'}, and since we have $\boldsymbol{\valA} = $ \inlinesql{'Los Angeles'} $= \boldsymbol{\valL}$, 
the logic in part (2)(b) dictates that the equivalent condition to be checked is \inlinesql{t.City} $\neq$ \inlinesql{'Los Angeles'}. We conjoin these conditions to obtain (note the last condition is implied by the second so we can omit it)
$$
(\inlinesql{t.Population $\neq$ 5}) \wedge   (\inlinesql{t.City = 'Los Angles'})
$$
Upon querying the database (\ref{tab:example-a}) with this condition, we find that one tuple is in conflict. $\square$
\end{example}



\iflong

\noindent We prove that the lack of the two types of conflict we discuss previously indicates that two \updates are commutative. We reiterate that the traditional transaction processing theory does not help here: although the end goal of determining commutativity is similar, the conflict types we propose are much ``narrower'' than the similar concepts from transaction processing theory.

\noindent \textbf{Lemma 1: } If $t \notin \mathcal{R}_D(\phi) \cup \mathcal{R}_{\psi(D)}(\phi)$ or $t \notin \mathcal{R}_D(\psi) \cup \mathcal{R}_{\phi(D)}(\psi)$, then $\phi(\psi(t)) = \psi(\phi(t)$. \\

\noindent \textbf{Proof: } This can be intuitively described as one of the updates never affecting $t$, regardless of whether the other update is executed before. More formally, $t \notin \mathcal{R}_D(\phi) \cup \mathcal{R}_{\psi(D)}(\phi)$ implies $\phi(\psi(t)) = \psi(t)$ and $\phi(t) = t$. These two, when put together, imply $\phi(\psi(t)) = \psi(t) = \psi(\phi(t))$. The same logic can be used to show that $t \notin \mathcal{R}_D(\psi) \cup \mathcal{R}_{\phi(D)}(\psi)$ also implies $\phi(\psi(t)) = \psi(\phi(t))$. $\square$ \\

\noindent \textbf{Theorem 2:} There is a write-write, read-write, or write-read conflict on tuple $t$ between $\phi: A = \alpha \to B = \beta$ and $\psi: K = \kappa \to L = \lambda$ if and only if $\phi(\psi(t)) \neq \psi(\phi(t))$. \\

\noindent \textbf{Proof:} Define three shorthand notations of sets that we will reference: Let $P = \mathcal{R}_{\psi(D)}(\phi) \cap \mathcal{R}_{\phi(D)}(\psi))$, $\Sigma = \mathcal{R}_D(\phi) \bigtriangleup \mathcal{R}_{\psi(D)}(\phi)$, and $T = \mathcal{R}_D(\psi) \bigtriangleup \mathcal{R}_{\phi(D)}(\psi)$. \\

\noindent $(\Rightarrow)$ This direction is outlined on the previous discussion.

\newcommand{\greenone}{\logicbox{green}{\cir{1}}{$(B \neq L)$}}
\newcommand{\greentwo}{\logicbox{green}{\cir{2}}{$(\beta = \lambda)$}}
\newcommand{\greenthree}{\logicbox{green}{\cir{3}}{$(t \notin P)$}}
\newcommand{\yellowone}{\logicbox{yellow}{\cir{1}}{$(t.B = \beta)$}}
\newcommand{\yellowtwo}{ \logicbox{yellow}{\cir{2}}{$(t \notin \Sigma)$}}
\newcommand{\purpleone}{\logicbox{purple}{\cir{1}}{$(t.L = \lambda)$}}
\newcommand{\purpletwo}{\logicbox{purple}{\cir{2}}{$(t \notin T)$}}
\newcommand{\aand}{$\wedge$}
\newcommand{\updatedyellow}{ \logicbox{yellow}{\cir{2}}{$(t \in \sigma)$}}
\newcommand{\updatedpurple}{ \logicbox{purple}{\cir{2}}{$(t \in \tau)$}}

\noindent $(\Leftarrow)$ We will prove the contrapositive of this statement, namely that the lack of WW, RW, and WR conflicts on $t$ implies $\phi(\psi(t)) = \psi(\phi(t))$. In first-order logic, the assumption of no WW, RW, or WR conflicts can be expressed as
$($\greenone $\vee$ \greentwo  $\vee$ \greenthree 
$) \wedge ($ \yellowone $\vee$ \yellowtwo 
$) \wedge ($ \purpleone $\vee$ \purpletwo $)$.
We begin by noting that if $\phi$ and $\psi$ are non-commutative on $t$, there must be some attributes of $t$ that are inconsistent. In fact, these attributes must be the subject of a write --- either of $\phi$ or of $\psi$ --- and are thus $B$ and $L$. With this fact, we will convert this logic into disjunctive normal form and consider each case, showing why it is sufficient in determining that there are no conflicts. In other words, we will examine all possible combination of \logicbox{green}{}{$X$} $\wedge$ \logicbox{yellow}{}{$Y$} $\wedge$ \logicbox{purple}{}{$Z$} while \logicbox{green}{}{$X$} is one of \logicbox{green}{\cirdown{1}}{ }, \logicbox{green}{\cirdown{2}}{ }, \logicbox{green}{\cirdown{3}}{ }, \logicbox{yellow}{}{$Y$} is one of \logicbox{yellow}{\cirdown{1}}{ }, \logicbox{yellow}{\cirdown{2}}{ }, and \logicbox{purple}{}{$Z$} is one of \logicbox{purple}{\cirdown{1}}{ }, \logicbox{purple}{\cirdown{2}}{ }.

Before proceeding, we note that the condition $t \notin \Sigma$ is equivalent to the condition $(t \in \mathcal{R}_D(\phi) \cap \mathcal{R}_{\psi(D)}(\phi)) \wedge (t \notin \mathcal{R}_D(\phi) \cup \mathcal{R}_{\psi(D)}(\phi))$. The result of Lemma 1 shows it is sufficient to consider the case where $t \in \sigma$, with $\sigma = \mathcal{R}_D(\phi) \cap \mathcal{R}_{\psi(D)}(\phi)$. Similar logic shows it is sufficient to consider the case where $t \in \tau$, with $\tau = \mathcal{R}_D(\psi) \cap \mathcal{R}_{\phi(D)}(\psi)$, in lieu of the broader condition $t \notin T$. 
\begin{enumerate}
    \item If \yellowone \aand \purpleone, then $\phi$ and $\psi$ never actually change the values of $t.B$ or $t.L$, regardless of whether $t$ satisfies the predicates of $\phi$ and $\psi$. Therefore, we are guaranteed that after applying $\phi$ and $\psi$ in whatever order, $t.B = \beta$ and $t.L = \lambda$.
    \item \greenone \aand \updatedyellow \aand \purpleone: $B$ and $L$ are different attributes, and from $(t \in \sigma)$, we know that whichever ordering we have, $t$ undergoes the update $\phi$. Therefore, at the end of applying both $\phi$ and $\psi$, we are guaranteed $t.B = \beta$. Using similar logic as case (1), we know that $t.L = \lambda$ implies that $\psi$ never actually changes the value of $t.L$; therefore, we conclude that regardless of whether $\psi$ was applied to $t$, we are guaranteed $t.L = \lambda$.
    \item \greenone \aand \yellowone \aand \purpletwo: symmetric case to (2).
    \item \greenone \aand \updatedyellow \aand \updatedpurple: From $(t \in \sigma)$, we know that whichever ordering we have, $t$ undergoes the update $\phi$. From $(t \in \tau)$, we know that whichever ordering we have, $t$ undergoes the update $\psi$. Therefore, $B$ and $L$ being different attributes guarantees that after application of both updates we have $t.B = \beta$ and $t.L = \lambda$.
    \item \greentwo \aand \updatedyellow \aand \purpleone: if we do not have $B = L$, we recover case (2). From $(t \in \Sigma)$ we know that $t$ must undergo the update $\phi$. If we choose the order $\psi \phi$, then we are guaranteed that $t.B = t.L = \beta$. If we choose the order $\phi \psi$, then immediately before $\psi$ is applied we have $(t.L = \lambda)$, so regardless of whether $\psi$ is applied, we are guaranteed that $t.L = \lambda$. But since $t.B = t.L$ and $\beta = \lambda$, there is no ambiguity and we are guaranteed $t.B = t.L = \beta = \lambda$.
    \item \greentwo \aand \yellowone \aand \updatedpurple: symmetric case to (5).
    \item \greentwo \aand \updatedyellow \aand \updatedpurple: from $(t \in \sigma)$ we know that $t$ must undergo the update $\phi$. Likewise, from $(t \in \tau)$, we know that whichever ordering we have, $t$ undergoes the update $\psi$. If $\phi$ is applied second, then it overwrites the update of $\psi$, and we are guaranteed $t.B = t.L = \beta$. Similarly, if $\psi$ is applied second, then it overwrites the update of $\phi$, and we are guaranteed $t.B = t.L = \lambda$. However, the condition $(\beta = \lambda)$ means there is no ambiguity, and guarantees $t.B = t.L = \beta = \lambda$.
    \item \greenthree \aand \updatedyellow \aand \purpleone: we must have $(B = L) \wedge (\beta \neq \lambda)$, otherwise we recover previous cases. The condition $(t \notin P)$ implies either $t \notin \mathcal{R}_{\psi(D)}(\phi)$ or $t \notin \mathcal{R}_{\phi(D)}(\psi)$.
    \begin{enumerate}
        \item If $t \notin \mathcal{R}_{\psi(D)}(\phi)$, this is a contradiction of $t \in \sigma$.
        \item If $t \notin \mathcal{R}_{\phi(D)}(\psi)$, then the update of $\psi$ is moot if applied after $\phi$: namely, $\psi(\phi(t))=\phi(t)$. Since $t \in \sigma$ implies $t \in \mathcal{R}_D(\phi)$, we conclude that after order of updates $\phi \psi$ is applied, we are guaranteed $t.B = t.L = \beta$. Additionally, we know $t \in \mathcal{R}_{\psi(D)}(\phi)$, and we can conclude that after the order of updates $\psi \phi$ is applied, $t.B = t.L = \beta$. Therefore, the updates must be commutative on $t$. 
    \end{enumerate}
    \item \greenthree \aand \yellowone \aand \updatedpurple: symmetric case to (8).
    \item \greenthree \aand \updatedyellow \aand \updatedpurple: from before, the condition $(t \notin P)$ implies either $t \notin \mathcal{R}_{\psi(D)}(\phi)$ or $t \notin \mathcal{R}_{\phi(D)}(\psi)$. If $t \notin \mathcal{R}_{\psi(D)}(\phi)$, this contradicts the condition $(t \in \sigma)$. Similarly, if $t \notin \mathcal{R}_{\phi(D)}(\psi)$, this contradicts the condition $(t \in \tau)$, so this case is not possible.
\end{enumerate}
As established before, $t.B$ and $t.L$ are the only possible inconsistent fields that a tuple $t$ may have. We have shown in all cases, the values of the $t.B$ and $t.L$ are unambiguous, implying that these conditions are sufficient for pairwise commutativity. $\square$ \\

\noindent To reconcile every record subject to the conflict $(\phi_i, \psi_j)$, we simply perform a union of the possible types of conflicts: 
\begin{align*} S_{\phi_i, \psi_j} = \texttt{conflictWW}(\phi_i, \psi_j) \cup \texttt{conflictRW}(\phi_i, \psi_j) \\ \cup \text{ } \texttt{conflictRW}(\psi_j, \phi_i) \end{align*}
where $S_{\phi_i, \psi_j}$ is the total set of records subject to the conflict $(\phi_i, \psi_j)$. The result of theorem 1 allows us to say with confidence that any records not flagged as conflicted must be automergeable. \\
\fi

\subsubsection{Condition Derivation via Backtracking}
\label{virtualizing}

\noindent Though we have a condition (i.e., predicate) for all non-commutative tuples from an intermediate database version $D_H$ for some $H$, we seek to avoid the costly process that executes $H$ on $D$ to materialize $D_H$. We ``backtrack'' the history $H$ as a sequence of \updates. At each step we consider the \update $\phi$: we have condition $C$ on data $D_{\phi}$, and our goal is to derive another condition $C'$ on $D$ such that $\phi(t)\in D_{\phi}[C]$ if and only if $t\in D[C']$. After the backtracking, we have a condition $C_0$ on the initial data $D_0$, and we evaluate $D_0[C_0]$.

This approach is significantly more efficient than the straightforward evaluation needed to materialize $D_H$. Evaluating $D_H$ requires updating all tuples involved in $H$ regardless whether these tuples are \amle or not. Instead, our approach defers evaluation on dataset as much as possible, and avoids the large overheads of updating data. Furthermore, if the final condition for conflict has a low selectivity, or even is unsatisfiable, we may further forego data evaluation costs. Our approach is especially efficient when data volumes are large---exactly the target scenario of \dvds.

Now we discuss the step of backtracking (building equivalent conditions on two versions of a database, separated by one \update). For the sake of exposition, we assume that we are trying to evaluate condition $C = f(c_1, c_2, \cdots)$, where all $c_i$ are simple equality conditions like $\colA = 10$, and $f$ is a logical function of the $c_i$ using \texttt{AND}, \texttt{OR}, and \texttt{NOT}. Note that this framework can be generalized to any complex logical conditions using whose axioms are state-independent; see Section \ref{more_general}.  Based on the semantics of $\phi$, we change all $c_i$ to corresponding $c'_i$. $c'_i$ could be simple equality conditions or logic clauses (to be discussed as follows). We keep $f(\cdots)$ unchanged and denote $C'= f(c'_1, c'_2, \cdots)$. Our goal is that for any tuple $t \in D$, $c'_i(t)=c_i(\phi(t))$. Thus, $C'(t)=C(\phi(t))$, in other words, $t\in D[C']$ if any only if $\phi(t)\in D_{\phi}[C]$.

We now discuss how to construct an equivalent condition $c'_i$ for corresponding axiom $c_i$. Suppose $\phi: (\colB = \valB) \to (\colA := \valA)$, and $c_i: (\colK = \valK)$. We discuss the possible scenarios: 
\begin{enumerate}
    \item $\phi$ does not write attribute $\colK$ ($\colA \neq \colK$), then for any tuple $t$, its $\colK$ attribute is not modified. Therefore, we know $c'_i=c_i$.
    \item $\phi$ writes attribute $\colA$ ($\colA = \colK$). In this case, there are two scenarios. 
    \begin{enumerate}
        \item If $\valA = \valK$, then $\phi(t)$ in $D_{\phi}$ may have either originally satisfied $c_i$ or its $\colK$ attribute may have been written by $\phi$, implying $(\colB = \valB)$. The new condition $c'_i$ is thus $(\colB = \valB) \vee c_i$.
        \item If $\valA \neq \valK$, then $\phi(t)$ must have satisfied $c_i$ already. Furthermore, it must not have been affected by $\phi$, necessarily meaning $(\colB \neq \valB)$. The new condition $c'_i$ is thus  $\lnot(\colB = \valB) \wedge c_i$.
    \end{enumerate}
\end{enumerate}

\begin{example}
\noindent We have a condition $C$: \inlinesql{t.Population = 2000}  $\wedge$ \inlinesql{t.Electricity = 43000} and wish to find the set $D_\phi[C]$ (i.e., all tuples which satisfy $C$ on the version of $D$ after application of \update $\phi$), where $\phi$ is  $(\text{\inlinesql{City = `Los Angeles'}}) \rightarrow (\text{\inlinesql{Electricity = 43000}})$.

Note that \inlinesql{t.Population = 2000} is unaffected and the condition \inlinesql{t.Electricity = 43000} becomes \inlinesql{t.Electricity = 43000} $\vee$ \inlinesql{t.City = `Los Angeles'} by applying rule (2)(a). Thus, the new condition $C'$ is (\inlinesql{t.Electricity = 43000} $\vee$ \inlinesql{t.City = `Los Angeles'}) $\wedge$ \inlinesql{t.Population = 2000}. $\square$
\end{example}

\subsection{More General Query Model}
\label{more_general}

\noindent Thus far, we have assumed queries of the form $Q: (\colA = \valA) \to (\colB = \valB)$ (in SQL, this can be represented as the query \lstinline[mathescape]{UPDATE SET $\colB = \valB$ WHERE $\colA = \valA$}). In this section, we consider generalizations to \reviewone{\lstinline{INSERT}s, \lstinline{DELETE}s, and complex \lstinline{UPDATE}s including \lstinline{JOIN}s}. 


\stitle{Inserts and Deletes}: We treat inserts and deletes as special updates that either read or write \lstinline{NULL} values. Specifically, we treat an insertion as an update writing a tuple with all attributes equal to \lstinline{NULL} (and a special hidden unique ID) in the initial version of database $D_0$. Similarly, in handling a delete, we consider it as an update writing \lstinline{NULL} values to an existing non-\lstinline{NULL} tuple. We can make simplifications, as we will demonstrate, since we know \lstinline{INSERT}s and \lstinline{DELETE}s do not touch remaining dataset.

We will demonstrate these accommodations on an insert as an example. Suppose we have an insert $I$ and an update $U$ of the form 
\begin{lstlisting}[mathescape=true]
$I$: INSERT VALUES ($\colA = \valA_1$, $\colB = \valB_1$, $\dots$)
$U$: UPDATE db SET $\colA = \valA_2$ WHERE $\colB = \valB_2$
\end{lstlisting}
To determine commutativity of $I$ and $U$ on version $V$, we note that only the ``write-write'' commutativity conditions must be checked on the tuple inserted by $I$, since $I$ does not touch other tuples. Applying the procedure from Section \ref{pairwise}, we find that the updates are not commutative on the inserted tuple if and only if $(\valA_1 \neq \valA_2) \wedge (\valB_1 = \valB_2)$. (These updates commute on any other tuple.)

During the process of Condition Derivation~(Section~\ref{virtualizing}), when the algorithm ``backtracks'' to an insert statement, it evaluates all conditions on the newly inserted tuples, reports them as \namle if applicable, and continues as if the insert statement does not exist. This is because we assume the inserted tuples are all \lstinline{NULL}s before the insertion, and all conditions evaluate to \lstinline{FALSE} on \lstinline{NULL} values. When the algorithm ``backtracks'' to a delete, it makes sure all deleted tuples are not being considered. Specifically, we note that $V_{\text{del}}[C'] = V[(C \wedge \neg C_D)]$ where $del$ is the delete and $C_{\text{del}}$ is the predicate of delete.

\stitle{Complex updates}: \reviewone{We discussed \lstinline{UPDATE}s with simple predicates (i.e., attribute equals literal) and constant update values. In this section, we explain how to accommodate more complex predicates and updates that are well-represented in everyday data analytics and also lend themselves to conflict backtracking (which is where most of \sys's efficiency gains lie).} We consider complex predicates that are \textit{state-independent}, meaning that whether a tuple satisfies the condition does not depend on the state of other tuples. For example, $\colA + \colB > \colC$ and $\colA \in \{2, 3, 5\}$ are state-independent, but $\colA \in \Pi_{\colB}(\sigma(D))$ is not if the database $D$ is not held constant. \reviewone{Without loss of generality, we consider a \update's predicate as a function $f(\colA, \colB, \cdots)$ defined on its attributes. This predicate could include \lstinline{JOIN}s (on read-only relations), range queries, functional expressions (including UDFs), and set operation queries (\lstinline{IN}, \lstinline{EXISTS}, \lstinline{ALL}, etc.) on read-only tables/sets.} Note that any semi-\lstinline{JOIN} can be made equivalent to an \lstinline{IN} clause involving the \lstinline{JOIN} attributes: for example, \lstinline[mathescape]{UPDATE R SET $\colA = \valA$ FROM R JOIN S on R.$\colB$ = S.$\colB$} is equivalent to \lstinline[mathescape]{UPDATE R SET $\colA = \valA$ WHERE R.$\colB$ IN (SELECT $\colB$ from S)}.

To determine commutativity between two such \updates, the procedure is no different from simple update. We still must check the same conditions for a read-write conflict and write-write conflict. Correctly evaluating conditions on versions of a database that have undergone \updates (e.g., $t \in D_\phi[\psi]$) depends on the following Condition Derivation mechanism. Suppose we want to evaluate the condition $C_2(\colA, \colB, \cdots)$ on database $D' = \phi(D)$, while $\phi$ is \lstinline[mathescape]{UPDATE db SET $\colB=\valB$ WHERE $C_1(\colA, \colB, \cdots)$}.
As long as $C_1$ and $C_2$ are state-independent conditions, we can consider two cases some tuple $t \in D'$ satisfying $C_2$ could have: (i) $t$ undergoes \update $\phi$, and $\phi(t)$ satisfies $C_2$, or (ii) $t$ does not undergo \update $\phi$, and $t$ satisfies $C_2$. In other words, our newly built condition is:
$$(\lnot C_1(\colA, \colB, \cdots) \wedge C_2(\colA, \colB, \cdots)) \vee (C_1(\colA, \colB, \cdots) \wedge C_2(\colA, \valB, \cdots))$$
\reviewone{Note that although \sys supports a wide range of complex queries in \lstinline{UPDATE}s, \lstinline{INSERT}s, and \lstinline{DELETE}s, their query evaluation time affects \sys's efficiency. Two special cases allow further optimizations.} First, if $C_2$ is independent of $B$, then evaluating $C_2$ on $D'$ is the same as evaluating $C_2$ on $D$. Second, if either of $C_1$ or $\lnot C_1$ is data-independent, we recover the simpler cases outlined in Section \ref{pairwise}. \vspace{-0.2cm}

\section{Conflict Resolution}
\label{resolution}

\noindent \reviewone{When \namle tuples exist, a user needs to resolve the conflicts and reconcile the versions. However, it is impractical for the user to manually specify a ``correct'' \inter of \updates: for example, there are more than $10^{17}$ possible \inters when both histories have $30$ \updates. \edit{Instead, we present the use with an easier question: \emph{given two \updates, which should precede the other? } \sys guides the user to a desired \inter by repeatedly posing this question, together with a set of specific records for which the two \updates conflict to provide a concrete example he/she can reason through.}  }

\edit{We present the formal algorithm in Algorithm~\ref{alg:resolution}. Similar with the conventional workflow from collaborative project management like Git, our framework merges two branches to form a new branch at one time.} The conflict resolution works by iteratively finding the first operation pairs that cause a conflict. At each iteration, the resolution determines the ``earliest'' pair of conflicting \updates, and asks the user to specify the order between them. After that, the earlier one (and all preceding ones from the same history) is removed from the history, and is appended to the ``finished'' list (i.e., variable $order$). It terminates when at least one history is empty. Similar to Section~\ref{virtualizing}, we virtually update $D$ by Condition Derivation. \reviewone{It is not hard to see the number of user interactions is no more than the combined size of histories:  each step of the loop  asks the user to specify order at most once, and removes at least one \update from two histories. Thus, the loop body is executed at most $|H_1|+|H_2|$ times, upper-bounding the number of user-interactions.} \edit{For comparison, the conventional algorithm that Git applies when merging two branches may ask the user to manually merge up to $|H_1|\times|H_2|$ commits, as it is possible for every commit in $H_1$ to require manual merging with every commit in $H_2$.  }

\begin{algorithm}[t!]\small
\SetAlgoLined
\SetAlgoNoEnd
\SetKwFunction{FMain}{resolution}
\SetKwFunction{Fchoose}{choose}
\SetKwProg{Fn}{function}{:}{}

\Fn{\Fchoose{$\gamma$, $H$, $sorted$, $D$}} {
    $D \gets \gamma(D)$ \\
    $sorted \gets sorted~\gamma$\\
    remove $\gamma$ from $H$
}

\Fn{\FMain{$H_1$, $H_2$, $D_0$}}{
    $D \gets D_0$ \\
    $order \gets \emptyset$\\
    \While{$H_1 \neq \emptyset$ and $H_2 \neq \emptyset$}{
        $\phi \gets H_1$'s first element\\
        find the first $\psi$ from $H_2$ that conflict($D$, $\phi$, $\psi$) is True\\
        \eIf{not found}{
            \Fchoose{$\{\phi\}$, $H_1$, $order$, $D$}
        }{
            prompt user to resolve $(\phi, \psi)$ \\
            \eIf{user specifies $\phi < \psi$}{
                \Fchoose{$\{\phi\}$, $H_1$, $order$, $D$}
            }{
                ${\psi}+ \gets$ all $H_2$'s op before $\psi$ (inclusive) \\
                \Fchoose{$\psi+$, $H_2$, $order$, $D$}
            }
        }
    }
    \KwRet{$order ~H_1 ~H_2$}
}
 \caption{Algorithm to resolve conflicts.}
 \label{alg:resolution}
\end{algorithm}


The resolution is expressible in the way that it never prohibits the user from resolving in a certain \inter.
Formally, assume the user has a valid partial order $\pi$ (of $H_1$ and $H_2$) in her mind, and always answers questions posed by \sys consistent with the desired ordering.
It can be shown that the resolution always returns an ordering $\pi'$ that equivalent to $\pi$ (i.e. $\pi'$ and $\pi$ yield the same final database state when applied).

\iflong
\noindent \textbf{Lemma 3: } At any time of this algorithm execution process,  $order$ satisfies that $\alpha \not> \beta$ (in $\pi$) for any $\alpha \in order$ and $\beta \not\in order$. \\

\noindent \textbf{Proof: } The proof is by induction. Before the \textit{while} loop this is certainly true since $order$ is empty. In one step of the \textit{while} loop, (i) if $\psi$ is not found, we have $\phi \not> \psi'$ for all $\psi'\in H_2 \cap order^c$, thus adding $\phi$ to $order$ still obeys the condition specified in Lemma 3; (ii) if $\psi$ is found and  $\phi < \psi$, we have $\phi \not> \rho$ for $\rho < \psi$ and $\phi < \rho$ for $\rho \geq \psi$, thus $\phi \not> \rho$ for all $\rho\in H_2 \cap order^c$, thus adding $\phi$ to $order$ still obeys the condition specified in Lemma 3; (iii) if $\psi$ is found and $\phi > \psi$, we have $\psi' \leq \psi < \phi \leq \phi'$ for any $\psi' \in \psi+$ and $\phi' \in H_1\cap order^c$, thus adding $\psi+$ to $order$ still obeys the condition specified in Lemma 3.  $\square$ \\

\noindent \textbf{Theorem 4:} The returned value $\pi'$ of Algorithm 1 is compatible with $\pi$ (i.e., $\pi$ is a subset of $\pi'$). \\

\noindent \textbf{Proof:} With Lemma 3, we prove the expressible claim by contradiction, then there exist two operations $\alpha$ and $\beta$ such that $\alpha<\beta$ in $\pi'$ and $\alpha>\beta$ in $\pi$. $\alpha<\beta$ in $\pi'$ means when $\alpha$ is appended to $order$, $\beta$ is not appended to $order$. Then the previous observation has $\alpha \not> \beta$ in $\pi$. This is the contradiction, so $\pi'$ is always compatible with $\pi$. \\

\fi

\stitle{Remarks:} \edit{It is important to emphasize that \sys is not a universal solution. Version reconciliation is a semantics-dependent problem: the correctness criteria depend on the application's (or the user's) specifications. As such, it is possible that \sys's conflict resolution framework cannot yield a satisfactory result, either because the user faces a dilemma upon a question (e.g., both orders of a pair of \updates are undesirable) or the final order fails additional constraints upon termination. \sys automates an otherwise-manual process, where the user intends to find a desirable \inter of two histories, but there is no guarantee that such an \inter exists. Even when \sys fails to merge two histories, though, it can still provide valuable information: for example, if an interleaving that results from \sys's version reconciliation violates integrity constraints, one can conclude that no \inter can correctly preserve them. In such cases, reordering modifications doesn't suffice, and users will have to resort to other merging mechanisms (e.g., removing or manually rewriting \updates), which are beyond the scope of this paper.  }

\section{Related Work}


\stitle{Existing Versioning Work}:
Systems that support versions have been developed for many different use cases and can take on many forms. DataHub \cite{bhardwaj2014datahub} provides a central data repository that incorporates relevant SQL and versioning capabilities. 
As part of DataHub, Decibel~\cite{maddox2016decibel} was a prototype centralized versioned-database to evaluate different physical designs and query execution strategies of versioned data.
OrpheusDB \cite{huang2017orpheus} is a ``bolt-on'' centralized versioning  system for relational databases, allowing users to store, track, and execute queries on different versions.
Databricks Delta~\cite{databricksdeltalake} is an industrial solution that can accommodate non-structured data, with special functionalities to streamline the analytics process. ForkBase \cite{wang2018forkbase} is a centralized \dvds that provides similar capabilities for applications that demand tamper evidence, and are of particular use in blockchains. \reviewthree{Versioned databases and ``GitHubs for Data'' have been widely applied in industrial practice~\cite{pachyderm,dolt,daff,noms,dvc}.}

Although these system support data versioning, their versioning control systems are all centralized in the way that users' \updates are synchronized with a server all the time. 
Their conflict identification and version reconciliation rely heavily on human intervene, which quickly becomes infeasible as data grow. OrpheusDB and Decibel track conflicts by materializing updates and performing \texttt{diff}s on different versions. \reviewtwo{As discussed in the introduction, they suffer from several drawbacks.} In the event of a conflict, they \cite{huang2017orpheus} offer some options for resolution, including a precedent ordering (a ranking of which updates are most important) or manual inspection of records. 
Databricks Delta~\cite{databricksdeltalake} also adopts a data-centric conflict identification/resolution scheme by tracking changes via delta logs and maintaining versions corresponding to individual users' \updates, in order to rollback \updates deemed to be in conflict with others. 



\stitle{Conflict Avoidance:} In general, two approaches can ensure that a set of operations are conflict-free (i.e., conflicts are logically impossible, regardless of ordering). 
One approach is for a system to be \textit{conflict-free by design}. \reviewone{Modifications expressed in these systems are always \amle.} Conflict-Free Replicated Data Types (CRDT)~\cite{crdt} are conflict-free data types in a replicated environment.  
CALM~\cite{Hellerstein:2010:DIE:1860702.1860704,AlvaroCHM11,Alvaro:2017:BCA:3155316.3110214} specifies a logical ``monotonic'' language that is guaranteed to be correct without coordination. 

\reviewone{When \updates are not expressible in above systems, they may or may not be \amle (regardless of the specific instance of data). One approach to figure out is \textit{static program analysis}, which} analyzes the operations to be executed in a concurrent environment, and reasons about the correctness guarantee. 
Invariant confluence~\cite{Bailis15,Bailis:2014:CAD:2735508.2735509,Whittaker:2018:ICC:3275536.3300966} and IPA~\cite{Balegas:2018:IIA:3297753.3316433} study when the transaction processing is not necessary to preserve invariants. These mechanisms  \cite{Aleen:2009:CAS:1508244.1508273, Rinard:1997:CAN:267959.269969,Kulkarni:2011:ECL:1993498.1993562,Tripp:2012:JEP:2254064.2254083,vonKoch:2018:TCA:3178372.3179513,9728} detect commutativity in two general programs. 
In the distributed environments, these  approaches~\cite{Clements:2015:SCR:2723895.2699681,Li:2014:ACC:2643634.2643664,Brutschy:2017:SEC:3009837.3009895,Gotsman:2016:CIS:2837614.2837625,Li:2012:MGS:2387880.2387906,ShangY17,Sivaramakrishnan:2015:DPO:2737924.2737981,Balegas:2018:IIA:3297753.3316433,Kraska:2009:CRC:1687627.1687657,216015,Burckhardt:2014:RDT:2535838.2535848} reason about when a strong consistency is necessary, based on the semantics of the operations.
Transaction chopping studies whether a transactions can be split into multiple smaller pieces that still guarantee serializability~\cite{chopping,Zhang:2013:TCA:2517349.2522729,Xie:2015:HAV:2815400.2815430,Wang:2016:SMD:2882903.2882934}.
Bernstein et al. \cite{839387} derive conditions that ensure the transactions are correct under each ANSI isolation level. \edit{Weihl \cite{10.1145/63264.63518} defines three properties on individual data that once enforced, will guarantee serializability.}

\onethree{Bounded staleness of divergent database replicas has also been explored} ~\cite{1319994,Guo:2004:SRC:1007568.1007706,Guo:2004:RCC:1007568.1007661,Guo:2005:CGE:1083592.1083647,Bernstein:2006:RSM:1142473.1142540, Olston:2000:OPT:645926.671877,Olston:2001:APS:375663.375710, Pu:1992:RLS:506378.506415,Wu:1992:DCE:645477.654631,Wu:1997:DCA:627309.627830, 5767927,Zellag:2012:CYC:2391229.2391235,Zellag:2014:CAM:2581628.2581630,Shang:2018:RRI:3183713.3196932}. These approaches do not detect conflicts specifically, nor do they reason about the effects of individual \updates. Moreover, they may not satisfy the demand for \dvds snapshots to be exactly consistent.

\reviewone{Two \updates that are not \amle on all possible data (which could be determined by, say, some variant of a commutativity test) could still be \amle on a specific instance of data. \sys focuses on evaluating \amty on a specific instance of data, significantly relaxing restrictions on allowable API operations.} Conversely, conflict-avoidance approaches often only offer a restrictive set of APIs for the user to express her \updates, and thus may be not practical in a \dvds setting, where it is likely that users will demand a richer set of capabilities when making their \updates.
As such, a conflict management method that relies on \reviewone{restricting operation capabilities to predict/avoid conflicts} is often not ideal.

\stitle{Merge and Fix:} Some merging and fixing approaches require user participation or only work on a situations that are naturally \amle, which in general we cannot assume to be the case in a collaborative setting. Unlike our approach, they cannot detect, nor do they reason about, the effects of conflicts. 
Doppel~\cite{Narula:2014:PRC:2685048.2685088} reconciles branches on CRDT-like data types.
Open nesting and transactional boosting~\cite{Gramoli:2014:DTP:2541883.2541900,Herlihy:2008:TBM:1345206.1345237} allows users to specify compensate/inverse operations to roll back if the two branches are not mergable.
Wu et al.~\cite{Wu:2016:THS:2882903.2915202} and Veldhuizen~\cite{DBLP:journals/corr/Veldhuizen14} speculatively execute the transactions without conflict detection/prevention, and then the fix the broken invariants if there is any exposed conflict.
TARDiS~\cite{Crooks:2016:TBA:2882903.2882951} and Burckhardt et al.~\cite{Burckhardt:2010:CPR:1869459.1869515} allows a user to branch and merge.
ConfluxDB~\cite{Chairunnanda:2014:CMR:2732967.2732970} merges transactions executed in multiple snapshot isolation servers.

\newcommand{\vg}{Vanilla Git}
\newcommand{\og}{Optimized Git}
\newcommand{\rltwd}{Record-Level 3-Way-Diff}

\newif\iftables
\tablesfalse

\iftables {
\begin{figure*}[!htb]
    \small
    \subcaptionbox{Varied Number of Attributes \label{fig:attr_accu}}
    {
    \begin{tabular}{r | c c c}
    & \multicolumn{3}{c}{Database Width (Columns)} \\ 
    & 15 & 30 & 60 \\ 
    \hline
    RL & 6.78 & 5.42 & 4.77 \\ 
    \hline
    CL & 5.37 & 3.01 & 1.95 \\ 
    \hline
    MP & 0.0569 & 0.00732 & 0.00239 \\ 
    \hline
    GT & 0.0567 & 0.00731 & 0.00235 \\ 
    \end{tabular}}
    \quad
    \subcaptionbox{Varied Attribute Biases \label{fig:column_accu}}{
    \begin{tabular}{r | c cc}
    & \multicolumn{3}{c}{Query Selection Biases} \\ 
    & Uniform & High Sel. & Low Sel. \\ 
    \hline
    RL & 5.42 & 16 & 0.0519 \\ 
    \hline
    CL & 3.01 & 11.6 & 0.0415 \\ 
    \hline
    MP & 0.00732 & 0.323 & 0 \\ 
    \hline
    GT & 0.00731 & 0.323 & 0 \\ 
    \end{tabular}}
    \quad
    \subcaptionbox{Varied Database Size \label{fig:size_accu}}{
    \begin{tabular}{r | c c c}
    & \multicolumn{3}{c}{Database Size} \\ 
    & 1 GB & 5 GB & 10 GB \\ 
    \hline
    RL & 5.42 & 5.42 & 5.6 \\ 
    \hline
    CL & 3.01 & 3.01 & 3.4 \\ 
    \hline
    MP & 0.00732 & 0.00746 & 0.00467 \\ 
    \hline
    GT & 0.00731 & 0.00745 & 0.00467 \\ 
    \end{tabular}}
    
    \vspace{0.3cm}
    
    \subcaptionbox{Varied Data Distributions \label{fig:dist_accu}}{
    \begin{tabular}{r | c c c c}
    & \multicolumn{4}{c}{Data Skewness} \\ 
    & Uniform & Slight & Skewed & Extreme \\ 
    \hline
    RL & 5.42 & 5.02 & 4.51 & 4.24 \\ 
    \hline
    CL & 3.01 & 2.82 & 2.63 & 2.52 \\ 
    \hline
    MP & 0.00732 & 0.00792 & 0.00454 & 0.00234 \\ 
    \hline
    GT & 0.00731 & 0.00792 & 0.00454 & 0.00234 \\ 
    \end{tabular}}
    \quad
    \subcaptionbox{Varied History Lengths \label{fig:history_accu}}{
    \begin{tabular}{r | c c c c c}
    & \multicolumn{5}{c}{Number of Queries in History} \\ 
    & 5 & 15 & 25 & 35 & 40 \\ 
    \hline
    RL & 1.38 & 2.41 & 5.42 & 7.44 & 7.56 \\ 
    \hline
    CL & 0.173 & 0.509 & 3.01 & 5.19 & 5.97 \\ 
    \hline
    MP & 0.000148 & 0.00131 & 0.00732 & 0.0119 & 0.063 \\ 
    \hline
    GT & 0.000148 & 0.00131 & 0.00731 & 0.0109 & 0.063 \\ 
    \end{tabular}}
    
    \caption{\onethree{Accuracy of various algorithms while varying data/query history configurations (RL = Record-Level Locking, CL = Cell-Level Locking, MP = \sys, GT = Ground Truth). Each cell of the above tables is the percentage of the database flagged as conflicting by the respective methods. 
    }
    }
    \label{fig:accuracy}
    \label{data1}
\end{figure*}
}

\else {


\begin{figure*}[!h]
    \centering
    \vspace*{-.2cm}
    \subcaptionbox*{}{\includegraphics[width=.55\linewidth]{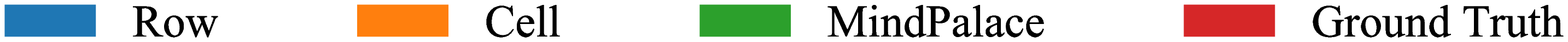}}%
    \vspace*{-.7cm}
    
    \subcaptionbox{Varied Data Distributions\label{fig:dist_accu}}
    {\includegraphics[height=3cm]{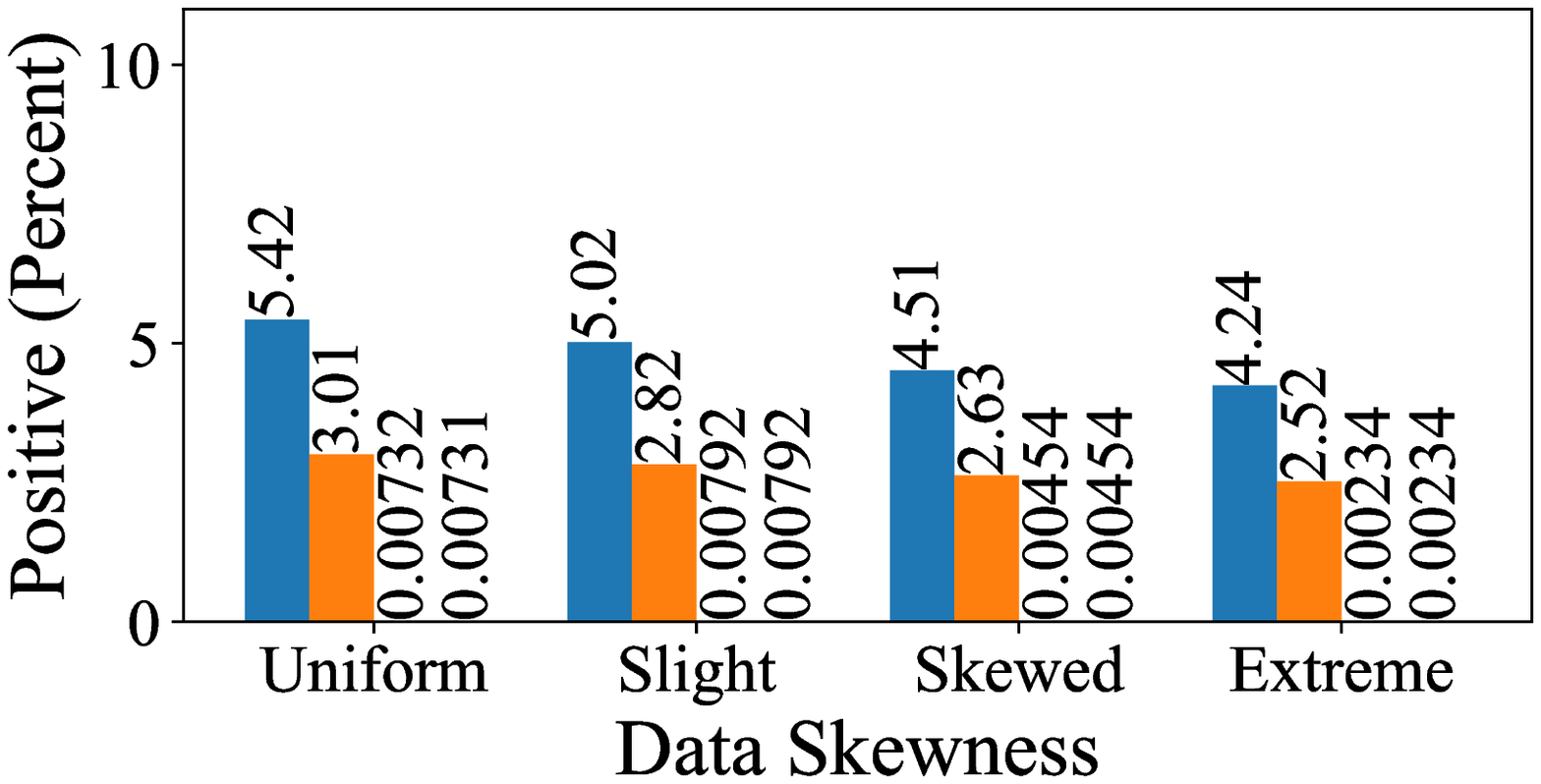}}%
    \subcaptionbox{Varied History Lengths\label{fig:history_accu}}
    {\includegraphics[height=3cm]{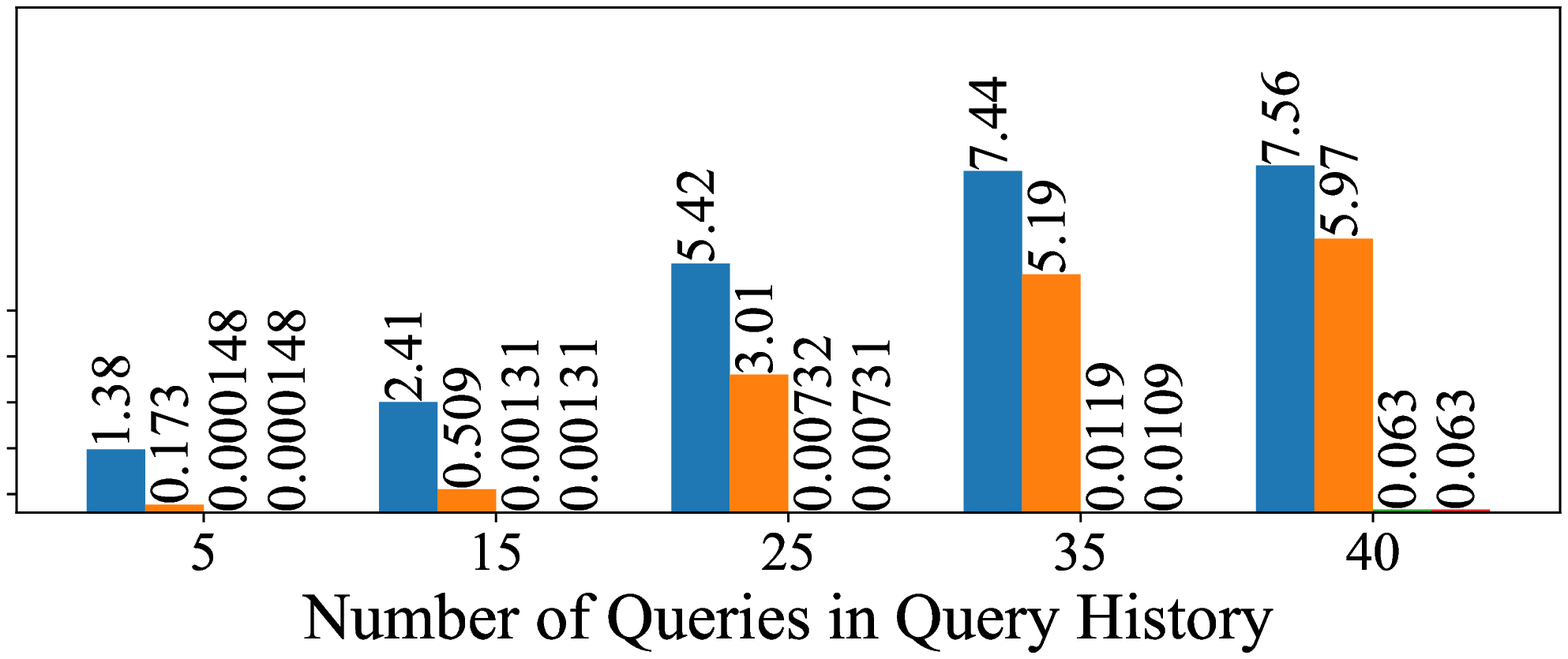}}%
    \subcaptionbox{Varied Attribute Biases\label{fig:column_accu}}
    {\includegraphics[height=3cm]{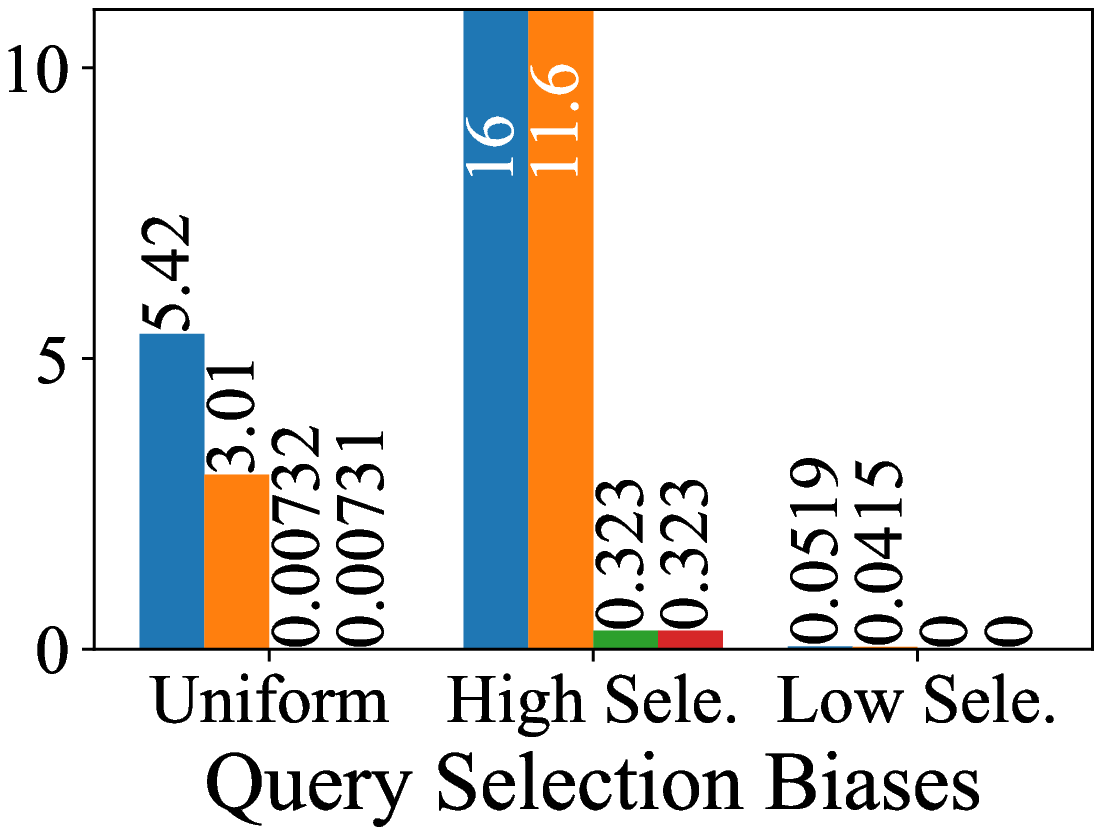}}%
    \vspace*{-0.25cm}
    \caption{Accuracy of various algorithm by various data/query history configurations. For each, we annotate the percent of the database flagged by the respective methods. \sys identifies tuples with $< 1 \%$ margin of error in each experiment, while the runner-up often identifies upward of $100$x the true number of conflicting records. }
    \vspace*{-0.25cm}
    \label{fig:accuracy}
    \label{data1}
\end{figure*}

}

\fi

\section{Experiments}
\label{expr}

\noindent Our experimental study addresses two questions:
(1) How does the accuracy and processing time of our proposed approach compare to alternatives in identifying \namle tuples, under various settings? (2) How reliant on human input is the version reconciliation algorithm for typical cases? 

\vspace{-0.2cm}
\subsection{Experimental Setting}
\label{sec:exp_setting}
\noindent Our evaluations were performed on machines with dual socket Intel Xeon-E5 2650 processors (10 cores, 2.3 GHz each), 64 GB of RAM, and running Ubuntu 18.04. Without otherwise specified, all numbers are averaged based on 10 executions.
As a simplification, all data is numerical, since the \sys's accuracy in identifying conflicts does not depend on the data type and only depends on whether a cell is concurrently modified. All values are generated independently, with all values in the same column following the same distribution. \reviewthree{The number of distinct values for each column range from $100$ to $10^6$, so as to incorporate a range of attribute selectivities.}

\stitle{Alternative Approaches:} We compare our approach with a na\"ive ground truth and some alternative approaches:

    \vspace{0.2cm}
    \noindent \textit{Ground Truth}: We use a dynamic programming algorithm that tracks all possible outcomes of \inters. For every tuple $t \in D$, we track the set of possible tuple $T_{i, j}(t)$ after a \inter of $\pref{i}{H_1}$ and $\pref{j}{H_2}$ applies on $t$. We have
    $T_{i, j}(t) = \phi_i(T_{i - 1, j}(t)) \cup \psi_j(T_{i, j - 1}(t))$. The tuple $t$ is \amle iff  $|T_{|H_1|,|H_2|}(t)| = 1$. The worst case incurs the same complexity as enumerating all \inters but this method runs faster in practice.
    
    \noindent 
    \textit{Compatible \onethree{(Virtual)} Locking}: We adopt locking schemes that help identifying conflicting tuples. The locks do not block \updates; they are for identifying \namle tuples.
    We report \namle tuples that are marked by non-compatible (i.e., read-write or write-write) requests from two histories.  
    \onethree{We have two locking schemes of different granularities, for the trade-off of costs and accuracy.
        For 
        {\textit{Record-Level}}, each tuple (record) has a lock; and
         for {\textit{Cell-Level}}, each cell (data value) has a lock.}
    
    \noindent
    \edit{\textit{Git/Git-Like \texttt{diff} Methods}: We benchmark three variants in Section \ref{sec:added_exp}. \textit{\vg} (VG) simply runs \texttt{git merge} on the final committed states of two versions (dumped to text files) and flags any records in a conflict block as conflicting. \textit{\og} (OG) improves on VG by ignoring records in conflict blocks whose final states were the same for both users. \textit{\rltwd} (RL3) executes three-way-diff logic on a record level, where a record is flagged as conflicting if its state in the original version and the two committed versions are each pairwise different. }

    \noindent 
    \edit{\textit{Online Dataset Versioning}: We benchmark OrpheusDB \cite{huang2017orpheus}, a centralized system for dataset versioning. Once versions are committed, we issue an SQL three-way \lstinline{JOIN} on the versioning metadata to identify conflicting records (using the same logical criterion as in the Record-Level 3-Way Diff method).
    \
    }

\stitle{Implementation:}
\sys is implemented in Python. We store the database in a PostgreSQL 10 server, which executes queries on the database.  
\onethree{The baselines are implemented by first tracking which physical tuples are read/written by each SQL update statement. Then, we flag those that were either written by both users or read by one and written by the other as conflicts.}

\edit{
Both the \vg~ and \og~ approaches use \texttt{git merge} on committed text versions of the dataset, annotating the database dump to identify conflicting blocks. A Python script parses the \texttt{git} annotations to identify \namle records (in the case of \og, it ignores records in conflict blocks with the same final state in both committed versions). To our knowledge, no existing \texttt{git} utility exactly implements the logic desired in the RL3 approach, so we use a Python script 
to identify conflicting records.
}

\edit{
The online dataset versioning experiments used only OrpheusDB's native commands (\texttt{checkout}, \texttt{commit}, \texttt{run}) to track conflicts. At the end, a SQL query was run on OrpheusDB's versioning metadata to identify conflicting records.
{Orpheus stores versioning metadata, including which branches wrote which records, alongside the original dataset in an SQL database.}
}




\subsection{Experiments: Simple Updates}
\label{sec:simple_query_exp}

\begin{table}[t!]
\small
\centering
\vspace*{-.1cm}
\caption{Parameters used in our experiments. Parameters were varied one at a time, with default values in bold.}
\label{tab:parameters}
\vspace*{-.3cm}
\begin{tabular}{r c}
\hline
Parameter & Values and Default (in \textbf{bold})     \\ \hline
{ Distribution Skew} &  \textbf{uniform}, $\beta = 4$, $\beta = 7$, and $\beta = 10$  \\ \hline
{History Length} &  5, 15, \textbf{25}, 35, 40  \\\hline
{Selectivity}  &  \textbf{uniform}, more selective,  less selective   \\\hline
 Width (Columns) & 15, \textbf{30}, 60     \\ \hline
{Volume}   & \textbf{1 GB}, 5 GB,  10 GB    \\ \hline
\end{tabular}
\vspace*{-.4cm}
\end{table}

\noindent We vary five data/history parameters: (1) data skew (by varying the parameter $\beta$ of a beta distribution), (2) history length, (3) selectivity, (4) number of columns, (5) size of the database. The options and default parameters are summarized in Table~\ref{tab:parameters}.
We focus on both accuracy and run time. For accuracy, we consider the relative amount of false positives and false negatives, as \onethree{percentages} of databases. All approaches (including \sys) are free of false negatives; in other words, they never report \namle tuples as \amle. We illustrate the reported positives (\edit{both true and false}) in Figure~\ref{fig:accuracy} \onethree{(we omit the experimental results of (4) number of attributes and (5) size of the database since \sys is mostly insensitive to these parameters). The ground-truth baseline (GT) reports the true set of \namle records \edit{and does not contain false positives}, meaning that \textit{the difference in positivity rates of a (non-ground truth) method and the ground truth represent the false positivity rate.} The lower the false positivity rate, the better.}

\begin{figure}

    \subcaptionbox*{}{\includegraphics[width=1\linewidth]{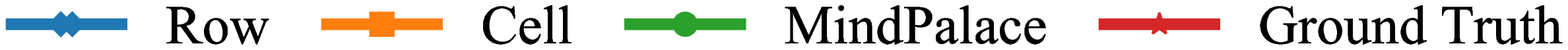}}%
    \vspace*{-.5cm}
    
    \subcaptionbox{Increasing History Length\label{fig:history_time}}
    {\includegraphics[width=.47\linewidth]{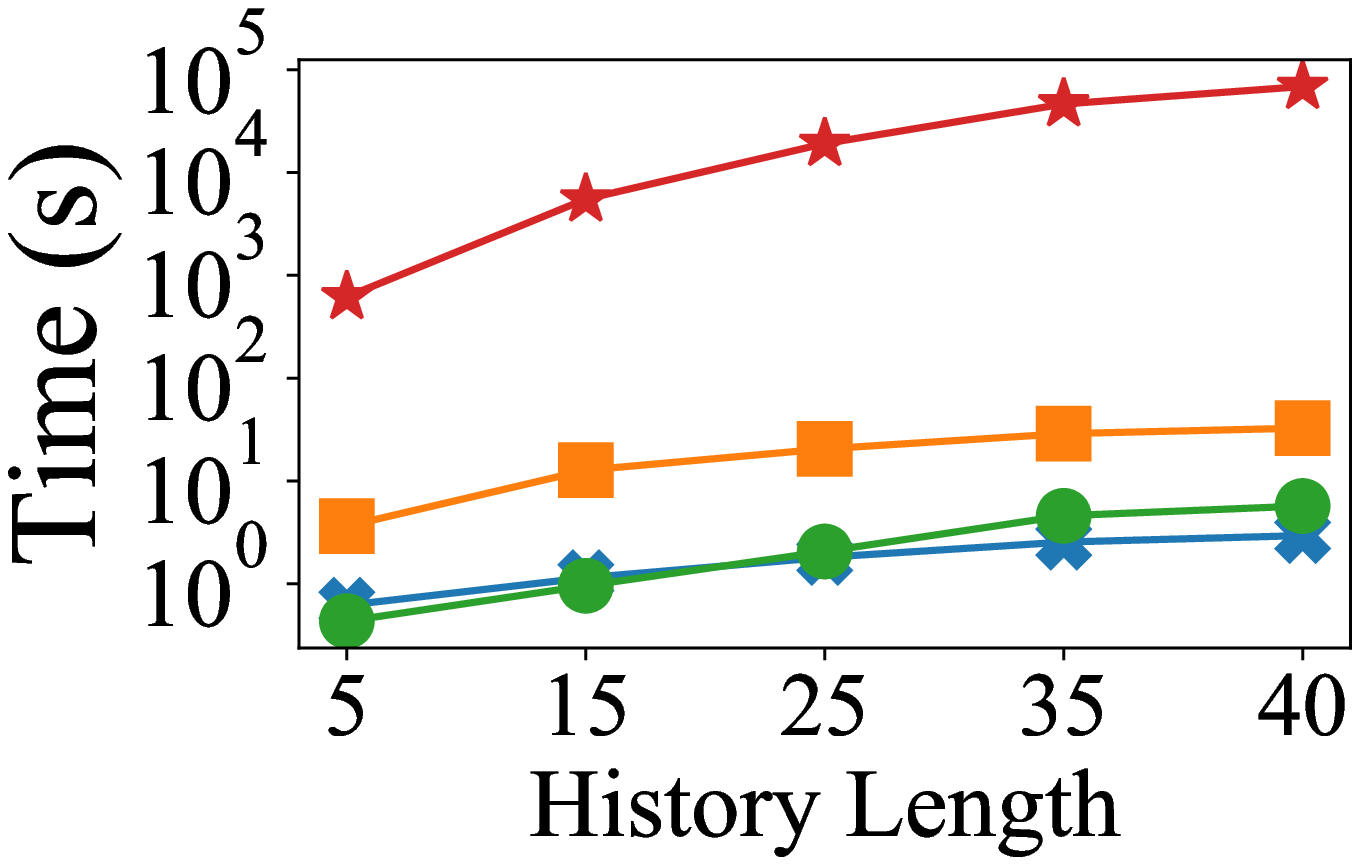}}%
    \hspace{0.2cm}
    \subcaptionbox{Increasing Database Size\label{fig:data_time}}
    {\includegraphics[width=.47\linewidth]{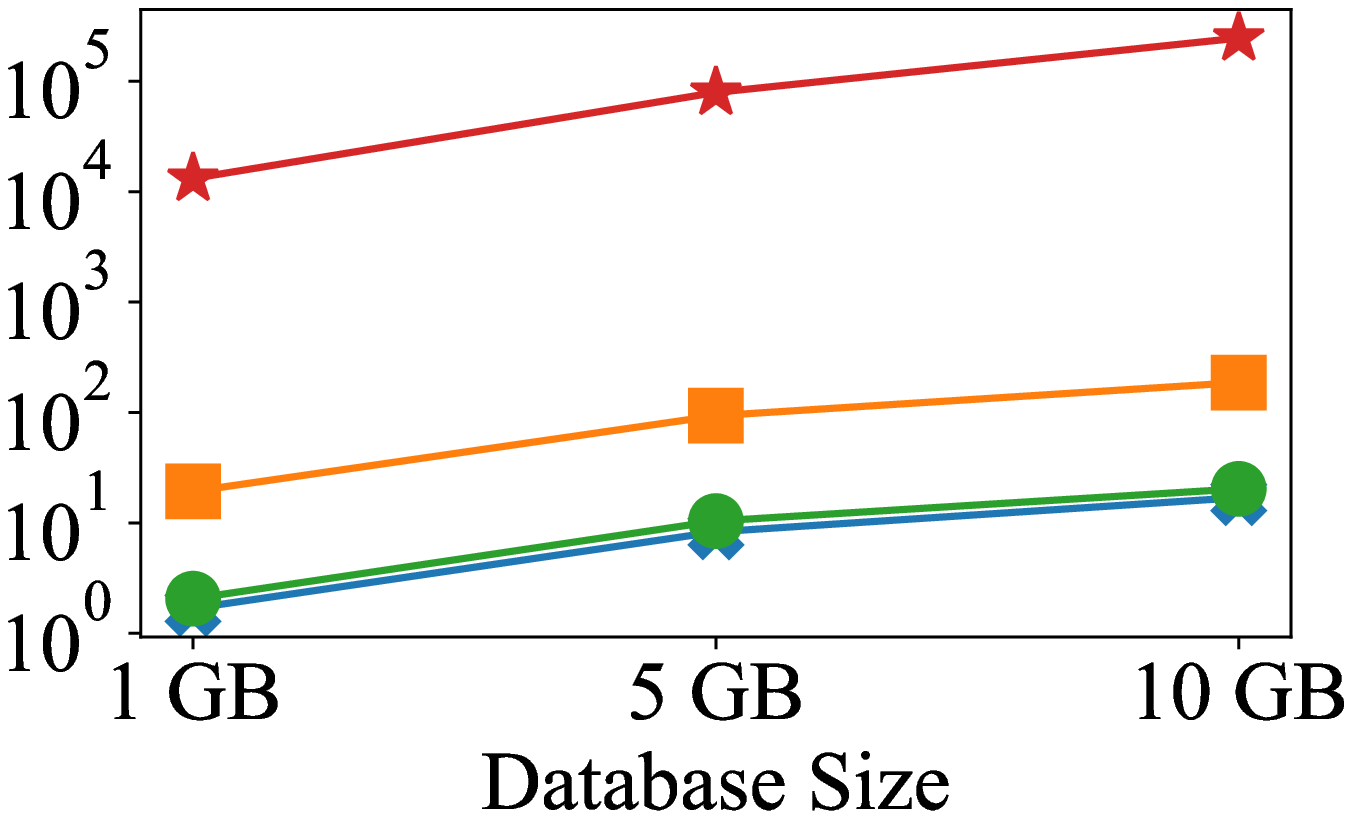}}%
    \vspace*{-0.3cm}
    \caption{Execution Time of each Conflict-Identification Algorithm with Simple Queries
    }\label{fig:time}
    \vspace*{-0.2cm}
\end{figure}

\begin{figure*}[!ht]
    \iftables
    \vspace{-0.3cm}
    \small
    \subcaptionbox{Accuracy: Varied Transaction Types \label{fig:idu_accuracy}}
    {
    \begin{tabular}{r | c c c c}
    & \multicolumn{4}{c}{Ratio of Updates:Inserts:Deletes} \\ 
    & 100:0:0 & 90:5:5 & 75:20:5 & 50:40:10 \\ 
    \hline
    RL & 69.6 & 70.6 & 64.5 & 55.6 \\ 
    \hline
    CL & 53 & 43.7 & 44.9 & 33.7 \\ 
    \hline
    MP & 5.99 & 5.72 & 4.63 & 5.78 \\ 
    \hline
    GT & 5.93 & 5.72 & 4.62 & 5.77 \\ 
    \end{tabular}
    }
    \qquad
    \subcaptionbox{Accuracy: Varied Predicate Types \label{fig:sc_accuracy}}
    {
    \begin{tabular}{r | c c c c}
    & \multicolumn{4}{c}{Ratio of Simple:Complex Preds.} \\ 
    & 100:0 & 90:10 & 80:20 & 70:30 \\ 
    \hline
    RL & 74.2 & 73.1 & 73 & 58.8 \\ 
    \hline
    CL & 56.9 & 57.9 & 54.2 & 36.6 \\ 
    \hline
    MP & 7.31 & 8.66 & 7.98 & 4.5 \\ 
    \hline
    GT & 7.29 & 8.63 & 7.97 & 4.49 \\ 
    \end{tabular}
    }
    \else
    \subcaptionbox*{}{\centering\includegraphics[width=.4\linewidth]{figures/legend.eps}}
    \vspace*{-.7cm}\\
    \subcaptionbox{Accuracy: Varied Transaction Types \label{fig:idu_accuracy}}
    {\includegraphics[width=.28\linewidth]{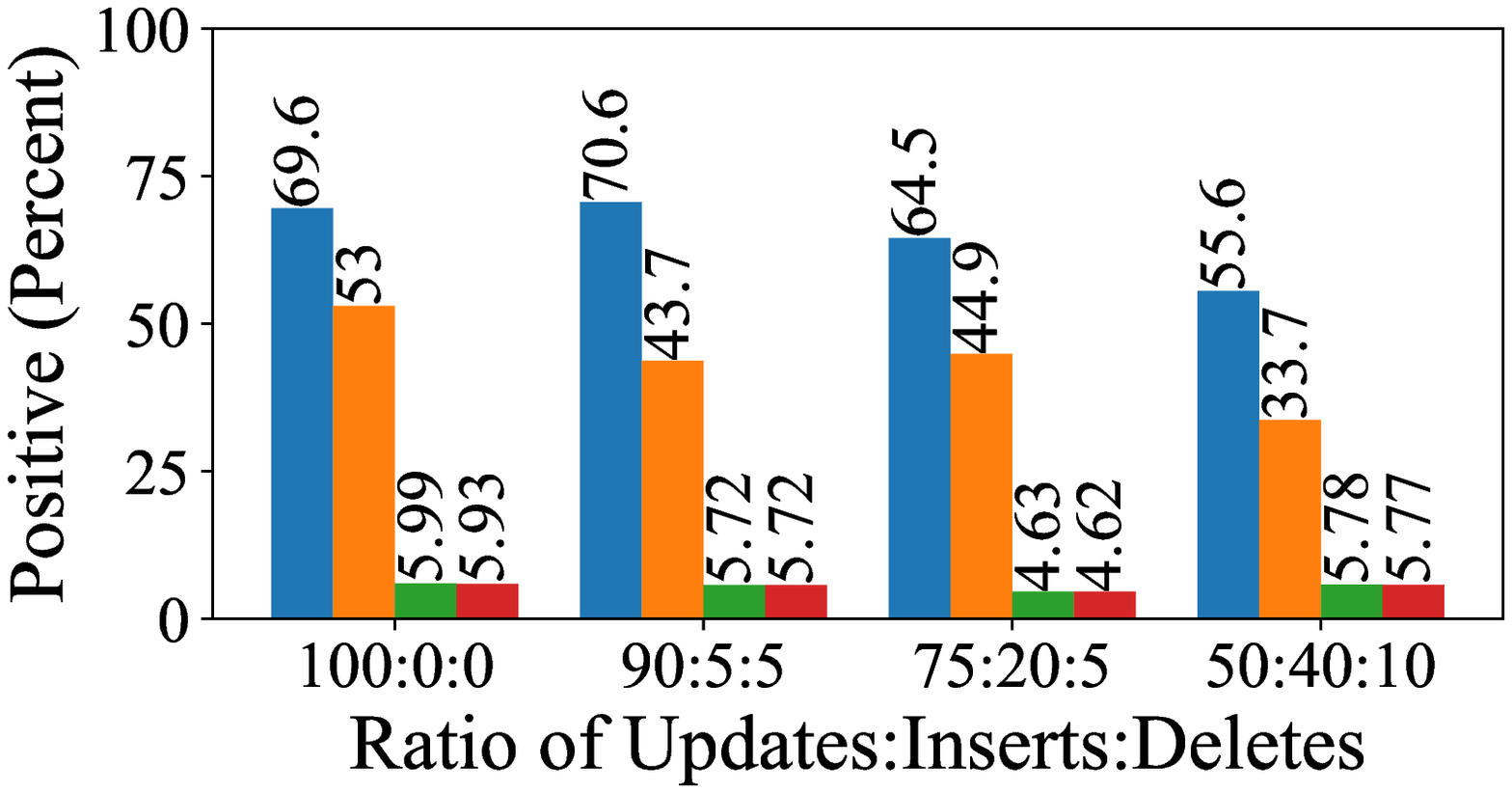}}%
    \subcaptionbox{Accuracy: Varied Predicate Types\label{fig:sc_accuracy}}
    {\includegraphics[width=.28\linewidth]{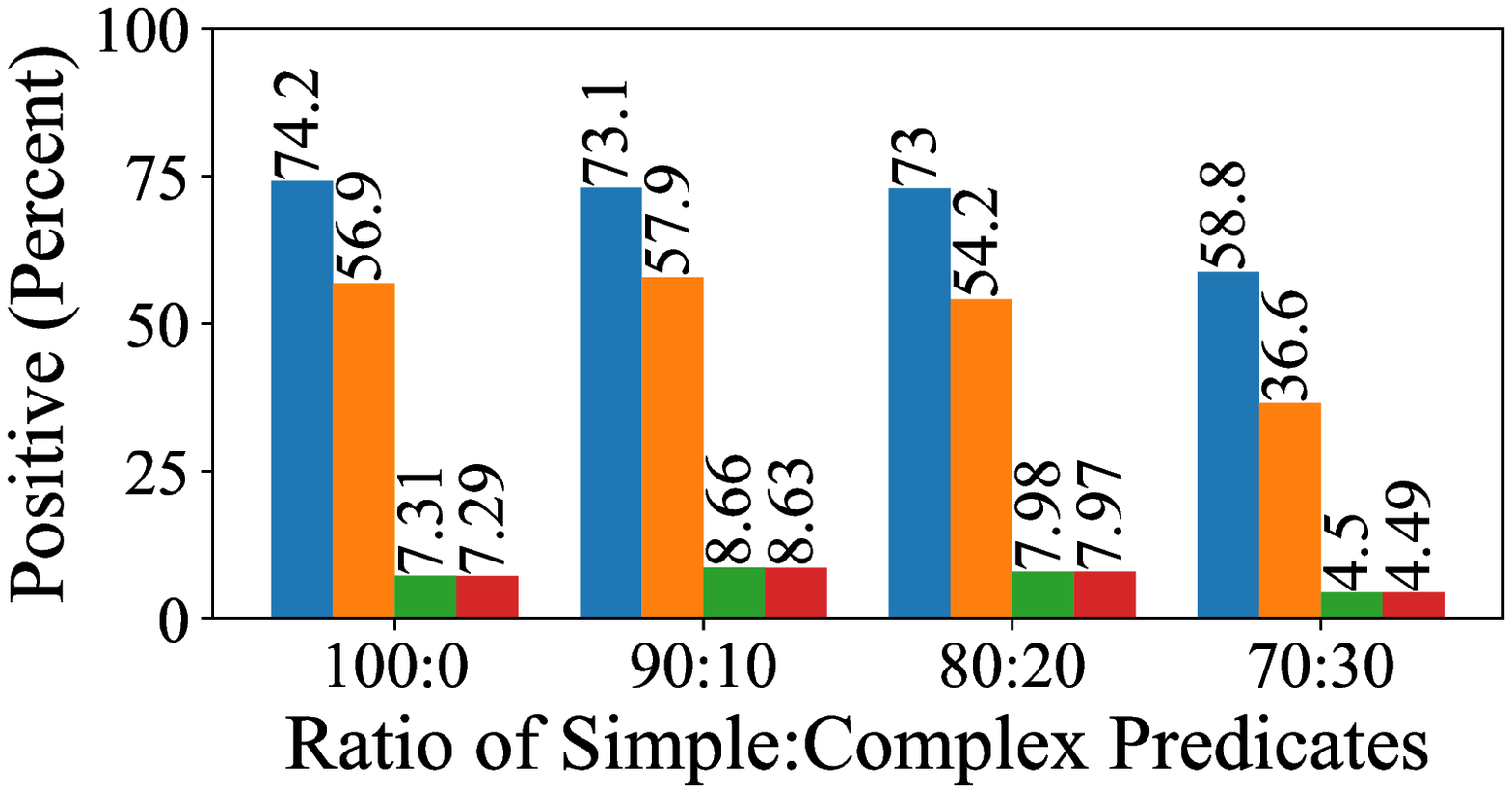}}
    \fi
    \subcaptionbox{Time: Varied Transaction \label{fig:idu_time}}
    {\includegraphics[width=.21\linewidth]{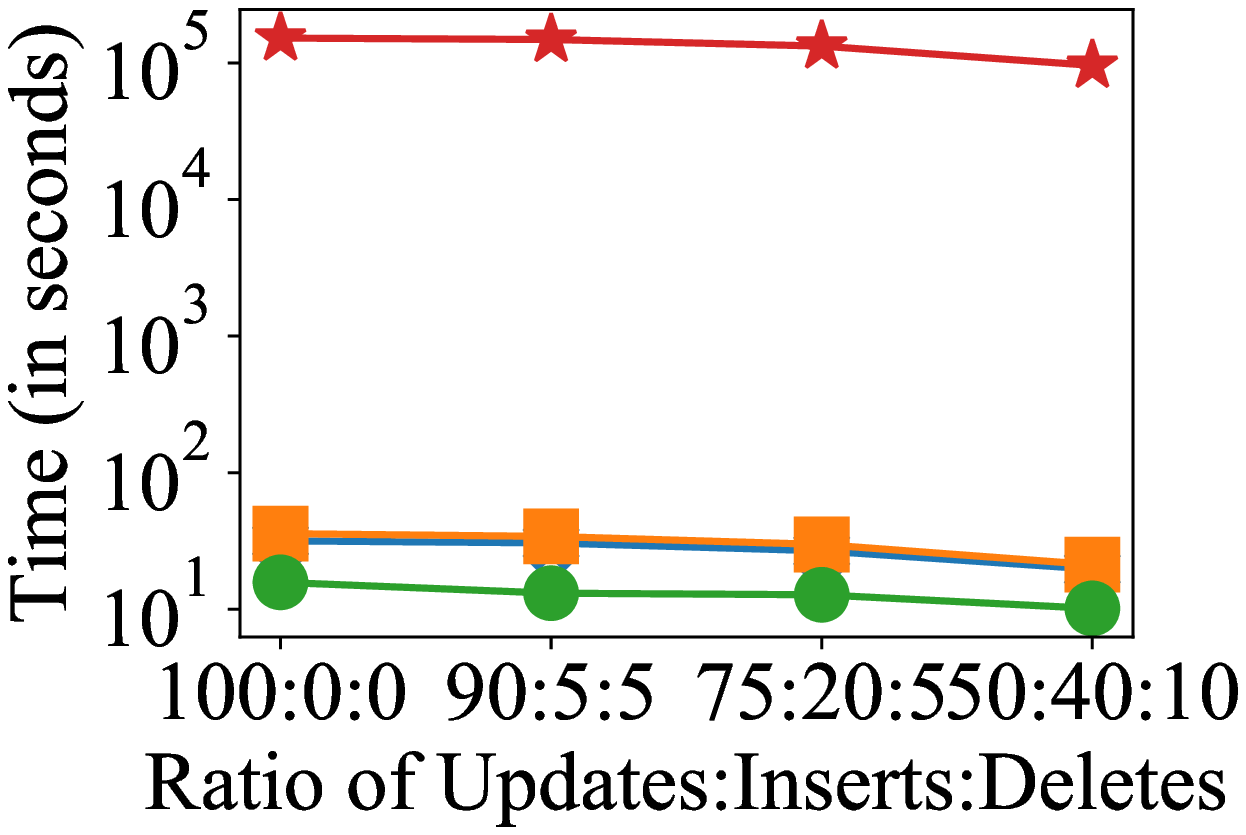}}%
    \subcaptionbox{Time: Varied Predicate\label{fig:sc_time}}
    {\includegraphics[width=.21\linewidth]{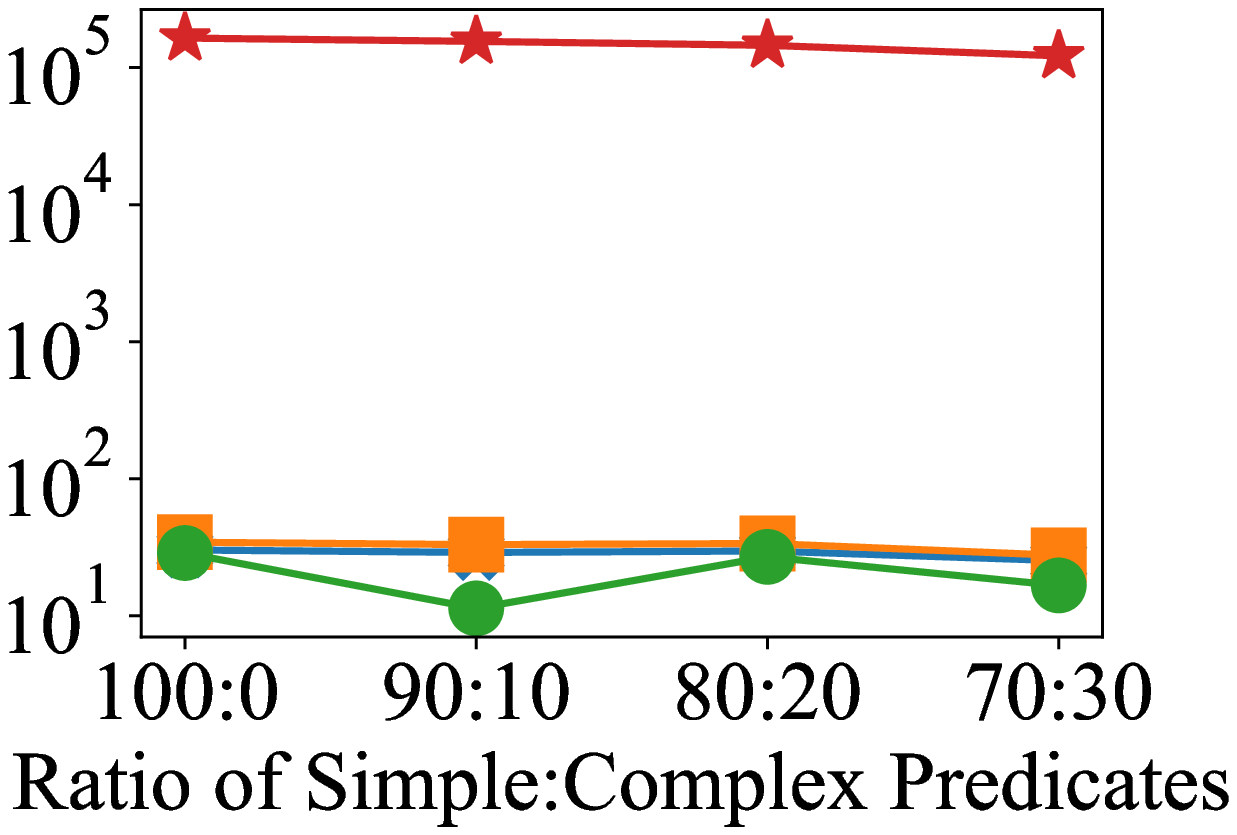}}%
    \vspace*{-0.3cm}
    \caption{Accuracy/run time collected on histories with more complex queries. 
    }\label{fig:complex}
\end{figure*}


In all these experiments, \sys outperforms other alternative approaches. While the amount of positives reported by \sys is close to the true amount of positives,
all locking-based approaches report a considerable portion of database as \namle incorrectly (anywhere from 100 to 10000 times the true number). \onethree{Note that in general fine-grained locking performs better than coarse-grained locking
though even the most fine-grained locking scheme (cell-level) reports orders-of-magnitude extra \namle records.} 

Figure \ref{fig:dist_accu} exhibits variation with different degrees of skewness. The decreasing proportion of conflicting tuples can be explained by the effective decrease in \update selectivity, as many data ranges are now more sparsely populated. 
In Figure \ref{fig:history_accu}, we can observe that the more data is manipulated (i.e., applying more \updates), conflict rates become higher. 
Nonetheless, \sys adjusts seemlessly to the variations in all of these parameters, always identifying a tight set (with <1\% relative margin of error) of tuples for reconciliation.  In Figure \ref{fig:column_accu}, we observe that skewing queries to select and modify more selective attributes (i.e., attributes with fewer distinct values in the domain) significantly increases the proportion of non-automergeable tuples in relation to the total database size, and vice-versa.

\onethree{We display run times in Figure \ref{fig:time} against both increases in database size and history length (both of which when increased will produce higher runtimes for all approaches).} \edit{The remaining parameters, when varied, do not impact runtime noticeably.} We observe that the dynamic programming ground-truth algorithm requires orders of magnitude more time than \sys, due to its higher computational complexity. It needs hours to finish on a 1GB dataset and two histories with just 25 \updates, and thus it is hard to employ such an approach in the real world. The virtual locking schemes employed are relatively lightweight, with associated time increasing as the granularity of locking gets smaller (intuitively explained by the fact that there are now more locks to manage). \sys's time-associated costs are comparable, if not better, than the locking-based approaches (which are already significantly faster than the ground-truth baseline and any ``physical locking'' approach). 

\subsection{Experiments: General Query Model}
\label{sec:general_query_exp}
\noindent We also consider general \updates, as discussed in Section \ref{more_general}. Specifically, each user history contains \lstinline{INSERT}s, \lstinline{DELETE}s, and \lstinline{UPDATE}s with complex predicates including range conditions, set containment, and \lstinline{JOIN} clauses. These experiments were performed on the default parameters as discussed in Section \ref{sec:simple_query_exp}.
Inserts were either generated to be single-tuple inserts, or bulk inserts of $150,000$ new tuples (each approx. 25 MB). Deletes were uniformly generated with equality or range predicates on randomly chosen attributes of the database. Joins with other tables were performed on primary key attributes. Complex predicates included complex logical conditions (e.g. conjunction), range conditions, and set containment queries (signified by keywords like \lstinline{IN}, \lstinline{EXISTS}, etc.). Each individual \update affects no more than 15\% of the  database. We illustrate the results in Figure \ref{fig:complex}./zo

In the first set of experiments, we vary ratios of query types while keeping the ratio of simple predicates to complex predicates at a constant $80/20$ percentage ratio. We choose the  percentage ratios of \lstinline{UPDATE}/\lstinline{INSERT}/\lstinline{DELETE} as 100/0/0, 90/5/5, 75/20/5, and 50/40/10. In the second set of experiments, we maintain the ratio of \lstinline{UPDATE}/\lstinline{INSERT}/\lstinline{DELETE} constant at 75/20/5, and vary the selection of simple predicates to complex predicates in the following ratios: 100/0, 90/10, 80/20, 70/30.
In both experiments, we observe that as with the simple query experiments, \sys identifies a tight set of conflicting tuples that strays less than $1\%$ from the true number. Note that the number of true positives has increased due to increased query selectivities of complex queries and that the number of false positives incurred by locking-based has grown disproportionately large. 
\sys's computational time is comparable to those of locking-based approaches, and is orders-of-magnitude faster than the dynamic programming approach, which has suffered approximately an order-of-magnitude increase in run time.

\sys runs more quickly as the number of \lstinline{INSERT}s and \lstinline{DELETE}s are increased. We believe that this occurs for two reasons: 1) querying inserted data is easier than querying already existing data since inserted data is smaller in size, and 2) the predicates that arise from handling conflicts with \lstinline{INSERT}s and \lstinline{DELETE}s are simpler than those involving \lstinline{UPDATE}s. Second, \sys's run time does not scale strongly with increases in the number of complex predicates. 
 
\subsection{Experiments: Diff-Based Methods}
\label{sec:added_exp}
\edit{
Next, we evaluate the accuracy and performance time of diff-based methods (i.e., the git-based methods and OrpheusDB). We use the parameters described in Section \ref{sec:simple_query_exp} (database width of 30, uniform selectivity, uniform data distributions, and a history length of 25).}

\edit{
As shown in Figure \ref{fig:diff_figure}, diff-based methods fail to accurately identify \amle tuples. \vg~ identifies more than 10 times the true number of positives and even still may fail to flag all \namle records. {\vg's accuracy is a result of Git's block-based conflict identification, which results in large blocks of conflicts being identified, even if most of the records in those blocks were untouched by both users. \og~ slightly improves upon \vg~but still identifies more than 9 times the true number.} While RL3 and OrpheusDB significantly improve upon \vg~ and \og~ false positive rate, the improvement comes at an unacceptable cost, as upwards of 75\% of \namle records may be falsely flagged as \amle. In general, accuracy in identifying conflicts is independent of any underlying implementation details.
}

\edit{Figure \ref{fig:3wd_exp} also benchmarks both the time needed to \textit{commit} a set of \updates and any tracking metadata, as well as the time needed to \textit{analyze} the committed data to find conflicts. The significant overhead with both recording and shipping data diffs is reflected in the increased commit times. Note that the commit times for the logical tracking methods are negligible, since only histories, at most a few kilobytes, need to be shipped. 
Figure \ref{fig:diff_figure} demonstrates that analyzing data diffs for conflicts is significantly more intensive than logical approaches like \sys and virtual locking, whose largest overhead is querying the original version. For the git-based methods, the extra overhead in the analysis step is largely due to the textual processing of the \texttt{git diff} output (which is at least as large as a textual dump of the database). For OrpheusDB, we believe the extra overhead is due to the view materialization needed to execute a three-way-\lstinline{JOIN} query on its versioning metadata to extract conflicting records.
}

\begin{figure}
    \footnotesize
    \small
    \centering
    \setlength\tabcolsep{4pt}
    \begin{tabular}{r | c c c || cc}
    & TP & FP & FN & Commit (s) & Analysis (s)\\
    \hline
    \vg & 3.50 & 38.32 & 0.16 & 146.60 & 76.90\\
    \hline
    \og & 3.50 & 32.48 & 0.16 & 143.40 & 76.70\\
    \hline
    RL3 & 0.65 & 6.27 & 3.01 & 60.90 & 22.00 \\
    \hline
    OrpheusDB & 0.65 & 6.27 & 3.01 & 150.90 & 176.80\\
    \hline
    Record Level & 3.66 & 49.24 & 0 & 0 & 18.70 \\
    \hline
    \sys & 3.66 & $\scriptstyle 3.74 \times 10^{-3}$ & 0 & 0 & 5.32  \\
    \end{tabular}
    \caption{\edit{Accuracy/run time of the diff-based methods (RL3 = \rltwd) with the parameters described in Section \ref{sec:added_exp}. TP, FP, and FN denote the number of true positives, false positives, and false negatives identified (as a \% of the database), respectively.  \label{fig:diff_figure}}
    }
    \label{fig:3wd_exp}
\end{figure}


\begin{figure}[t]
\centering
\includegraphics[width=.6 \linewidth]{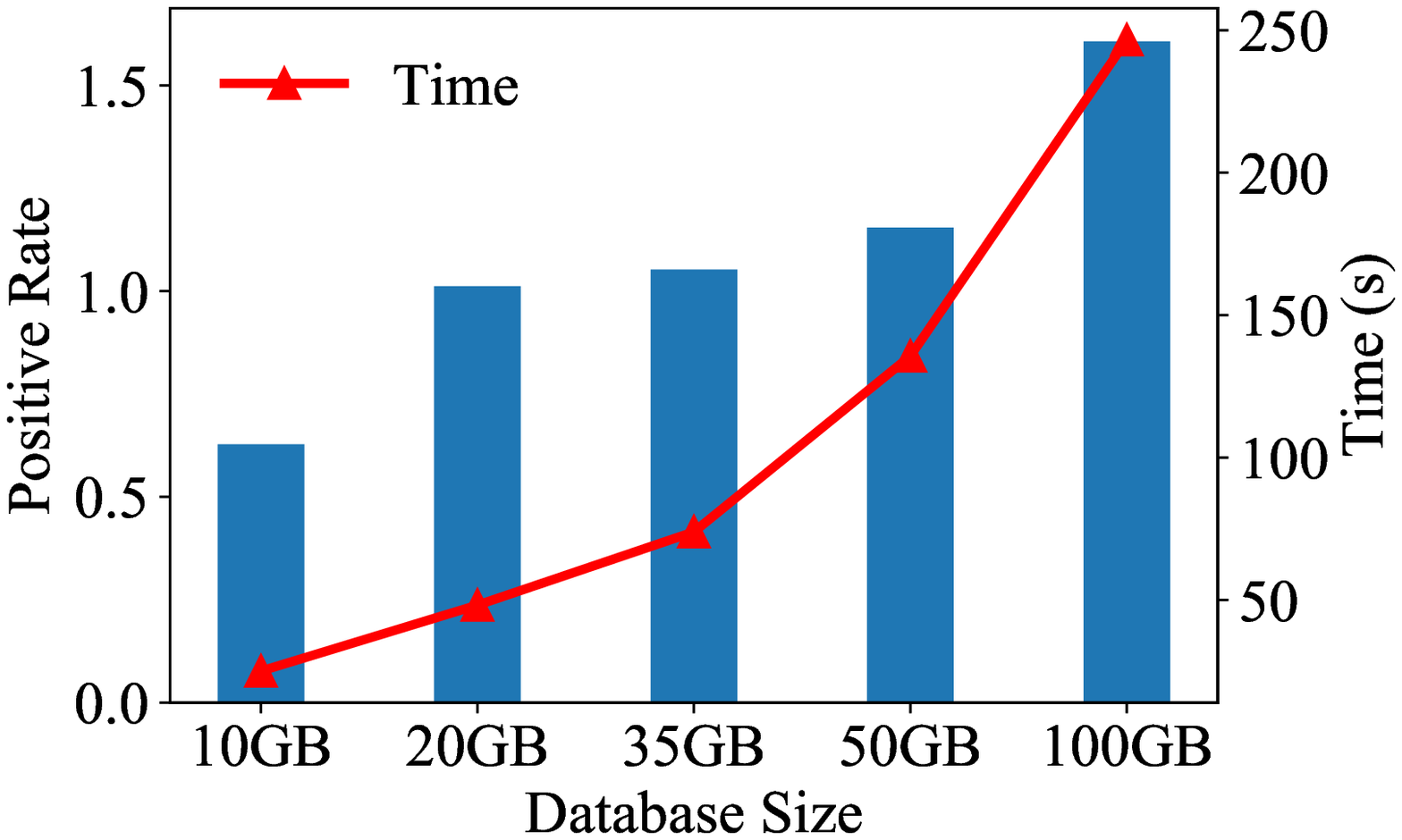}
\vspace*{-0.2cm}
\caption{Evaluating Performance with Scaling Size}  \label{fig:scalability}
\vspace*{-0.3cm}
\end{figure}

\begin{figure}[t]
\centering
\includegraphics[width=.58\linewidth]{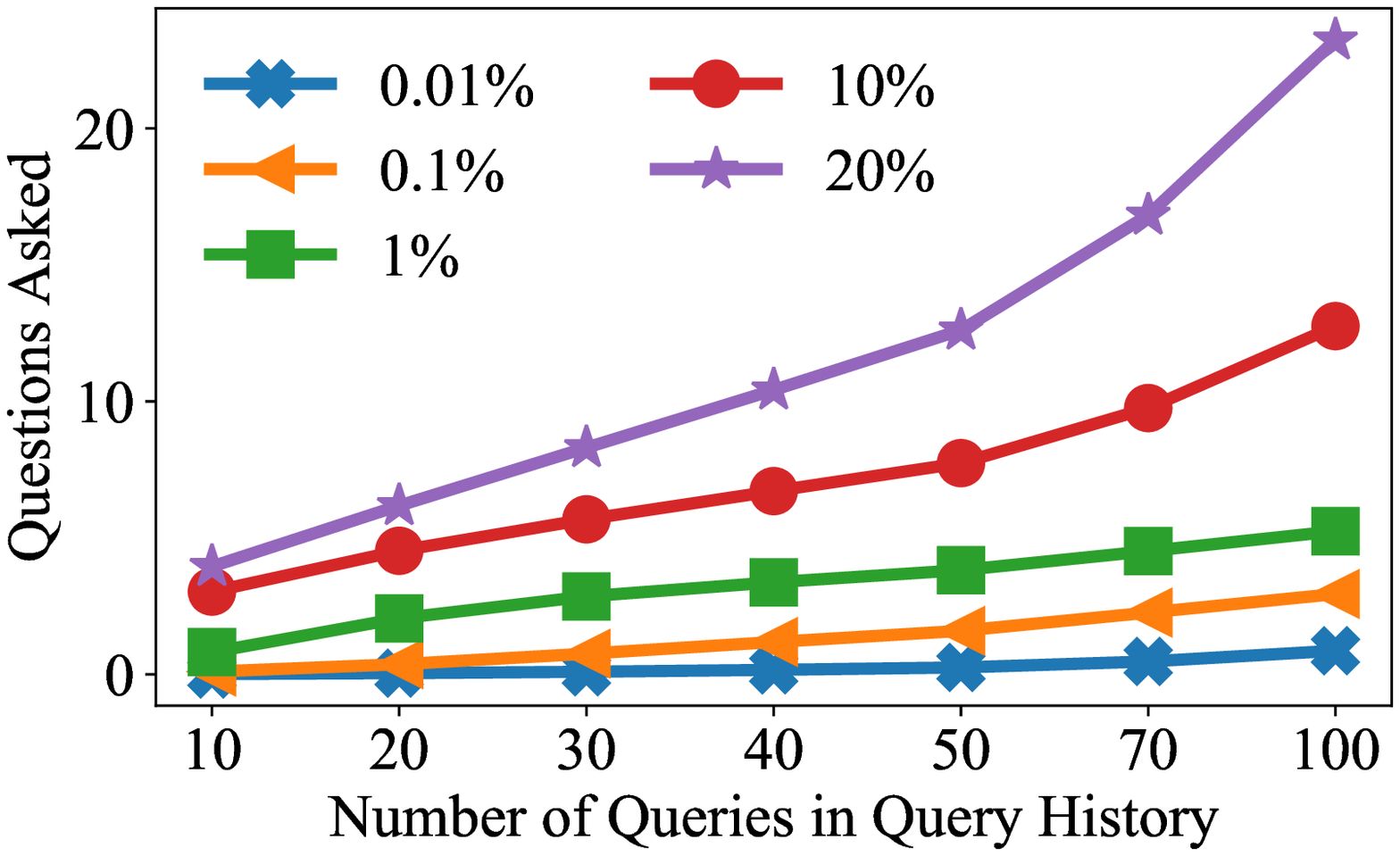}
\vspace*{-0.2cm}
\caption{Human Interactions during Version Reconciliation}  \label{fig:merge}
\vspace*{-0.5cm}
\end{figure}

\subsection{Scalability of \sys}
\noindent The ground-truth algorithm, even with optimizations, simply becomes infeasible at these scales in terms of running time. In contrast, \sys not only scales well but also has roughly thee same accuracy as size increases.
We vary larger data sizes from 10 GB to 100 GB, and demonstrate that performance times remain practical. In each trial, we report the average running time of a size $15$ history with \lstinline{UPDATES/INSERTS/DELETES} ratio = 75/20/5 and simple/complex predicates ratio = 80/20. We illustrate the results in Figure~\ref{fig:scalability}, maintaining the rest of the default data parameters. 

\subsection{Version Reconciliation}
\noindent We measure the performance of version reconciliation. We consider different history sizes ranging from 10 to 100, and probabilities of conflicting ranging from 0.01\% to 20\% for each pair of \updates (a workflow with higher query selectivities may not be fit for a \dvds). 
For each parameter set, we generate 100k different randomly generated partial orders as the user's desired \inter, simulate our user interaction framework, and report the average number of questions asked by \sys in Figure~\ref{fig:merge}. The figure shows that the number of question asked increase as the size of history and/or the probability of conflict increase, but the absolute number remains low. For example, when the size of histories is 100 and conflict probability is 1\%, \sys asks less than 5 questions on average \reviewone{(note that in this case, the expected number of conflicting \update pairs is $100\times100\times1\%=100$). This not only verifies our theoretical analysis that the number of user-interactions is bounded by the (sum of) sizes of the two histories, but also demonstrates the efficiency of our version reconciliation framework.}
\section{Conclusion}
\noindent We present \sys, a lightweight method for version reconciliation in distributed versioned database systems. 
\sys identifies when disconnected histories of operations can safely merge together, and if not identifies the records and conflicting operations that need to be manually reconciled. 
We believe \sys is a critical component for building disconnected collaborative data systems, and opens up interesting subsequent research questions.


\bibliographystyle{ACM-Reference-Format}
\bibliography{references}


\begin{thebibliography}{65}


\ifx \showCODEN    \undefined \def \showCODEN     #1{\unskip}     \fi
\ifx \showDOI      \undefined \def \showDOI       #1{#1}\fi
\ifx \showISBNx    \undefined \def \showISBNx     #1{\unskip}     \fi
\ifx \showISBNxiii \undefined \def \showISBNxiii  #1{\unskip}     \fi
\ifx \showISSN     \undefined \def \showISSN      #1{\unskip}     \fi
\ifx \showLCCN     \undefined \def \showLCCN      #1{\unskip}     \fi
\ifx \shownote     \undefined \def \shownote      #1{#1}          \fi
\ifx \showarticletitle \undefined \def \showarticletitle #1{#1}   \fi
\ifx \showURL      \undefined \def \showURL       {\relax}        \fi
\providecommand\bibfield[2]{#2}
\providecommand\bibinfo[2]{#2}
\providecommand\natexlab[1]{#1}
\providecommand\showeprint[2][]{arXiv:#2}

\bibitem[\protect\citeauthoryear{Aleen and Clark}{Aleen and Clark}{2009}]%
        {Aleen:2009:CAS:1508244.1508273}
\bibfield{author}{\bibinfo{person}{Farhana Aleen} {and} \bibinfo{person}{Nathan
  Clark}.} \bibinfo{year}{2009}\natexlab{}.
\newblock \showarticletitle{Commutativity Analysis for Software
  Parallelization: Letting Program Transformations See the Big Picture}. In
  \bibinfo{booktitle}{\emph{Proceedings of the 14th International Conference on
  Architectural Support for Programming Languages and Operating Systems}}
  (Washington, DC, USA) \emph{(\bibinfo{series}{ASPLOS XIV})}.
  \bibinfo{publisher}{ACM}, \bibinfo{address}{New York, NY, USA},
  \bibinfo{pages}{241--252}.
\newblock
\showISBNx{978-1-60558-406-5}
\urldef\tempurl%
\url{https://doi.org/10.1145/1508244.1508273}
\showDOI{\tempurl}


\bibitem[\protect\citeauthoryear{Alvaro, Conway, Hellerstein, and Maier}{Alvaro
  et~al\mbox{.}}{2017}]%
        {Alvaro:2017:BCA:3155316.3110214}
\bibfield{author}{\bibinfo{person}{Peter Alvaro}, \bibinfo{person}{Neil
  Conway}, \bibinfo{person}{Joseph~M. Hellerstein}, {and}
  \bibinfo{person}{David Maier}.} \bibinfo{year}{2017}\natexlab{}.
\newblock \showarticletitle{Blazes: Coordination Analysis and Placement for
  Distributed Programs}.
\newblock \bibinfo{journal}{\emph{ACM Trans. Database Syst.}}
  \bibinfo{volume}{42}, \bibinfo{number}{4}, Article \bibinfo{articleno}{23}
  (\bibinfo{date}{Oct.} \bibinfo{year}{2017}), \bibinfo{numpages}{31}~pages.
\newblock
\showISSN{0362-5915}
\urldef\tempurl%
\url{https://doi.org/10.1145/3110214}
\showDOI{\tempurl}


\bibitem[\protect\citeauthoryear{Alvaro, Conway, Hellerstein, and
  Marczak}{Alvaro et~al\mbox{.}}{2011}]%
        {AlvaroCHM11}
\bibfield{author}{\bibinfo{person}{Peter Alvaro}, \bibinfo{person}{Neil
  Conway}, \bibinfo{person}{Joseph~M. Hellerstein}, {and}
  \bibinfo{person}{William~R. Marczak}.} \bibinfo{year}{2011}\natexlab{}.
\newblock \showarticletitle{Consistency Analysis in Bloom: a {CALM} and
  Collected Approach}. In \bibinfo{booktitle}{\emph{{CIDR} 2011, Fifth Biennial
  Conference on Innovative Data Systems Research, Asilomar, CA, USA, January
  9-12, 2011, Online Proceedings}}. \bibinfo{publisher}{www.cidrdb.org},
  \bibinfo{address}{Asilomar, California}, \bibinfo{pages}{249--260}.
\newblock
\urldef\tempurl%
\url{http://cidrdb.org/cidr2011/Papers/CIDR11\_Paper35.pdf}
\showURL{%
\tempurl}


\bibitem[\protect\citeauthoryear{{Attic Labs}.}{{Attic Labs}.}{2021}]%
        {noms}
\bibfield{author}{\bibinfo{person}{{Attic Labs}.}}
  \bibinfo{year}{2021}\natexlab{}.
\newblock \bibinfo{title}{noms}.
\newblock \bibinfo{howpublished}{\url{https://github.com/attic-labs/noms}}.
\newblock


\bibitem[\protect\citeauthoryear{Bailis}{Bailis}{2015}]%
        {Bailis15}
\bibfield{author}{\bibinfo{person}{Peter Bailis}.}
  \bibinfo{year}{2015}\natexlab{}.
\newblock \emph{\bibinfo{title}{Coordination Avoidance in Distributed
  Databases}}.
\newblock \bibinfo{thesistype}{Ph.D. Dissertation}. \bibinfo{school}{University
  of California, Berkeley, {USA}}.
\newblock
\urldef\tempurl%
\url{http://www.escholarship.org/uc/item/8k8359g2}
\showURL{%
\tempurl}


\bibitem[\protect\citeauthoryear{Bailis, Fekete, Franklin, Ghodsi, Hellerstein,
  and Stoica}{Bailis et~al\mbox{.}}{2014}]%
        {Bailis:2014:CAD:2735508.2735509}
\bibfield{author}{\bibinfo{person}{Peter Bailis}, \bibinfo{person}{Alan
  Fekete}, \bibinfo{person}{Michael~J. Franklin}, \bibinfo{person}{Ali Ghodsi},
  \bibinfo{person}{Joseph~M. Hellerstein}, {and} \bibinfo{person}{Ion Stoica}.}
  \bibinfo{year}{2014}\natexlab{}.
\newblock \showarticletitle{Coordination Avoidance in Database Systems}.
\newblock \bibinfo{journal}{\emph{Proc. VLDB Endow.}} \bibinfo{volume}{8},
  \bibinfo{number}{3} (\bibinfo{date}{Nov.} \bibinfo{year}{2014}),
  \bibinfo{pages}{185--196}.
\newblock
\showISSN{2150-8097}
\urldef\tempurl%
\url{https://doi.org/10.14778/2735508.2735509}
\showDOI{\tempurl}


\bibitem[\protect\citeauthoryear{Balegas, Duarte, Ferreira, Rodrigues, and
  Pregui\c{c}a}{Balegas et~al\mbox{.}}{2018}]%
        {Balegas:2018:IIA:3297753.3316433}
\bibfield{author}{\bibinfo{person}{Valter Balegas}, \bibinfo{person}{S{\'e}rgio
  Duarte}, \bibinfo{person}{Carla Ferreira}, \bibinfo{person}{Rodrigo
  Rodrigues}, {and} \bibinfo{person}{Nuno Pregui\c{c}a}.}
  \bibinfo{year}{2018}\natexlab{}.
\newblock \showarticletitle{IPA: Invariant-preserving Applications for Weakly
  Consistent Replicated Databases}.
\newblock \bibinfo{journal}{\emph{Proc. VLDB Endow.}} \bibinfo{volume}{12},
  \bibinfo{number}{4} (\bibinfo{date}{Dec.} \bibinfo{year}{2018}),
  \bibinfo{pages}{404--418}.
\newblock
\showISSN{2150-8097}
\urldef\tempurl%
\url{https://doi.org/10.14778/3297753.3297760}
\showDOI{\tempurl}


\bibitem[\protect\citeauthoryear{{Bernstein}, {Lewis}, and {Lu}}{{Bernstein}
  et~al\mbox{.}}{2000}]%
        {839387}
\bibfield{author}{\bibinfo{person}{A.~J. {Bernstein}}, \bibinfo{person}{P.~M.
  {Lewis}}, {and} \bibinfo{person}{S. {Lu}}.} \bibinfo{year}{2000}\natexlab{}.
\newblock \showarticletitle{Semantic conditions for correctness at different
  isolation levels}. In \bibinfo{booktitle}{\emph{Proceedings of 16th
  International Conference on Data Engineering (Cat. No.00CB37073)}}.
  \bibinfo{publisher}{IEEE Computer Society}, \bibinfo{address}{Washington, DC,
  USA}, \bibinfo{pages}{57--66}.
\newblock
\showISSN{1063-6382}
\urldef\tempurl%
\url{https://doi.org/10.1109/ICDE.2000.839387}
\showDOI{\tempurl}


\bibitem[\protect\citeauthoryear{Bernstein, Fekete, Guo, Ramakrishnan, and
  Tamma}{Bernstein et~al\mbox{.}}{2006}]%
        {Bernstein:2006:RSM:1142473.1142540}
\bibfield{author}{\bibinfo{person}{Philip~A. Bernstein}, \bibinfo{person}{Alan
  Fekete}, \bibinfo{person}{Hongfei Guo}, \bibinfo{person}{Raghu Ramakrishnan},
  {and} \bibinfo{person}{Pradeep Tamma}.} \bibinfo{year}{2006}\natexlab{}.
\newblock \showarticletitle{Relaxed-currency Serializability for Middle-tier
  Caching and Replication}. In \bibinfo{booktitle}{\emph{Proceedings of the
  2006 ACM SIGMOD International Conference on Management of Data}} (Chicago,
  IL, USA) \emph{(\bibinfo{series}{SIGMOD '06})}. \bibinfo{publisher}{ACM},
  \bibinfo{address}{New York, NY, USA}, \bibinfo{pages}{599--610}.
\newblock
\showISBNx{1-59593-434-0}
\urldef\tempurl%
\url{https://doi.org/10.1145/1142473.1142540}
\showDOI{\tempurl}


\bibitem[\protect\citeauthoryear{Bhardwaj, Bhattacherjee, Chavan, Deshpande,
  Elmore, Madden, and Parameswaran}{Bhardwaj et~al\mbox{.}}{2015}]%
        {bhardwaj2014datahub}
\bibfield{author}{\bibinfo{person}{Anant~P. Bhardwaj}, \bibinfo{person}{Souvik
  Bhattacherjee}, \bibinfo{person}{Amit Chavan}, \bibinfo{person}{Amol
  Deshpande}, \bibinfo{person}{Aaron~J. Elmore}, \bibinfo{person}{Samuel
  Madden}, {and} \bibinfo{person}{Aditya~G. Parameswaran}.}
  \bibinfo{year}{2015}\natexlab{}.
\newblock \showarticletitle{DataHub: Collaborative Data Science {\&} Dataset
  Version Management at Scale}. In \bibinfo{booktitle}{\emph{{CIDR} 2015,
  Seventh Biennial Conference on Innovative Data Systems Research, Asilomar,
  CA, USA, January 4-7, 2015, Online Proceedings}}.
  \bibinfo{publisher}{www.cidrdb.org}.
\newblock
\urldef\tempurl%
\url{http://cidrdb.org/cidr2015/Papers/CIDR15\_Paper18.pdf}
\showURL{%
\tempurl}


\bibitem[\protect\citeauthoryear{Brutschy, Dimitrov, M\"{u}ller, and
  Vechev}{Brutschy et~al\mbox{.}}{2017}]%
        {Brutschy:2017:SEC:3009837.3009895}
\bibfield{author}{\bibinfo{person}{Lucas Brutschy}, \bibinfo{person}{Dimitar
  Dimitrov}, \bibinfo{person}{Peter M\"{u}ller}, {and} \bibinfo{person}{Martin
  Vechev}.} \bibinfo{year}{2017}\natexlab{}.
\newblock \showarticletitle{Serializability for Eventual Consistency:
  Criterion, Analysis, and Applications}. In
  \bibinfo{booktitle}{\emph{Proceedings of the 44th ACM SIGPLAN Symposium on
  Principles of Programming Languages}} (Paris, France)
  \emph{(\bibinfo{series}{POPL 2017})}. \bibinfo{publisher}{ACM},
  \bibinfo{address}{New York, NY, USA}, \bibinfo{pages}{458--472}.
\newblock
\showISBNx{978-1-4503-4660-3}
\urldef\tempurl%
\url{https://doi.org/10.1145/3009837.3009895}
\showDOI{\tempurl}


\bibitem[\protect\citeauthoryear{Burckhardt, Baldassin, and Leijen}{Burckhardt
  et~al\mbox{.}}{2010}]%
        {Burckhardt:2010:CPR:1869459.1869515}
\bibfield{author}{\bibinfo{person}{Sebastian Burckhardt},
  \bibinfo{person}{Alexandro Baldassin}, {and} \bibinfo{person}{Daan Leijen}.}
  \bibinfo{year}{2010}\natexlab{}.
\newblock \showarticletitle{Concurrent Programming with Revisions and Isolation
  Types}. In \bibinfo{booktitle}{\emph{Proceedings of the ACM International
  Conference on Object Oriented Programming Systems Languages and
  Applications}} (Reno/Tahoe, Nevada, USA) \emph{(\bibinfo{series}{OOPSLA
  '10})}. \bibinfo{publisher}{ACM}, \bibinfo{address}{New York, NY, USA},
  \bibinfo{pages}{691--707}.
\newblock
\showISBNx{978-1-4503-0203-6}
\urldef\tempurl%
\url{https://doi.org/10.1145/1869459.1869515}
\showDOI{\tempurl}


\bibitem[\protect\citeauthoryear{Burckhardt, Gotsman, Yang, and
  Zawirski}{Burckhardt et~al\mbox{.}}{2014}]%
        {Burckhardt:2014:RDT:2535838.2535848}
\bibfield{author}{\bibinfo{person}{Sebastian Burckhardt},
  \bibinfo{person}{Alexey Gotsman}, \bibinfo{person}{Hongseok Yang}, {and}
  \bibinfo{person}{Marek Zawirski}.} \bibinfo{year}{2014}\natexlab{}.
\newblock \showarticletitle{Replicated Data Types: Specification, Verification,
  Optimality}. In \bibinfo{booktitle}{\emph{Proceedings of the 41st ACM
  SIGPLAN-SIGACT Symposium on Principles of Programming Languages}} (San Diego,
  California, USA) \emph{(\bibinfo{series}{POPL '14})}.
  \bibinfo{publisher}{ACM}, \bibinfo{address}{New York, NY, USA},
  \bibinfo{pages}{271--284}.
\newblock
\showISBNx{978-1-4503-2544-8}
\urldef\tempurl%
\url{https://doi.org/10.1145/2535838.2535848}
\showDOI{\tempurl}


\bibitem[\protect\citeauthoryear{Chairunnanda, Daudjee, and
  \"{O}zsu}{Chairunnanda et~al\mbox{.}}{2014}]%
        {Chairunnanda:2014:CMR:2732967.2732970}
\bibfield{author}{\bibinfo{person}{Prima Chairunnanda},
  \bibinfo{person}{Khuzaima Daudjee}, {and} \bibinfo{person}{M.~Tamer
  \"{O}zsu}.} \bibinfo{year}{2014}\natexlab{}.
\newblock \showarticletitle{ConfluxDB: Multi-master Replication for Partitioned
  Snapshot Isolation Databases}.
\newblock \bibinfo{journal}{\emph{Proc. VLDB Endow.}} \bibinfo{volume}{7},
  \bibinfo{number}{11} (\bibinfo{date}{July} \bibinfo{year}{2014}),
  \bibinfo{pages}{947--958}.
\newblock
\showISSN{2150-8097}
\urldef\tempurl%
\url{https://doi.org/10.14778/2732967.2732970}
\showDOI{\tempurl}


\bibitem[\protect\citeauthoryear{Clements, Kaashoek, Zeldovich, Morris, and
  Kohler}{Clements et~al\mbox{.}}{2015}]%
        {Clements:2015:SCR:2723895.2699681}
\bibfield{author}{\bibinfo{person}{Austin~T. Clements},
  \bibinfo{person}{M.~Frans Kaashoek}, \bibinfo{person}{Nickolai Zeldovich},
  \bibinfo{person}{Robert~T. Morris}, {and} \bibinfo{person}{Eddie Kohler}.}
  \bibinfo{year}{2015}\natexlab{}.
\newblock \showarticletitle{The Scalable Commutativity Rule: Designing Scalable
  Software for Multicore Processors}.
\newblock \bibinfo{journal}{\emph{ACM Trans. Comput. Syst.}}
  \bibinfo{volume}{32}, \bibinfo{number}{4}, Article \bibinfo{articleno}{10}
  (\bibinfo{date}{Jan.} \bibinfo{year}{2015}), \bibinfo{numpages}{47}~pages.
\newblock
\showISSN{0734-2071}
\urldef\tempurl%
\url{https://doi.org/10.1145/2699681}
\showDOI{\tempurl}


\bibitem[\protect\citeauthoryear{Crooks, Pu, Alvisi, and Clement}{Crooks
  et~al\mbox{.}}{2017}]%
        {natacha2017}
\bibfield{author}{\bibinfo{person}{Natacha Crooks}, \bibinfo{person}{Youer Pu},
  \bibinfo{person}{Lorenzo Alvisi}, {and} \bibinfo{person}{Allen Clement}.}
  \bibinfo{year}{2017}\natexlab{}.
\newblock \showarticletitle{Seeing is Believing: A Client-Centric Specification
  of Database Isolation}. In \bibinfo{booktitle}{\emph{Proceedings of the ACM
  Symposium on Principles of Distributed Computing}} (Washington, DC, USA)
  \emph{(\bibinfo{series}{PODC '17})}. \bibinfo{publisher}{Association for
  Computing Machinery}, \bibinfo{address}{New York, NY, USA},
  \bibinfo{pages}{73–82}.
\newblock
\showISBNx{9781450349925}
\urldef\tempurl%
\url{https://doi.org/10.1145/3087801.3087802}
\showDOI{\tempurl}


\bibitem[\protect\citeauthoryear{Crooks, Pu, Estrada, Gupta, Alvisi, and
  Clement}{Crooks et~al\mbox{.}}{2016}]%
        {Crooks:2016:TBA:2882903.2882951}
\bibfield{author}{\bibinfo{person}{Natacha Crooks}, \bibinfo{person}{Youer Pu},
  \bibinfo{person}{Nancy Estrada}, \bibinfo{person}{Trinabh Gupta},
  \bibinfo{person}{Lorenzo Alvisi}, {and} \bibinfo{person}{Allen Clement}.}
  \bibinfo{year}{2016}\natexlab{}.
\newblock \showarticletitle{TARDiS: A Branch-and-Merge Approach To Weak
  Consistency}. In \bibinfo{booktitle}{\emph{Proceedings of the 2016
  International Conference on Management of Data}} (San Francisco, California,
  USA) \emph{(\bibinfo{series}{SIGMOD '16})}. \bibinfo{publisher}{ACM},
  \bibinfo{address}{New York, NY, USA}, \bibinfo{pages}{1615--1628}.
\newblock
\showISBNx{978-1-4503-3531-7}
\urldef\tempurl%
\url{https://doi.org/10.1145/2882903.2882951}
\showDOI{\tempurl}


\bibitem[\protect\citeauthoryear{Databricks}{Databricks}{[n.d.]}]%
        {databricksdeltalake}
\bibfield{author}{\bibinfo{person}{Databricks}.}
  \bibinfo{year}{[n.d.]}\natexlab{}.
\newblock \bibinfo{title}{Introduction to Delta Lake}.
\newblock
  \bibinfo{howpublished}{\url{https://docs.databricks.com/delta/delta-intro.html?_ga=2.85364377.12421025.1582418006-552522563.1582418006}}.
\newblock
\newblock
\shownote{Accessed: 2020-09-22.}


\bibitem[\protect\citeauthoryear{{DoltHub, Inc.}}{{DoltHub, Inc.}}{2021}]%
        {dolt}
\bibfield{author}{\bibinfo{person}{{DoltHub, Inc.}}}
  \bibinfo{year}{2021}\natexlab{}.
\newblock \bibinfo{title}{Dolt}.
\newblock \bibinfo{howpublished}{\url{https://www.dolthub.com/}}.
\newblock


\bibitem[\protect\citeauthoryear{Fitzpatrick}{Fitzpatrick}{2021}]%
        {daff}
\bibfield{author}{\bibinfo{person}{Paul Fitzpatrick}.}
  \bibinfo{year}{2021}\natexlab{}.
\newblock \bibinfo{title}{daff}.
\newblock \bibinfo{howpublished}{\url{https://paulfitz.github.io/daff/}}.
\newblock


\bibitem[\protect\citeauthoryear{Gotsman, Yang, Ferreira, Najafzadeh, and
  Shapiro}{Gotsman et~al\mbox{.}}{2016}]%
        {Gotsman:2016:CIS:2837614.2837625}
\bibfield{author}{\bibinfo{person}{Alexey Gotsman}, \bibinfo{person}{Hongseok
  Yang}, \bibinfo{person}{Carla Ferreira}, \bibinfo{person}{Mahsa Najafzadeh},
  {and} \bibinfo{person}{Marc Shapiro}.} \bibinfo{year}{2016}\natexlab{}.
\newblock \showarticletitle{'Cause I'm Strong Enough: Reasoning About
  Consistency Choices in Distributed Systems}. In
  \bibinfo{booktitle}{\emph{Proceedings of the 43rd Annual ACM SIGPLAN-SIGACT
  Symposium on Principles of Programming Languages}} (St. Petersburg, FL, USA)
  \emph{(\bibinfo{series}{POPL '16})}. \bibinfo{publisher}{ACM},
  \bibinfo{address}{New York, NY, USA}, \bibinfo{pages}{371--384}.
\newblock
\showISBNx{978-1-4503-3549-2}
\urldef\tempurl%
\url{https://doi.org/10.1145/2837614.2837625}
\showDOI{\tempurl}


\bibitem[\protect\citeauthoryear{Gramoli and Guerraoui}{Gramoli and
  Guerraoui}{2014}]%
        {Gramoli:2014:DTP:2541883.2541900}
\bibfield{author}{\bibinfo{person}{Vincent Gramoli} {and}
  \bibinfo{person}{Rachid Guerraoui}.} \bibinfo{year}{2014}\natexlab{}.
\newblock \showarticletitle{Democratizing Transactional Programming}.
\newblock \bibinfo{journal}{\emph{Commun. ACM}} \bibinfo{volume}{57},
  \bibinfo{number}{1} (\bibinfo{date}{Jan.} \bibinfo{year}{2014}),
  \bibinfo{pages}{86--93}.
\newblock
\showISSN{0001-0782}
\urldef\tempurl%
\url{https://doi.org/10.1145/2541883.2541900}
\showDOI{\tempurl}


\bibitem[\protect\citeauthoryear{Guo, Larson, and Ramakrishnan}{Guo
  et~al\mbox{.}}{2005}]%
        {Guo:2005:CGE:1083592.1083647}
\bibfield{author}{\bibinfo{person}{Hongfei Guo}, \bibinfo{person}{Per-Ake
  Larson}, {and} \bibinfo{person}{Raghu Ramakrishnan}.}
  \bibinfo{year}{2005}\natexlab{}.
\newblock \showarticletitle{Caching with "Good Enough" Currency, Consistency,
  and Completeness}. In \bibinfo{booktitle}{\emph{Proceedings of the 31st
  International Conference on Very Large Data Bases}}
  \emph{(\bibinfo{series}{VLDB '05})}. \bibinfo{publisher}{VLDB Endowment},
  \bibinfo{address}{Trondheim, Norway}, \bibinfo{pages}{457--468}.
\newblock
\showISBNx{1-59593-154-6}
\urldef\tempurl%
\url{http://dl.acm.org/citation.cfm?id=1083592.1083647}
\showURL{%
\tempurl}


\bibitem[\protect\citeauthoryear{Guo, Larson, Ramakrishnan, and Goldstein}{Guo
  et~al\mbox{.}}{2004a}]%
        {Guo:2004:RCC:1007568.1007661}
\bibfield{author}{\bibinfo{person}{Hongfei Guo}, \bibinfo{person}{Per-Ake
  Larson}, \bibinfo{person}{Raghu Ramakrishnan}, {and}
  \bibinfo{person}{Jonathan Goldstein}.} \bibinfo{year}{2004}\natexlab{a}.
\newblock \showarticletitle{Relaxed Currency and Consistency: How to Say "Good
  Enough" in SQL}. In \bibinfo{booktitle}{\emph{Proceedings of the 2004 ACM
  SIGMOD International Conference on Management of Data}} (Paris, France)
  \emph{(\bibinfo{series}{SIGMOD '04})}. \bibinfo{publisher}{ACM},
  \bibinfo{address}{New York, NY, USA}, \bibinfo{pages}{815--826}.
\newblock
\showISBNx{1-58113-859-8}
\urldef\tempurl%
\url{https://doi.org/10.1145/1007568.1007661}
\showDOI{\tempurl}


\bibitem[\protect\citeauthoryear{Guo, Larson, Ramakrishnan, and Goldstein}{Guo
  et~al\mbox{.}}{2004b}]%
        {Guo:2004:SRC:1007568.1007706}
\bibfield{author}{\bibinfo{person}{Hongfei Guo}, \bibinfo{person}{Per-Ake
  Larson}, \bibinfo{person}{Raghu Ramakrishnan}, {and}
  \bibinfo{person}{Jonathan Goldstein}.} \bibinfo{year}{2004}\natexlab{b}.
\newblock \showarticletitle{Support for Relaxed Currency and Consistency
  Constraints in MTCache}. In \bibinfo{booktitle}{\emph{Proceedings of the 2004
  ACM SIGMOD International Conference on Management of Data}} (Paris, France)
  \emph{(\bibinfo{series}{SIGMOD '04})}. \bibinfo{publisher}{ACM},
  \bibinfo{address}{New York, NY, USA}, \bibinfo{pages}{937--938}.
\newblock
\showISBNx{1-58113-859-8}
\urldef\tempurl%
\url{https://doi.org/10.1145/1007568.1007706}
\showDOI{\tempurl}


\bibitem[\protect\citeauthoryear{Hellerstein}{Hellerstein}{2010}]%
        {Hellerstein:2010:DIE:1860702.1860704}
\bibfield{author}{\bibinfo{person}{Joseph~M. Hellerstein}.}
  \bibinfo{year}{2010}\natexlab{}.
\newblock \showarticletitle{The Declarative Imperative: Experiences and
  Conjectures in Distributed Logic}.
\newblock \bibinfo{journal}{\emph{SIGMOD Rec.}} \bibinfo{volume}{39},
  \bibinfo{number}{1} (\bibinfo{date}{Sept.} \bibinfo{year}{2010}),
  \bibinfo{pages}{5--19}.
\newblock
\showISSN{0163-5808}
\urldef\tempurl%
\url{https://doi.org/10.1145/1860702.1860704}
\showDOI{\tempurl}


\bibitem[\protect\citeauthoryear{Herlihy and Koskinen}{Herlihy and
  Koskinen}{2008}]%
        {Herlihy:2008:TBM:1345206.1345237}
\bibfield{author}{\bibinfo{person}{Maurice Herlihy} {and} \bibinfo{person}{Eric
  Koskinen}.} \bibinfo{year}{2008}\natexlab{}.
\newblock \showarticletitle{Transactional Boosting: A Methodology for
  Highly-concurrent Transactional Objects}. In
  \bibinfo{booktitle}{\emph{Proceedings of the 13th ACM SIGPLAN Symposium on
  Principles and Practice of Parallel Programming}} (Salt Lake City, UT, USA)
  \emph{(\bibinfo{series}{PPoPP '08})}. \bibinfo{publisher}{ACM},
  \bibinfo{address}{New York, NY, USA}, \bibinfo{pages}{207--216}.
\newblock
\showISBNx{978-1-59593-795-7}
\urldef\tempurl%
\url{https://doi.org/10.1145/1345206.1345237}
\showDOI{\tempurl}


\bibitem[\protect\citeauthoryear{Huang, Xu, Liu, Elmore, and
  Parameswaran}{Huang et~al\mbox{.}}{2017}]%
        {huang2017orpheus}
\bibfield{author}{\bibinfo{person}{Silu Huang}, \bibinfo{person}{Liqi Xu},
  \bibinfo{person}{Jialin Liu}, \bibinfo{person}{Aaron~J Elmore}, {and}
  \bibinfo{person}{Aditya Parameswaran}.} \bibinfo{year}{2017}\natexlab{}.
\newblock \showarticletitle{Orpheus DB: bolt-on versioning for relational
  databases}.
\newblock \bibinfo{journal}{\emph{Proceedings of the VLDB Endowment}}
  \bibinfo{volume}{10}, \bibinfo{number}{10} (\bibinfo{year}{2017}),
  \bibinfo{pages}{1130--1141}.
\newblock


\bibitem[\protect\citeauthoryear{{Iterative, Inc.}}{{Iterative, Inc.}}{2021}]%
        {dvc}
\bibfield{author}{\bibinfo{person}{{Iterative, Inc.}}}
  \bibinfo{year}{2021}\natexlab{}.
\newblock \bibinfo{title}{Dava Version Control}.
\newblock \bibinfo{howpublished}{\url{https://dvc.org/}}.
\newblock


\bibitem[\protect\citeauthoryear{Jordan, Banerjee, and Batman}{Jordan
  et~al\mbox{.}}{1981}]%
        {semanticlock}
\bibfield{author}{\bibinfo{person}{J.~R. Jordan}, \bibinfo{person}{J.
  Banerjee}, {and} \bibinfo{person}{R.~B. Batman}.}
  \bibinfo{year}{1981}\natexlab{}.
\newblock \showarticletitle{Precision Locks}. In
  \bibinfo{booktitle}{\emph{Proceedings of the 1981 ACM SIGMOD International
  Conference on Management of Data}} (Ann Arbor, Michigan)
  \emph{(\bibinfo{series}{SIGMOD '81})}. \bibinfo{publisher}{Association for
  Computing Machinery}, \bibinfo{address}{New York, NY, USA},
  \bibinfo{pages}{143–147}.
\newblock
\showISBNx{0897910400}
\urldef\tempurl%
\url{https://doi.org/10.1145/582318.582340}
\showDOI{\tempurl}


\bibitem[\protect\citeauthoryear{Kraska, Hentschel, Alonso, and
  Kossmann}{Kraska et~al\mbox{.}}{2009}]%
        {Kraska:2009:CRC:1687627.1687657}
\bibfield{author}{\bibinfo{person}{Tim Kraska}, \bibinfo{person}{Martin
  Hentschel}, \bibinfo{person}{Gustavo Alonso}, {and} \bibinfo{person}{Donald
  Kossmann}.} \bibinfo{year}{2009}\natexlab{}.
\newblock \showarticletitle{Consistency Rationing in the Cloud: Pay Only when
  It Matters}.
\newblock \bibinfo{journal}{\emph{Proc. VLDB Endow.}} \bibinfo{volume}{2},
  \bibinfo{number}{1} (\bibinfo{date}{Aug.} \bibinfo{year}{2009}),
  \bibinfo{pages}{253--264}.
\newblock
\showISSN{2150-8097}
\urldef\tempurl%
\url{https://doi.org/10.14778/1687627.1687657}
\showDOI{\tempurl}


\bibitem[\protect\citeauthoryear{Kulkarni, Nguyen, Prountzos, Sui, and
  Pingali}{Kulkarni et~al\mbox{.}}{2011}]%
        {Kulkarni:2011:ECL:1993498.1993562}
\bibfield{author}{\bibinfo{person}{Milind Kulkarni}, \bibinfo{person}{Donald
  Nguyen}, \bibinfo{person}{Dimitrios Prountzos}, \bibinfo{person}{Xin Sui},
  {and} \bibinfo{person}{Keshav Pingali}.} \bibinfo{year}{2011}\natexlab{}.
\newblock \showarticletitle{Exploiting the Commutativity Lattice}. In
  \bibinfo{booktitle}{\emph{Proceedings of the 32Nd ACM SIGPLAN Conference on
  Programming Language Design and Implementation}} (San Jose, California, USA)
  \emph{(\bibinfo{series}{PLDI '11})}. \bibinfo{publisher}{ACM},
  \bibinfo{address}{New York, NY, USA}, \bibinfo{pages}{542--555}.
\newblock
\showISBNx{978-1-4503-0663-8}
\urldef\tempurl%
\url{https://doi.org/10.1145/1993498.1993562}
\showDOI{\tempurl}


\bibitem[\protect\citeauthoryear{{Larson}, {Goldstein}, and {Zhou}}{{Larson}
  et~al\mbox{.}}{2004}]%
        {1319994}
\bibfield{author}{\bibinfo{person}{P.~Ake. {Larson}}, \bibinfo{person}{J.
  {Goldstein}}, {and} \bibinfo{person}{J. {Zhou}}.}
  \bibinfo{year}{2004}\natexlab{}.
\newblock \showarticletitle{MTCache: transparent mid-tier database caching in
  SQL server}. In \bibinfo{booktitle}{\emph{Proceedings. 20th International
  Conference on Data Engineering}}. \bibinfo{publisher}{IEEE Computer Society},
  \bibinfo{address}{Washington, DC, USA}, \bibinfo{pages}{177--188}.
\newblock
\showISSN{1063-6382}
\urldef\tempurl%
\url{https://doi.org/10.1109/ICDE.2004.1319994}
\showDOI{\tempurl}


\bibitem[\protect\citeauthoryear{Li, Leit\~{a}o, Clement, Pregui\c{c}a,
  Rodrigues, and Vafeiadis}{Li et~al\mbox{.}}{2014}]%
        {Li:2014:ACC:2643634.2643664}
\bibfield{author}{\bibinfo{person}{Cheng Li}, \bibinfo{person}{Jo\~{a}o
  Leit\~{a}o}, \bibinfo{person}{Allen Clement}, \bibinfo{person}{Nuno
  Pregui\c{c}a}, \bibinfo{person}{Rodrigo Rodrigues}, {and}
  \bibinfo{person}{Viktor Vafeiadis}.} \bibinfo{year}{2014}\natexlab{}.
\newblock \showarticletitle{Automating the Choice of Consistency Levels in
  Replicated Systems}. In \bibinfo{booktitle}{\emph{Proceedings of the 2014
  USENIX Conference on USENIX Annual Technical Conference}} (Philadelphia, PA)
  \emph{(\bibinfo{series}{USENIX ATC'14})}. \bibinfo{publisher}{USENIX
  Association}, \bibinfo{address}{Berkeley, CA, USA},
  \bibinfo{pages}{281--292}.
\newblock
\showISBNx{978-1-931971-10-2}
\urldef\tempurl%
\url{http://dl.acm.org/citation.cfm?id=2643634.2643664}
\showURL{%
\tempurl}


\bibitem[\protect\citeauthoryear{Li, Porto, Clement, Gehrke, Pregui\c{c}a, and
  Rodrigues}{Li et~al\mbox{.}}{2012}]%
        {Li:2012:MGS:2387880.2387906}
\bibfield{author}{\bibinfo{person}{Cheng Li}, \bibinfo{person}{Daniel Porto},
  \bibinfo{person}{Allen Clement}, \bibinfo{person}{Johannes Gehrke},
  \bibinfo{person}{Nuno Pregui\c{c}a}, {and} \bibinfo{person}{Rodrigo
  Rodrigues}.} \bibinfo{year}{2012}\natexlab{}.
\newblock \showarticletitle{Making Geo-replicated Systems Fast As Possible,
  Consistent when Necessary}. In \bibinfo{booktitle}{\emph{Proceedings of the
  10th USENIX Conference on Operating Systems Design and Implementation}}
  (Hollywood, CA, USA) \emph{(\bibinfo{series}{OSDI'12})}.
  \bibinfo{publisher}{USENIX Association}, \bibinfo{address}{Berkeley, CA,
  USA}, \bibinfo{pages}{265--278}.
\newblock
\showISBNx{978-1-931971-96-6}
\urldef\tempurl%
\url{http://dl.acm.org/citation.cfm?id=2387880.2387906}
\showURL{%
\tempurl}


\bibitem[\protect\citeauthoryear{Li, Pregui{\c c}a, and Rodrigues}{Li
  et~al\mbox{.}}{2018}]%
        {216015}
\bibfield{author}{\bibinfo{person}{Cheng Li}, \bibinfo{person}{Nuno Pregui{\c
  c}a}, {and} \bibinfo{person}{Rodrigo Rodrigues}.}
  \bibinfo{year}{2018}\natexlab{}.
\newblock \showarticletitle{Fine-grained consistency for geo-replicated
  systems}. In \bibinfo{booktitle}{\emph{2018 {USENIX} Annual Technical
  Conference ({USENIX} {ATC} 18)}}. \bibinfo{publisher}{{USENIX} Association},
  \bibinfo{address}{Boston, MA}, \bibinfo{pages}{359--372}.
\newblock
\showISBNx{978-1-931971-44-7}
\urldef\tempurl%
\url{https://www.usenix.org/conference/atc18/presentation/li-cheng}
\showURL{%
\tempurl}


\bibitem[\protect\citeauthoryear{Maddox, Goehring, Elmore, Madden,
  Parameswaran, and Deshpande}{Maddox et~al\mbox{.}}{2016}]%
        {maddox2016decibel}
\bibfield{author}{\bibinfo{person}{Michael Maddox}, \bibinfo{person}{David
  Goehring}, \bibinfo{person}{Aaron~J Elmore}, \bibinfo{person}{Samuel Madden},
  \bibinfo{person}{Aditya Parameswaran}, {and} \bibinfo{person}{Amol
  Deshpande}.} \bibinfo{year}{2016}\natexlab{}.
\newblock \showarticletitle{Decibel: The relational dataset branching system}.
\newblock \bibinfo{journal}{\emph{Proceedings of the VLDB Endowment}}
  \bibinfo{volume}{9}, \bibinfo{number}{9} (\bibinfo{year}{2016}),
  \bibinfo{pages}{624--635}.
\newblock


\bibitem[\protect\citeauthoryear{Narula, Cutler, Kohler, and Morris}{Narula
  et~al\mbox{.}}{2014}]%
        {Narula:2014:PRC:2685048.2685088}
\bibfield{author}{\bibinfo{person}{Neha Narula}, \bibinfo{person}{Cody Cutler},
  \bibinfo{person}{Eddie Kohler}, {and} \bibinfo{person}{Robert Morris}.}
  \bibinfo{year}{2014}\natexlab{}.
\newblock \showarticletitle{Phase Reconciliation for Contended In-memory
  Transactions}. In \bibinfo{booktitle}{\emph{Proceedings of the 11th USENIX
  Conference on Operating Systems Design and Implementation}} (Broomfield, CO)
  \emph{(\bibinfo{series}{OSDI'14})}. \bibinfo{publisher}{USENIX Association},
  \bibinfo{address}{Berkeley, CA, USA}, \bibinfo{pages}{511--524}.
\newblock
\showISBNx{978-1-931971-16-4}
\urldef\tempurl%
\url{http://dl.acm.org/citation.cfm?id=2685048.2685088}
\showURL{%
\tempurl}


\bibitem[\protect\citeauthoryear{Olston, Loo, and Widom}{Olston
  et~al\mbox{.}}{2001}]%
        {Olston:2001:APS:375663.375710}
\bibfield{author}{\bibinfo{person}{Chris Olston}, \bibinfo{person}{Boon~Thau
  Loo}, {and} \bibinfo{person}{Jennifer Widom}.}
  \bibinfo{year}{2001}\natexlab{}.
\newblock \showarticletitle{Adaptive Precision Setting for Cached Approximate
  Values}. In \bibinfo{booktitle}{\emph{Proceedings of the 2001 ACM SIGMOD
  International Conference on Management of Data}} (Santa Barbara, California,
  USA) \emph{(\bibinfo{series}{SIGMOD '01})}. \bibinfo{publisher}{ACM},
  \bibinfo{address}{New York, NY, USA}, \bibinfo{pages}{355--366}.
\newblock
\showISBNx{1-58113-332-4}
\urldef\tempurl%
\url{https://doi.org/10.1145/375663.375710}
\showDOI{\tempurl}


\bibitem[\protect\citeauthoryear{Olston and Widom}{Olston and Widom}{2000}]%
        {Olston:2000:OPT:645926.671877}
\bibfield{author}{\bibinfo{person}{Chris Olston} {and}
  \bibinfo{person}{Jennifer Widom}.} \bibinfo{year}{2000}\natexlab{}.
\newblock \showarticletitle{Offering a Precision-Performance Tradeoff for
  Aggregation Queries over Replicated Data}. In
  \bibinfo{booktitle}{\emph{Proceedings of the 26th International Conference on
  Very Large Data Bases}} \emph{(\bibinfo{series}{VLDB '00})}.
  \bibinfo{publisher}{Morgan Kaufmann Publishers Inc.}, \bibinfo{address}{San
  Francisco, CA, USA}, \bibinfo{pages}{144--155}.
\newblock
\showISBNx{1-55860-715-3}
\urldef\tempurl%
\url{http://dl.acm.org/citation.cfm?id=645926.671877}
\showURL{%
\tempurl}


\bibitem[\protect\citeauthoryear{{Pachyderm Inc.}}{{Pachyderm Inc.}}{2021}]%
        {pachyderm}
\bibfield{author}{\bibinfo{person}{{Pachyderm Inc.}}}
  \bibinfo{year}{2021}\natexlab{}.
\newblock \bibinfo{title}{Pachyderm}.
\newblock \bibinfo{howpublished}{\url{https://www.pachyderm.com/}}.
\newblock


\bibitem[\protect\citeauthoryear{Pu}{Pu}{1992}]%
        {Pu:1992:RLS:506378.506415}
\bibfield{author}{\bibinfo{person}{Calton Pu}.}
  \bibinfo{year}{1992}\natexlab{}.
\newblock \showarticletitle{Relaxing the Limitations of Serializable
  Transactions in Distributed Systems}. In
  \bibinfo{booktitle}{\emph{Proceedings of the 5th Workshop on ACM SIGOPS
  European Workshop: Models and Paradigms for Distributed Systems Structuring}}
  (Mont Saint-Michel, France) \emph{(\bibinfo{series}{EW 5})}.
  \bibinfo{publisher}{ACM}, \bibinfo{address}{New York, NY, USA},
  \bibinfo{pages}{1--6}.
\newblock
\urldef\tempurl%
\url{https://doi.org/10.1145/506378.506415}
\showDOI{\tempurl}


\bibitem[\protect\citeauthoryear{Rinard and Diniz}{Rinard and Diniz}{1997}]%
        {Rinard:1997:CAN:267959.269969}
\bibfield{author}{\bibinfo{person}{Martin~C. Rinard} {and}
  \bibinfo{person}{Pedro~C. Diniz}.} \bibinfo{year}{1997}\natexlab{}.
\newblock \showarticletitle{Commutativity Analysis: A New Analysis Technique
  for Parallelizing Compilers}.
\newblock \bibinfo{journal}{\emph{ACM Trans. Program. Lang. Syst.}}
  \bibinfo{volume}{19}, \bibinfo{number}{6} (\bibinfo{date}{Nov.}
  \bibinfo{year}{1997}), \bibinfo{pages}{942--991}.
\newblock
\showISSN{0164-0925}
\urldef\tempurl%
\url{https://doi.org/10.1145/267959.269969}
\showDOI{\tempurl}


\bibitem[\protect\citeauthoryear{Shang and Yu}{Shang and Yu}{2017}]%
        {ShangY17}
\bibfield{author}{\bibinfo{person}{Zechao Shang} {and}
  \bibinfo{person}{Jeffrey~Xu Yu}.} \bibinfo{year}{2017}\natexlab{}.
\newblock \showarticletitle{My Weak Consistency is Strong}. In
  \bibinfo{booktitle}{\emph{{CIDR} 2017, 8th Biennial Conference on Innovative
  Data Systems Research, Chaminade, CA, USA, January 8-11, 2017, Online
  Proceedings}}. \bibinfo{publisher}{www.cidrdb.org}.
\newblock
\urldef\tempurl%
\url{http://cidrdb.org/cidr2017/papers/p115-shang-cidr17.pdf}
\showURL{%
\tempurl}


\bibitem[\protect\citeauthoryear{Shang, Yu, and Elmore}{Shang
  et~al\mbox{.}}{2018}]%
        {Shang:2018:RRI:3183713.3196932}
\bibfield{author}{\bibinfo{person}{Zechao Shang}, \bibinfo{person}{Jeffrey~Xu
  Yu}, {and} \bibinfo{person}{Aaron~J. Elmore}.}
  \bibinfo{year}{2018}\natexlab{}.
\newblock \showarticletitle{RushMon: Real-time Isolation Anomalies Monitoring}.
  In \bibinfo{booktitle}{\emph{Proceedings of the 2018 International Conference
  on Management of Data}} (Houston, TX, USA) \emph{(\bibinfo{series}{SIGMOD
  '18})}. \bibinfo{publisher}{ACM}, \bibinfo{address}{New York, NY, USA},
  \bibinfo{pages}{647--662}.
\newblock
\showISBNx{978-1-4503-4703-7}
\urldef\tempurl%
\url{https://doi.org/10.1145/3183713.3196932}
\showDOI{\tempurl}


\bibitem[\protect\citeauthoryear{Shapiro, Pregui\c{c}a, Baquero, and
  Zawirski}{Shapiro et~al\mbox{.}}{2011}]%
        {crdt}
\bibfield{author}{\bibinfo{person}{Marc Shapiro}, \bibinfo{person}{Nuno
  Pregui\c{c}a}, \bibinfo{person}{Carlos Baquero}, {and} \bibinfo{person}{Marek
  Zawirski}.} \bibinfo{year}{2011}\natexlab{}.
\newblock \showarticletitle{Conflict-free Replicated Data Types}. In
  \bibinfo{booktitle}{\emph{13th International Conference on Stabilization,
  Safety, and Security of Distributed Systems}} \emph{(\bibinfo{series}{SSS
  2011})}. \bibinfo{publisher}{Springer LNCS volume 6976},
  \bibinfo{pages}{386--400}.
\newblock
\urldef\tempurl%
\url{https://doi.org/10.1007/978-3-642-24550-3_29}
\showDOI{\tempurl}


\bibitem[\protect\citeauthoryear{Shasha, Simon, and Valduriez}{Shasha
  et~al\mbox{.}}{1992}]%
        {chopping}
\bibfield{author}{\bibinfo{person}{Dennis Shasha}, \bibinfo{person}{Eric
  Simon}, {and} \bibinfo{person}{Patrick Valduriez}.}
  \bibinfo{year}{1992}\natexlab{}.
\newblock \showarticletitle{Simple Rational Guidance for Chopping Up
  Transactions}.
\newblock \bibinfo{journal}{\emph{SIGMOD Rec.}} \bibinfo{volume}{21},
  \bibinfo{number}{2} (\bibinfo{date}{June} \bibinfo{year}{1992}),
  \bibinfo{pages}{298--307}.
\newblock
\showISSN{0163-5808}
\urldef\tempurl%
\url{https://doi.org/10.1145/141484.130328}
\showDOI{\tempurl}


\bibitem[\protect\citeauthoryear{Sivaramakrishnan, Kaki, and
  Jagannathan}{Sivaramakrishnan et~al\mbox{.}}{2015}]%
        {Sivaramakrishnan:2015:DPO:2737924.2737981}
\bibfield{author}{\bibinfo{person}{KC Sivaramakrishnan},
  \bibinfo{person}{Gowtham Kaki}, {and} \bibinfo{person}{Suresh Jagannathan}.}
  \bibinfo{year}{2015}\natexlab{}.
\newblock \showarticletitle{Declarative Programming over Eventually Consistent
  Data Stores}. In \bibinfo{booktitle}{\emph{Proceedings of the 36th ACM
  SIGPLAN Conference on Programming Language Design and Implementation}}
  (Portland, OR, USA) \emph{(\bibinfo{series}{PLDI '15})}.
  \bibinfo{publisher}{ACM}, \bibinfo{address}{New York, NY, USA},
  \bibinfo{pages}{413--424}.
\newblock
\showISBNx{978-1-4503-3468-6}
\urldef\tempurl%
\url{https://doi.org/10.1145/2737924.2737981}
\showDOI{\tempurl}


\bibitem[\protect\citeauthoryear{Terry, Theimer, Petersen, Demers, Spreitzer,
  and Hauser}{Terry et~al\mbox{.}}{1995}]%
        {terry1995managing}
\bibfield{author}{\bibinfo{person}{Douglas~B Terry}, \bibinfo{person}{Marvin~M
  Theimer}, \bibinfo{person}{Karin Petersen}, \bibinfo{person}{Alan~J Demers},
  \bibinfo{person}{Mike~J Spreitzer}, {and} \bibinfo{person}{Carl~H Hauser}.}
  \bibinfo{year}{1995}\natexlab{}.
\newblock \showarticletitle{Managing update conflicts in Bayou, a weakly
  connected replicated storage system}.
\newblock \bibinfo{journal}{\emph{ACM SIGOPS Operating Systems Review}}
  \bibinfo{volume}{29}, \bibinfo{number}{5} (\bibinfo{year}{1995}),
  \bibinfo{pages}{172--182}.
\newblock


\bibitem[\protect\citeauthoryear{Tripp, Manevich, Field, and Sagiv}{Tripp
  et~al\mbox{.}}{2012}]%
        {Tripp:2012:JEP:2254064.2254083}
\bibfield{author}{\bibinfo{person}{Omer Tripp}, \bibinfo{person}{Roman
  Manevich}, \bibinfo{person}{John Field}, {and} \bibinfo{person}{Mooly
  Sagiv}.} \bibinfo{year}{2012}\natexlab{}.
\newblock \showarticletitle{JANUS: Exploiting Parallelism via Hindsight}. In
  \bibinfo{booktitle}{\emph{Proceedings of the 33rd ACM SIGPLAN Conference on
  Programming Language Design and Implementation}} (Beijing, China)
  \emph{(\bibinfo{series}{PLDI '12})}. \bibinfo{publisher}{ACM},
  \bibinfo{address}{New York, NY, USA}, \bibinfo{pages}{145--156}.
\newblock
\showISBNx{978-1-4503-1205-9}
\urldef\tempurl%
\url{https://doi.org/10.1145/2254064.2254083}
\showDOI{\tempurl}


\bibitem[\protect\citeauthoryear{Veldhuizen}{Veldhuizen}{2014}]%
        {DBLP:journals/corr/Veldhuizen14}
\bibfield{author}{\bibinfo{person}{Todd~L. Veldhuizen}.}
  \bibinfo{year}{2014}\natexlab{}.
\newblock \showarticletitle{Transaction Repair: Full Serializability Without
  Locks}.
\newblock \bibinfo{journal}{\emph{CoRR}}  \bibinfo{volume}{abs/1403.5645}
  (\bibinfo{year}{2014}).
\newblock
\showeprint[arxiv]{1403.5645}
\urldef\tempurl%
\url{http://arxiv.org/abs/1403.5645}
\showURL{%
\tempurl}


\bibitem[\protect\citeauthoryear{von Koch, Manilov, Vasiladiotis, Cole, and
  Franke}{von Koch et~al\mbox{.}}{2018}]%
        {vonKoch:2018:TCA:3178372.3179513}
\bibfield{author}{\bibinfo{person}{Tobias J. K.~Edler von Koch},
  \bibinfo{person}{Stanislav Manilov}, \bibinfo{person}{Christos Vasiladiotis},
  \bibinfo{person}{Murray Cole}, {and} \bibinfo{person}{Bj\"{o}rn Franke}.}
  \bibinfo{year}{2018}\natexlab{}.
\newblock \showarticletitle{Towards a Compiler Analysis for Parallel
  Algorithmic Skeletons}. In \bibinfo{booktitle}{\emph{Proceedings of the 27th
  International Conference on Compiler Construction}} (Vienna, Austria)
  \emph{(\bibinfo{series}{CC 2018})}. \bibinfo{publisher}{ACM},
  \bibinfo{address}{New York, NY, USA}, \bibinfo{pages}{174--184}.
\newblock
\showISBNx{978-1-4503-5644-2}
\urldef\tempurl%
\url{https://doi.org/10.1145/3178372.3179513}
\showDOI{\tempurl}


\bibitem[\protect\citeauthoryear{Wang, Dinh, Lin, Xie, Zhang, Cai, Chen, Ooi,
  and Ruan}{Wang et~al\mbox{.}}{2018}]%
        {wang2018forkbase}
\bibfield{author}{\bibinfo{person}{Sheng Wang}, \bibinfo{person}{Tien Tuan~Anh
  Dinh}, \bibinfo{person}{Qian Lin}, \bibinfo{person}{Zhongle Xie},
  \bibinfo{person}{Meihui Zhang}, \bibinfo{person}{Qingchao Cai},
  \bibinfo{person}{Gang Chen}, \bibinfo{person}{Beng~Chin Ooi}, {and}
  \bibinfo{person}{Pingcheng Ruan}.} \bibinfo{year}{2018}\natexlab{}.
\newblock \showarticletitle{Forkbase: An efficient storage engine for
  blockchain and forkable applications}.
\newblock \bibinfo{journal}{\emph{Proceedings of the VLDB Endowment}}
  \bibinfo{volume}{11}, \bibinfo{number}{10} (\bibinfo{year}{2018}),
  \bibinfo{pages}{1137--1150}.
\newblock


\bibitem[\protect\citeauthoryear{Wang, Mu, Cui, Yi, Chen, and Li}{Wang
  et~al\mbox{.}}{2016}]%
        {Wang:2016:SMD:2882903.2882934}
\bibfield{author}{\bibinfo{person}{Zhaoguo Wang}, \bibinfo{person}{Shuai Mu},
  \bibinfo{person}{Yang Cui}, \bibinfo{person}{Han Yi}, \bibinfo{person}{Haibo
  Chen}, {and} \bibinfo{person}{Jinyang Li}.} \bibinfo{year}{2016}\natexlab{}.
\newblock \showarticletitle{Scaling Multicore Databases via Constrained
  Parallel Execution}. In \bibinfo{booktitle}{\emph{Proceedings of the 2016
  International Conference on Management of Data}} (San Francisco, California,
  USA) \emph{(\bibinfo{series}{SIGMOD '16})}. \bibinfo{publisher}{ACM},
  \bibinfo{address}{New York, NY, USA}, \bibinfo{pages}{1643--1658}.
\newblock
\showISBNx{978-1-4503-3531-7}
\urldef\tempurl%
\url{https://doi.org/10.1145/2882903.2882934}
\showDOI{\tempurl}


\bibitem[\protect\citeauthoryear{{Weihl}}{{Weihl}}{1988}]%
        {9728}
\bibfield{author}{\bibinfo{person}{W.~E. {Weihl}}.}
  \bibinfo{year}{1988}\natexlab{}.
\newblock \showarticletitle{Commutativity-based concurrency control for
  abstract data types}.
\newblock \bibinfo{journal}{\emph{IEEE Trans. Comput.}} \bibinfo{volume}{37},
  \bibinfo{number}{12} (\bibinfo{date}{Dec} \bibinfo{year}{1988}),
  \bibinfo{pages}{1488--1505}.
\newblock
\showISSN{0018-9340}
\urldef\tempurl%
\url{https://doi.org/10.1109/12.9728}
\showDOI{\tempurl}


\bibitem[\protect\citeauthoryear{Weihl}{Weihl}{1989}]%
        {10.1145/63264.63518}
\bibfield{author}{\bibinfo{person}{W.~E. Weihl}.}
  \bibinfo{year}{1989}\natexlab{}.
\newblock \showarticletitle{Local Atomicity Properties: Modular Concurrency
  Control for Abstract Data Types}.
\newblock \bibinfo{journal}{\emph{ACM Trans. Program. Lang. Syst.}}
  \bibinfo{volume}{11}, \bibinfo{number}{2} (\bibinfo{date}{April}
  \bibinfo{year}{1989}), \bibinfo{pages}{249–282}.
\newblock
\showISSN{0164-0925}
\urldef\tempurl%
\url{https://doi.org/10.1145/63264.63518}
\showDOI{\tempurl}


\bibitem[\protect\citeauthoryear{Whittaker and Hellerstein}{Whittaker and
  Hellerstein}{2018}]%
        {Whittaker:2018:ICC:3275536.3300966}
\bibfield{author}{\bibinfo{person}{Michael Whittaker} {and}
  \bibinfo{person}{Joseph~M. Hellerstein}.} \bibinfo{year}{2018}\natexlab{}.
\newblock \showarticletitle{Interactive Checks for Coordination Avoidance}.
\newblock \bibinfo{journal}{\emph{Proc. VLDB Endow.}} \bibinfo{volume}{12},
  \bibinfo{number}{1} (\bibinfo{date}{Sept.} \bibinfo{year}{2018}),
  \bibinfo{pages}{14--27}.
\newblock
\showISSN{2150-8097}
\urldef\tempurl%
\url{https://doi.org/10.14778/3275536.3275538}
\showDOI{\tempurl}


\bibitem[\protect\citeauthoryear{Wu, Yu, and Pu}{Wu et~al\mbox{.}}{1992}]%
        {Wu:1992:DCE:645477.654631}
\bibfield{author}{\bibinfo{person}{Kun-Lung Wu}, \bibinfo{person}{Philip~S.
  Yu}, {and} \bibinfo{person}{Calton Pu}.} \bibinfo{year}{1992}\natexlab{}.
\newblock \showarticletitle{Divergence Control for Epsilon-Serializability}. In
  \bibinfo{booktitle}{\emph{Proceedings of the Eighth International Conference
  on Data Engineering}}. \bibinfo{publisher}{IEEE Computer Society},
  \bibinfo{address}{Washington, DC, USA}, \bibinfo{pages}{506--515}.
\newblock
\showISBNx{0-8186-2545-7}
\urldef\tempurl%
\url{http://dl.acm.org/citation.cfm?id=645477.654631}
\showURL{%
\tempurl}


\bibitem[\protect\citeauthoryear{Wu, Yu, and Pu}{Wu et~al\mbox{.}}{1997}]%
        {Wu:1997:DCA:627309.627830}
\bibfield{author}{\bibinfo{person}{Kun-Lung Wu}, \bibinfo{person}{Philip~S.
  Yu}, {and} \bibinfo{person}{Calton Pu}.} \bibinfo{year}{1997}\natexlab{}.
\newblock \showarticletitle{Divergence Control Algorithms for Epsilon
  Serializability}.
\newblock \bibinfo{journal}{\emph{IEEE Trans. on Knowl. and Data Eng.}}
  \bibinfo{volume}{9}, \bibinfo{number}{2} (\bibinfo{date}{March}
  \bibinfo{year}{1997}), \bibinfo{pages}{262--274}.
\newblock
\showISSN{1041-4347}
\urldef\tempurl%
\url{https://doi.org/10.1109/69.591451}
\showDOI{\tempurl}


\bibitem[\protect\citeauthoryear{Wu, Chan, and Tan}{Wu et~al\mbox{.}}{2016}]%
        {Wu:2016:THS:2882903.2915202}
\bibfield{author}{\bibinfo{person}{Yingjun Wu}, \bibinfo{person}{Chee-Yong
  Chan}, {and} \bibinfo{person}{Kian-Lee Tan}.}
  \bibinfo{year}{2016}\natexlab{}.
\newblock \showarticletitle{Transaction Healing: Scaling Optimistic Concurrency
  Control on Multicores}. In \bibinfo{booktitle}{\emph{Proceedings of the 2016
  International Conference on Management of Data}} (San Francisco, California,
  USA) \emph{(\bibinfo{series}{SIGMOD '16})}. \bibinfo{publisher}{ACM},
  \bibinfo{address}{New York, NY, USA}, \bibinfo{pages}{1689--1704}.
\newblock
\showISBNx{978-1-4503-3531-7}
\urldef\tempurl%
\url{https://doi.org/10.1145/2882903.2915202}
\showDOI{\tempurl}


\bibitem[\protect\citeauthoryear{Xie, Su, Littley, Alvisi, Kapritsos, and
  Wang}{Xie et~al\mbox{.}}{2015}]%
        {Xie:2015:HAV:2815400.2815430}
\bibfield{author}{\bibinfo{person}{Chao Xie}, \bibinfo{person}{Chunzhi Su},
  \bibinfo{person}{Cody Littley}, \bibinfo{person}{Lorenzo Alvisi},
  \bibinfo{person}{Manos Kapritsos}, {and} \bibinfo{person}{Yang Wang}.}
  \bibinfo{year}{2015}\natexlab{}.
\newblock \showarticletitle{High-performance ACID via Modular Concurrency
  Control}. In \bibinfo{booktitle}{\emph{Proceedings of the 25th Symposium on
  Operating Systems Principles}} (Monterey, California)
  \emph{(\bibinfo{series}{SOSP '15})}. \bibinfo{publisher}{ACM},
  \bibinfo{address}{New York, NY, USA}, \bibinfo{pages}{279--294}.
\newblock
\showISBNx{978-1-4503-3834-9}
\urldef\tempurl%
\url{https://doi.org/10.1145/2815400.2815430}
\showDOI{\tempurl}


\bibitem[\protect\citeauthoryear{{Zellag} and {Kemme}}{{Zellag} and
  {Kemme}}{2011}]%
        {5767927}
\bibfield{author}{\bibinfo{person}{K. {Zellag}} {and} \bibinfo{person}{B.
  {Kemme}}.} \bibinfo{year}{2011}\natexlab{}.
\newblock \showarticletitle{Real-time quantification and classification of
  consistency anomalies in multi-tier architectures}. In
  \bibinfo{booktitle}{\emph{2011 IEEE 27th International Conference on Data
  Engineering}}. \bibinfo{publisher}{IEEE Computer Society},
  \bibinfo{address}{Washington, DC, USA}, \bibinfo{pages}{613--624}.
\newblock
\showISSN{2375-026X}
\urldef\tempurl%
\url{https://doi.org/10.1109/ICDE.2011.5767927}
\showDOI{\tempurl}


\bibitem[\protect\citeauthoryear{Zellag and Kemme}{Zellag and Kemme}{2012}]%
        {Zellag:2012:CYC:2391229.2391235}
\bibfield{author}{\bibinfo{person}{Kamal Zellag} {and} \bibinfo{person}{Bettina
  Kemme}.} \bibinfo{year}{2012}\natexlab{}.
\newblock \showarticletitle{How Consistent is Your Cloud Application?}. In
  \bibinfo{booktitle}{\emph{Proceedings of the Third ACM Symposium on Cloud
  Computing}} (San Jose, California) \emph{(\bibinfo{series}{SoCC '12})}.
  \bibinfo{publisher}{ACM}, \bibinfo{address}{New York, NY, USA}, Article
  \bibinfo{articleno}{6}, \bibinfo{numpages}{14}~pages.
\newblock
\showISBNx{978-1-4503-1761-0}
\urldef\tempurl%
\url{https://doi.org/10.1145/2391229.2391235}
\showDOI{\tempurl}


\bibitem[\protect\citeauthoryear{Zellag and Kemme}{Zellag and Kemme}{2014}]%
        {Zellag:2014:CAM:2581628.2581630}
\bibfield{author}{\bibinfo{person}{Kamal Zellag} {and} \bibinfo{person}{Bettina
  Kemme}.} \bibinfo{year}{2014}\natexlab{}.
\newblock \showarticletitle{Consistency Anomalies in Multi-tier Architectures:
  Automatic Detection and Prevention}.
\newblock \bibinfo{journal}{\emph{The VLDB Journal}} \bibinfo{volume}{23},
  \bibinfo{number}{1} (\bibinfo{date}{Feb.} \bibinfo{year}{2014}),
  \bibinfo{pages}{147--172}.
\newblock
\showISSN{1066-8888}
\urldef\tempurl%
\url{https://doi.org/10.1007/s00778-013-0318-x}
\showDOI{\tempurl}


\bibitem[\protect\citeauthoryear{Zhang, Power, Zhou, Sovran, Aguilera, and
  Li}{Zhang et~al\mbox{.}}{2013}]%
        {Zhang:2013:TCA:2517349.2522729}
\bibfield{author}{\bibinfo{person}{Yang Zhang}, \bibinfo{person}{Russell
  Power}, \bibinfo{person}{Siyuan Zhou}, \bibinfo{person}{Yair Sovran},
  \bibinfo{person}{Marcos~K. Aguilera}, {and} \bibinfo{person}{Jinyang Li}.}
  \bibinfo{year}{2013}\natexlab{}.
\newblock \showarticletitle{Transaction Chains: Achieving Serializability with
  Low Latency in Geo-distributed Storage Systems}. In
  \bibinfo{booktitle}{\emph{Proceedings of the Twenty-Fourth ACM Symposium on
  Operating Systems Principles}} (Farminton, Pennsylvania)
  \emph{(\bibinfo{series}{SOSP '13})}. \bibinfo{publisher}{ACM},
  \bibinfo{address}{New York, NY, USA}, \bibinfo{pages}{276--291}.
\newblock
\showISBNx{978-1-4503-2388-8}
\urldef\tempurl%
\url{https://doi.org/10.1145/2517349.2522729}
\showDOI{\tempurl}


\end{thebibliography}

\end{document}